\documentclass[12pt]{book}

\usepackage{delarray}
\usepackage{rotating}
\usepackage{general_cite}
\usepackage{caption}
\usepackage{psfig}
\usepackage{bm}       
\usepackage{setspace}
\usepackage{amssymb}
\usepackage{amsbsy}   
\usepackage{amsmath}
\usepackage{threeparttable}
\usepackage{lscape}
\usepackage{graphics}
\usepackage{footmisc}

\newcommand{\degr}{\mbox{$^\circ$}}

\newcommand\fs{\mbox{$.\!\!^{\mathrm s}$}}
\newcommand\arcmin{\mbox{$^\prime$}}

\newcommand\farcs{\mbox{$.\!\!^{\prime\prime}$}}

\newcommand\la\lesssim
\newcommand\ga\gtrsim

\newcommand{\captionfonts}{\footnotesize\bf}
\makeatletter  
\long\def\@makecaption#1#2{%
  \vskip\abovecaptionskip
  \sbox\@tempboxa{{\captionfonts #1: #2}}%
  \ifdim \wd\@tempboxa >\hsize
    {\captionfonts #1: #2\par}
  \else
    \hbox to\hsize{\hfil\box\@tempboxa\hfil}%
  \fi
  \vskip\belowcaptionskip}
\makeatother   

\newcommand{\citep}[1]{\cite{#1}}
\newcommand{\citei}[2]{(#1 \pcite{#2})}

\newcommand{\citet}[1]{\scite{#1}}
\newcommand{\citealt}[1]{\pcite{#1}}

\newcommand\nodata\ldots

\begin{document}

\pagestyle{empty}


\frontmatter

\begin{titlepage}
\begin{center}

{\textsc{\Huge\bf
Long-Term Timing \\
of Millisecond Pulsars \\
and Gravitational Wave Detection \\ }}

\vspace{4cm}

{\Large\bf Joris Paul Wim Verbiest} \\

\vspace{2cm}

\centerline{\psfig{figure=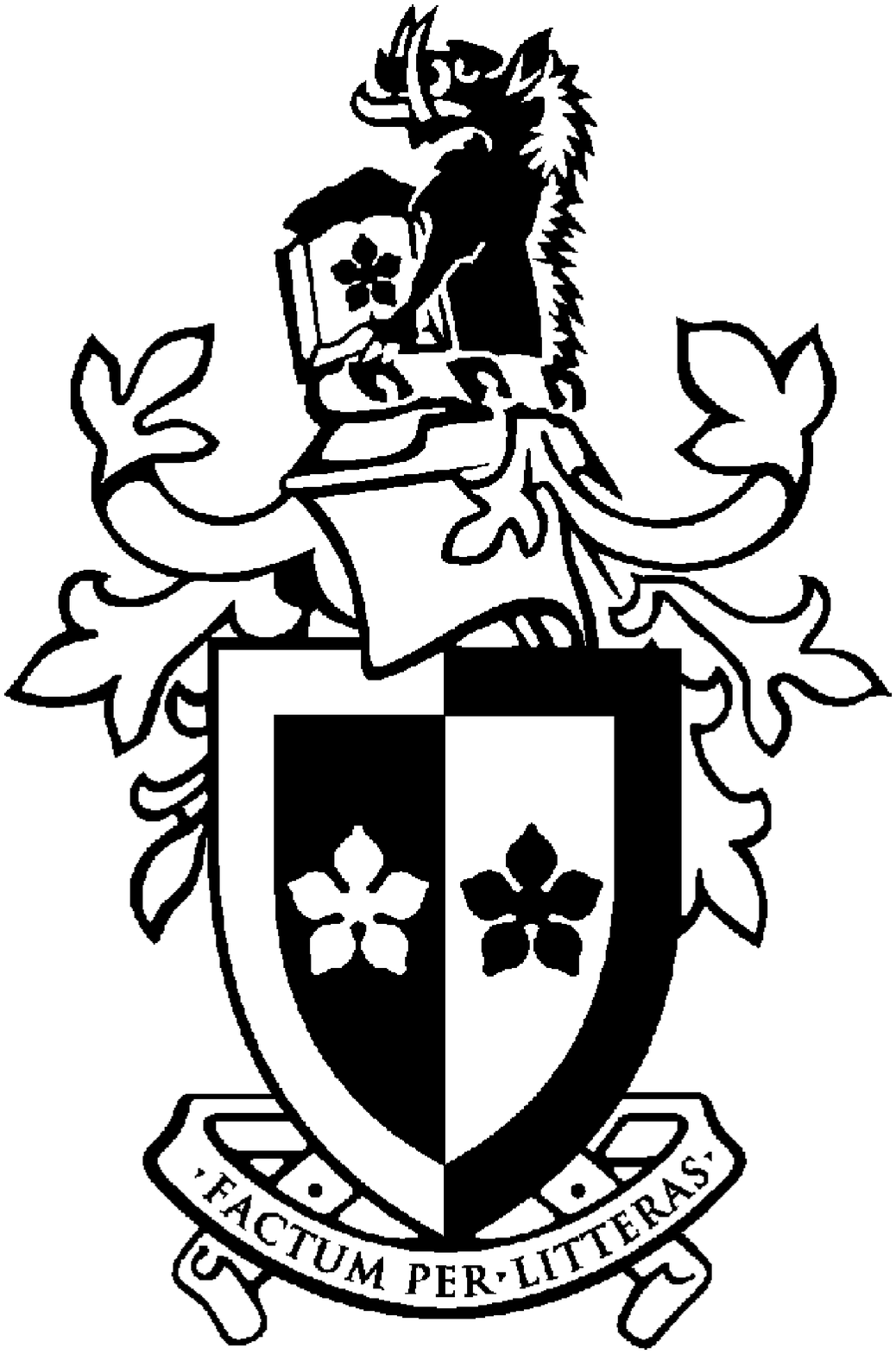,width=4cm}}

\vspace{2cm}

{\it A dissertation \\
Presented in fulfillment of the requirements\\
for the degree of Doctor of Philosophy\\
at Swinburne University of Technology\\
}

\vspace{1cm} {\it January 2009}

\end{center} 
\end{titlepage}

\chapter*{ }
\vspace*{4cm}

\noindent \textsf{``I know nothing, except the fact of my ignorance.''}

\vspace*{1.5 cm}

\textit{Socrates}, as cited by \textit{Diogenes Laertius} in \textsf{``The
Lives and Opinions of Eminent Philosophers''}


\chapter*{Abstract}

This thesis presents the results from a long-term timing campaign on
20 millisecond pulsars (MSPs). The stability of these pulsars is
analysed in order to allow assessment of gravitational wave (GW)
detection efforts through pulsar timing. In addition, we present a new
method of limiting the amplitude of a stochastic background of GWs and
derive a strong limit from applying this method to our data.

GWs are a prediction of general relativity (GR) that has thus far only
been confirmed indirectly. While a direct detection could give
important evidence of GW properties and provide insight into the
processes that are predicted to generate these waves, a detection that
contradicts GR might herald a breakthrough in gravitational theory
and fundamental science. Two types of projects are currently being
undertaken to make the first direct detection of GWs. One of these
uses ground-based interferometers to detect the GW-induced space-time
curvature, the other uses pulsar timing. This thesis is concerned with
the latter: the Pulsar Timing Arrays (PTAs).

The high stability of some MSPs, along with ever increasing levels of
timing precision, has been predicted to enable detection of GW effects
on the Earth. Specifically, it has been shown that if the timing
precision on 20 MSPs can be maintained at levels of $\sim$100\,ns
during five years to a decade, a correlated effect owing to GWs from
predicted cosmic origins, can be detected. However, no timing at a
precision of 100\,ns has been maintained for more than a few years -
and only on a few pulsars.

After combining archival data and employing state-of-the-art
calibration methods, we achieved 200\,ns timing precision over
10\,years on PSR J0437$-$4715 - which is a record at such time
scales. This high stability in itself provides several interesting
measurements, for example of the variation of Newton's gravitational
constant and of the pulsar mass.

We also present long-term timing results on 19 other pulsars that
constitute the Parkes PTA. Our results show that most pulsars in our
sample are stable and dominated by receiver noise. The potential for
sub-100\,ns timing is demonstrated on two of our brightest
sources. These timing results are used to estimate timescales for GW
detection of potential PTAs worldwide and to limit the amplitude of
GWs in the data. Our limit of $A< 1.0\times 10^{-14}$ for a
background with $\alpha = -2/3$ is slightly more stringent than the
best limit published yet.

\chapter*{Acknowledgments}

This thesis would not have been if Matthew and Dick hadn't trusted me
to bring it to a good end. I'm not sure how I deserved that trust
nearly four years ago, but I hope to have earned it by now, because
both of my supervisors have provided me with just about anything a
student could hope for. First and foremost, it was an honour and a
pleasure to find myself steeped into a project as exciting and
fascinating as P140 and the PPTA. At the outset I didn't know enough
about pulsar timing to fully grasp it, but through the years I have
continuously increased my appreciation of the project and
collaborators I have been offered - and for which I may not have shown
enough gratitude at the time. Beyond the greatness of the project, I
should of course acknowledge the supervision and general support I
have received throughout the past 42 months. The patient explanations,
discussions that often went on for much longer than they needed to and
the freedom to let me find my own way - all of it is very much
appreciated. As my primary supervisor, I thank Matthew for frequently
pulling away from the busy schedule he has, to provide the pearls of
wisdom and daring guidance I needed to progress.

Academically speaking, the post-docs come after the
professors. Invaluable to students like myself, they taught me all I
needed to know with endless patience and have as such provided me with
a great example of what I should aspire to become after graduation. In
addition to a whole lot more scientific and less-than-scientific
stuff, George (who doesn't believe in acknowledgements, so I'll be
brief) taught me C in just about a day - it sounds like a miracle and
I'm not sure how he managed it. Willem was never annoyed or irritated
when I came up with yet another PSRCHIVE bug that turned out to be a
typo on my behalf - or when I asked him the same question over and
over again, fundamentally coming down to ``explain me polarisation and
don't use math''. Ramesh's knowledge about the ISM turned out to be a
boundless source of information which helped me understand the
background to what I was actually doing. He would - like just about
every pulsar astronomer I've come to know - take as much time as I
could possibly ask for to get the final tiny misunderstandings I might
have ironed out, no matter how boring or basic my problems may have
seemed to him.

I only met Bill Coles halfway through my thesis but his influence is
undeniable. It took me a while to get used to his engineering approach
and his e-mails full of mathematical, spectral and statistical jargon
generally took me three hours to read and three more hours to (mostly)
understand, but they were always useful and never dull. Rick Jenet,
Simon Johnston, David Champion, Andrea Lommen, Barney Rickett, Russell
Edwards, Dan Stinebring, Mark Walker and Steve Ord should also be
mentioned in the category of great people who sadly live far away (at
times) but are always willing to help a simple student wrap his mind
around something finicky.

Another group of people that has been of vital importance to my thesis
and my appreciation of pulsar astronomy in specific and radio
astronomy in general, is the Parkes staff. I haven't calculated what
percentage of ``my'' data was obtained by John Sarkissian, but it must
be heaps. John was also almost always there when I was observing -
sacrificing sleep and (more importantly) family time so that I could
leave for lunch. John and I may disagree on a few worldly issues, but
that doesn't deny the fact that I will longingly look back at the
lively debates we had in the control room - and the great 3D pictures
of just about any interesting place in the Solar System. In the past
decade ``Parkes'' has nearly become synonymous with John Reynolds. The
indefatigable attitude JR displayed in trying to fix anything that
\emph{might} negatively affect observations was astonishing and I,
like all other observers, am utterly grateful. Beyond the two Johns,
the entire Parkes staff has been more than helpful and accommodating
in attempting to create a flawless scientific instrument and a home
away from home. Thanks therefore to all of you: Andrew, Anne,
Brett$^2$, Geoff, Gina, Janette, Julia, Mal, Shirley, Simon, Stacy and
everyone I've forgotton to mention.

At Swinburne as in Sydney, work only goes so far in turning
``survival'' into ``life''. Luckily there were many friends along the
way, some long gone, others just arrived, who provide some comic
relief to give my brain the occasional rest. Thanks therefore to my
housemates: Simon, Nadia, Meg, Paul, Elaine, Thomas and Lenneke for
introducing me to the best TV series I know and for the discussions
and insights in subjects as varied as sub-atomic physics, Australian
culture and German cuisine (though mostly sub-atomic physics). Thanks
also to my fellow students and office-mates (both in Swinburne and at
the ATNF): Xiao Peng, Trevor, Tim, Sarah, Paul, Nick, Natasha,
Meredith, Max$^2$, Lina, Lee, Kathryn, Jeremy, Haydon, Emily, Emil,
Daniel, Chris, Caroline, Berkeley, Annie, Anneke, Andy, Alyson,
Albert, Adrian and Adam for both encouraging and preventing
procrastination, for lunch and dinner, for the Age Superquiz, for
telling me more than I needed to hear about AFL and rugby. For poker,
movies, music. For drinks and laughter. For taking me seriously, but
not too seriously. For camping and hikes. For sleep-deprived comedy at
4am in the Parkes control room. And for pointing out the
obvious. Apart from housemates and fellow students, the frisbee crowd
provided me with an energy release that money cannot buy. Thanks for
putting up with my galloping across your field, guys.

Last but not least, thanks to the people I've been neglecting most of
all: my father, who predicted twenty years ago that I'll become a
particle physicist; my mother, who keeps on defying the Universe in
her claim that Australia is a long way from Belgium; and my siblings,
Kathleen \& Maarten, who are in their own way close while
distant. It's the four of you - and the eighteen years I've lived with
you in Belgium, that have provided me with the dauntless international
vision that brought me to the other end of the world and with the
scientific intrigue that made me want to understand the Universe and,
above all, gravity.

\chapter*{Declaration}

This thesis contains no material that has been accepted for the award
of any other degree or diploma. To the best of my knowledge, this
thesis contains no material previously published or written by another
author, except where due reference is made in the text of the
thesis. All work presented is primarily that of the author. Chapters
\ref{chap:0437} and \ref{chap:20PSRS} have appeared in or been
submitted to refereed journals. In each case, I authored the majority
of the text, receiving limited assistance with the style and content
of some sections. My supervisors, close colleagues and a number of
anonymous referees helped to refine each manuscript. Appendix
\ref{app:Sens} is submitted to a refereed journal as appendix to the
article containing Chapter \ref{chap:20PSRS}. The mathematical
derivation presented in this Appendix was constructed with guidance
from William Coles. The research in Chapter \ref{chap:GWBLimit} was
performed in close collaboration with George Hobbs, William Coles and
Andrea Lommen and most of its contents will be duplicated in a paper
to be written by one of these three collaborators. Chapter
\ref{chap:GWBLimit} itself, though, is written independently and
authored by myself alone.

This thesis analyses data recorded at the Parkes radio telescope
during a 15\,year period. The relevant observing proposals (P140 and
P456) are both still ongoing. During my time at Swinburne,
observations for these projects were made by a team of astronomers
including mostly Matthew Bailes, Willem van Straten, Ramesh Bhat,
Albert Teoh, Sarah Burke-Spolaor and myself from Swinburne University
and Richard Manchester, George Hobbs, John Sarkissian, Russell
Edwards, David Champion and Daniel Yardley from the Australia
Telescope National Facility (ATNF). John Reynolds, Brett Dawson and
Brett Preisig have assisted with telescope operations. Aidan Hotan and
Steve Ord have supported the running of CPSR2 when required and
provided occasional observing assistance.

Support with debugging and development of the PSRCHIVE software scheme
was provided by Willem van Straten and assistance with the
\textsc{Tempo2} software was provided by George Hobbs.

\vspace{2cm}
\noindent Joris Paul Wim Verbiest, January 2009


\tableofcontents
\listoffigures
\listoftables


\mainmatter

\pagestyle{headings}

\chapter[Introduction]{Introduction: On Pulsars and Gravity}
\label{chap:intro}
\noindent \textsf{With all reserve we advance the view that a
  super-nova represents the transition of an ordinary star into a
  \emph{neutron star}, consisting mainly of neutrons.\\}
\vspace{0.25cm}
\textit{Baade \& Zwicky, ``Cosmic Rays from Super-novae'', Proceedings
  of the National Academy of Sciences, 1934}
\vspace{1.5cm}

When \citet{hbp+68} serendipitously discovered the first pulsar in
1967, they hypothesised these new objects might be related to neutron
stars - dense remnants of supernova explosions, first proposed by
\citet{bz34c}. \citet{gol68} and \citet{pac68} further investigated
the properties of pulsar emission and first proposed the ``lighthouse
model'' as an explanation for the regularity of the pulses. According
to this model, the radio waves originate from near the magnetic poles,
which are offset from the rotation axis of the neutron star - causing
the emission to sweep through space like that generated by a
lighthouse. While this model provides a geometric explanation to many
characteristics, the actual mechanism that generates the pulsar
radiation remained unexplained. \citet{gj69} and \citet{rs75}
analysed several possibilities, leading to a simplified
electrodynamic model of pulsar radiation. This allowed the now
standard classification of pulsars according to characteristic age and
surface magnetic field strength, based on the assumption of magnetic
dipole radiation \citep{og69,cr93a}.

Within a decade of the initial pulsar discovery, \citet{ht75a}
identified the first pulsar in a binary system. This particular
system, PSR B1913+16, consists of two neutron stars in close orbit
around each other. It consequentially exhibits many gravitational
effects that are predicted by general relativity but had thus far been
impossible to measure \citep{sb76}. One of these effects was the
emission of gravitational radiation. \citet{tw82} used the PSR
B1913+16 binary to make the first indirect detection of gravitational
waves caused by the centripetal acceleration in the binary
system. Also in 1982, \citet{bkh+82} discovered yet another type of
radio pulsar: the first millisecond pulsar PSR B1937+21, with a
rotational frequency of well over 600\,Hz. Since evolutionary
scenarios for this highly ``spun-up'' type of pulsar required a binary
system \citep{bv91}, the fact that PSR B1937+21 is a single pulsar
defies common theories. Potential alternatives were proposed by
\citet{hv83a} and \citet{rs83}, but are hard to verify. Yet another
class of related objects was discovered a decade later by
\citet{dt92a}: the magnetars. These more slowly rotating objects
(pulse periods up to 10\,s or more) have higher magnetic field
strengths than radio pulsars. Accepted models had suggested that radio
wave production would not occur at such long periods and high magnetic
field strengths, which was in agreement with the fact that magnetars
were only observable in X-rays and $\gamma$-rays. Recently, though,
\citet{crh+06} did detect radio pulses from magnetar AXP XTE
J1810$-$197, in contradiction to models for radio pulsar emission
mechanisms. Another few recent discoveries that demonstrate the
complexity of the pulsar emission mechanism, are the intermittent
pulsars \citep{klo+06} and the rotating radio transients
\citei{RRATs;}{mll+06}, both of which behave like normal pulsars at
some times, but are entirely undetectable at others.

This chapter will provide an introduction into pulsars, their basic
characteristics and how they can be used to study, amongst other
things, gravity. Section \ref{intro:psrs} gives an overview of the
origins and typical characteristics of both common and millisecond
pulsars. The technique of pulsar timing is described in Section
\ref{intro:pt}, along with its application to measuring properties of
pulsars, the interstellar medium and the Solar System alike. In
Section \ref{intro:pta}, we will look more closely at the effect
gravitational waves have on pulsar timing and how this effect may be
detected. Finally, Section \ref{intro:struct} provides the outline for
the rest of the thesis.

\section{Pulsars and Millisecond Pulsars}
\label{intro:psrs}
\subsection{Birth of a Neutron Star}\label{intro:Birth}
Most stars are in a state of hydrostatic equilibrium provided by the
gravity of their own mass on one hand and nuclear synthesis on the
other. A fundamental requirement for nuclear synthesis is that atoms
come close enough together for the nuclear binding force (the
\emph{strong nuclear force}) to overtake the electromagnetic repulsive
force. Since the nuclear charge of heavier atoms is higher, this
condition requires more pressure and higher temperatures for fusion of
heavier elements. Therefore, as ever heavier elements are fused, the
core of the star is expected to gradually contract, allowing a
pressure gradient to form, resulting in different stages of nuclear
fusion at different distances from the centre of the star. If the star
is massive enough \citei{$\gtrsim10\,{\rm M}_{\odot}$;}{pri00}, this
chain reaction should eventually result in the production of
iron. Since iron is the element with the highest binding energy per
nucleon, fusion beyond iron will require energy rather than release
energy, meaning that the equilibrium cannot be sustained by further
fusion. As a result, an iron core is predicted to grow at the centre
of the star, sustained by electron degeneracy pressure (which is a
quantum-mechanical pressure that follows from the Pauli exclusion
principle). This pressure can, however, only sustain bodies of masses
up to the Chandrasekhar mass ($\sim1.4\,{\rm M}_{\odot}$;
\citealt{cha31}). Once the core grows beyond this mass, it collapses
under its own weight, causing dramatic rises in temperature and
pressure, which causes the iron to dissociate back into single
nucleons - and after that, fusing protons and electrons into
neutrons. According to present theories, this collapse either
continues until a body of infinite density is created (a \emph{black
hole}) or until it is halted by neutron degeneracy pressure (which is
similar in principle to electron degeneracy pressure, but exists at
higher densities). In this latter case the implosion is suddenly
halted, causing the mantle and outer layers of the star to be pushed
outward in a giant shockwave. This cataclysmic collapse is known as a
core-collapse supernova explosion; the material that is expelled in
the shock wave evolves into a supernova remnant (a well-known example
being the Crab nebula) and the core that stays behind, is a neutron
star or - if observable - a pulsar \citep{bz34c}.

\subsection{Discovery and Fundamental Properties}

When \citet{bz34c} predicted the existence of neutron stars,
observational evidence in support of this work was not expected
because no mechanism to create radiation was known, besides blackbody
radiation which would vanish quickly as the star cooled. However, the
discovery of radio pulsations by \citet{hbp+68}, was soon linked to
neutron stars by \citet{gol68} and \citet{pac68}. They argued that,
given the short and extremely accurate periodicities of the observed
radio pulsations, neutron stars were the only likely source of this
radiation. Furthermore, they hypothesized that the radio emission
would originate from compact regions within the pulsar magnetosphere
which rotates with the pulsar, thus creating a rotating beam of
radiation. If the radiation beam intersects our line of sight at any
point of the pulsar's rotation, we will see periodic pulses.
Proposed in 1968, this model (known as the lighthouse model) still
lies at the basis of our understanding of pulsars, but fails to
explain the physical process that generates the radiation. Many
attempts have been made to identify the mechanism that creates the
radiation \citei{see e.g.}{gj69,rs75}, or to pin down the precise
region from which the radiation emanates, but no conclusive argument
has been presented \citep{mel04}. The main difficulties for any
comprehensive explanation of pulsar emission are the similarity of
emission characteristics for wide ranges of pulse period and magnetic
field strength - as evidenced by the largely comparable pulse profiles
and polarisation properties of most pulsars. Other problems are the
coherency of the radio emission and the broad spectrum of the emission
- with coherent radio emission at frequencies as low as several
hundred MHz (or lower) and up to 10\,GHz or higher. Non-standard
behaviour seen in many of the slower and some of the fast pulsars,
such as drifting subpulses, periodic nulling, giant pulses or sudden
spin-ups, introduce further constraints on this complicated analysis.

Gold also predicted that a slight decrease in spin frequency would be
detected as a consequence of energy loss. Pulsars do indeed lose
energy due to magnetic dipole radiation (caused by the rotation of the
pulsar's magnetic field) and classical electromagnetic theory can
therefore relate the loss of angular momentum to the strength of the
pulsar's magnetic field, as derived in \citet{mt77} after
\citet{gj69}:
\begin{equation}\label{eq:Binit}
  B_{\rm 0} \approx \sqrt{\frac{3 I c^3 P \dot{P}}{8
  \pi^2R^6}},
\end{equation}
where $B_{\rm 0}$ is the characteristic magnetic field strength at the
pulsar surface (in Gauss), $I$ is the moment of inertia for the
pulsar, $c = 3\times 10^8$\,m/s is the speed of light in vacuum, $R$
is the radius of the pulsar and $P$ and $\dot{P}$ are the pulsar spin
period and period derivative, respectively. With all pulsar masses
determined to date varying between one and two solar masses and since
the radius of a pulsar is expected to be around $10\,$km (as derived
from equations of state for dense nuclear matter), we can calculate
the moment of inertia, assuming the pulsar approximates a solid
sphere:
$$I \approx 0.4 M R^2 \approx 10^{45}\,{\rm g\, cm^2}.$$
This can be substituted in turn, reducing Equation \ref{eq:Binit} to:
\begin{equation}\label{eq:B}
  B_{\rm 0} \approx 3.2\times10^{19} \sqrt{P \dot{P}}
\end{equation}
with $B_{\rm 0}$ in Gauss.

An alternative means of using the spindown is to analyse the energy
loss due to magnetic dipole radiation. Equating the loss of rotational
energy as observed through the spin period derivative to the energy
loss predicted from magnetic dipole radiation, provides a relationship
between the spin period and its first time derivative \citep{lk05}:
\begin{equation}\label{eq:Ageinit}
  \dot{P} = \frac{2 (2\pi)^{n-1}\,m\,sin^2\alpha}{3 I c^3}P^{2-n},
\end{equation}
with $m$ the magnetic dipole moment, $\alpha$ the angle between the
magnetic and rotation axes and $n$ the braking index; $n = 3$
for pure magnetic dipole emission. Differentiating this Equation again
and replacing the constant fraction with $\dot{P}P^{n-2}$ allows
solving for $n$:
\begin{equation}
  n = 2-\frac{P\ddot{P}}{\dot{P}^2} = \frac{\nu\ddot{\nu}}{\dot{\nu}^2}
\end{equation}
in which $\nu=1/P$ is the spin frequency and $\dot{\nu}$, $\ddot{\nu}$
are its first and second time-derivatives. This Equation allows $n$ to
be determined, as has been done for six pulsars to date, with values
ranging from 2.14 for PSR B0540$-$69 \citep{lkg+07} to 2.91 for PSR
J1119$-$6127 \citep{ckl+00}. This suggests a spindown mechanism other
than pure magnetic dipole radiation is at work, though this could be
due to the small number statistics. \citet{jg99} performed a different
analysis which was applicable to many more pulsars and found a large
variation in braking indices. However, timing irregularities in the
pulsars of their sample may have corrupted their results.

Equation \ref{eq:Ageinit} can be integrated to give:
\begin{equation}
  {\rm Age} = \frac{P^{n-1}-P_{\rm 0}^{n-1}}{(n-1)K},
\end{equation}
where $P_{\rm 0}$ is the initial spin period and $K$ is the constant
fraction in Equation \ref{eq:Ageinit}, which can therefore be replaced
by $K = \dot{P} P^{n-2}$. Rewriting provides:
\begin{equation}
  {\rm Age} = \frac{P}{\dot{P}(n-1)} \left(1-\left(\frac{P_{\rm
  0}}{P}\right)^{n-1}\right).
\end{equation}
Assuming $P_{\rm 0} \ll P$ and $n = 3$, results in the simple relation 
\begin{equation}\label{eq:Age}
  {\rm Age} = \tau_{\rm c} = \frac{P}{2\dot{P}}
\end{equation}
\citei{a more rigorous derivation of which is presented
by}{og69}. While this value is easily derived from observations, the
assumptions that entered into its derivation must be noted, as well as
the fact that much of the pulsar emission mechanism is still badly
understood.

\subsection{The $P-\dot{P}$ diagram}
\begin{figure}
  \centerline{\psfig{figure=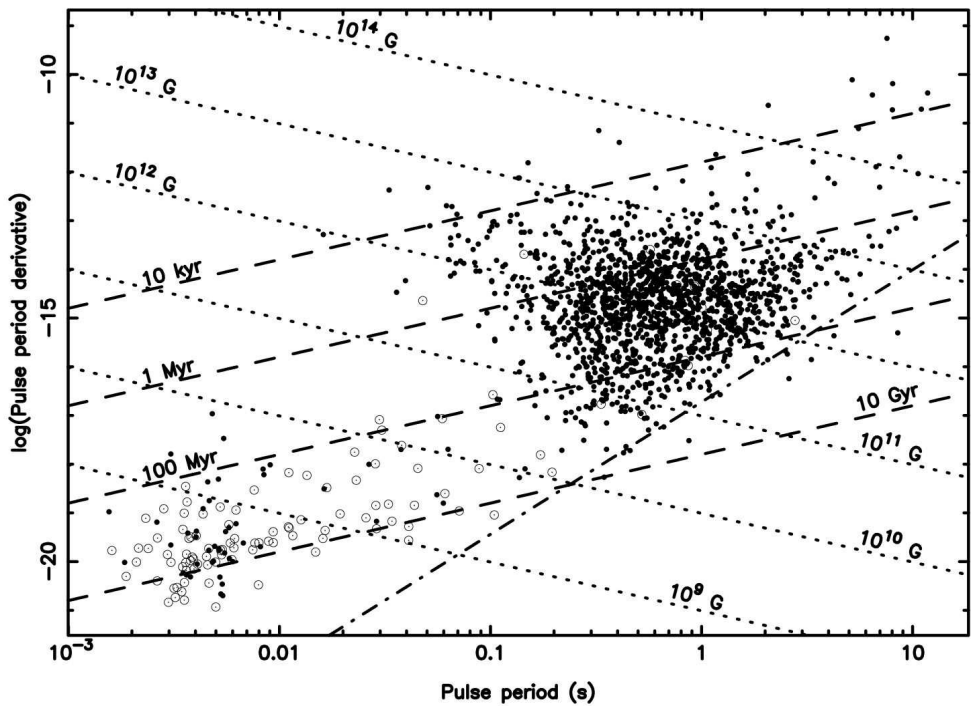,width=13cm,angle=0}}
  \caption[$P-\dot{P}$ diagram]{$P-\dot{P}$ diagram of all pulsars
	currently available in the ATNF pulsar
	catalogue. 
	Inclined dashed lines show the characteristic age ($\tau_{\rm c}$)
	of the pulsars, the dotted lines give the surface magnetic field
	strength ($B_{\rm 0}$) and the dash-dotted line is a death line
	proposed by \citet{cr93a}. Dots represent single pulsars; circled
	dots represent pulsars in binary systems.  }
  \label{fig:PPdot}
\end{figure}

Since period ($P$) and spindown ($\dot{P}$) are amongst the most
readily determined parameters pertaining to pulsars, a straightforward
tool to compare and analyse pulsar properties is the $P-\dot{P}$
diagram (Figure \ref{fig:PPdot}).
The clearest feature in this plot is the ``island'' where most pulsars
reside, with spin periods between 0.1 and a few seconds and magnetic
field strengths of typically $10^{11}$ to $10^{13}$ Gauss. These
pulsars are generally called \emph{normal pulsars}. Also, while the
diagram is reasonably well filled, the right-hand edge is remarkably
empty, even at characteristic ages well below a Hubble time. This is
because as pulsars spin down, they eventually have too little energy
left to produce radio waves, causing them to ``turn off'' at longer
periods. The virtual line across which this happens, is called the
\emph{death line} - while there are still neutron stars beyond this
line, they can no longer produce radio waves and are therefore
invisible to us. The few pulsars that do show up beyond the death line
show that this phenomenon is not strictly a function of the surface
magnetic field strength and pulse period, but also depends on the more
general geometry of the magnetic field, as more fully described by
\citet{cr93a}.

Besides the normal pulsars, there are two categories with somewhat
different characteristics. One fairly small group is positioned in the
top right-hand corner, with extremely high magnetic fields and long
pulse periods. These sources are called magnetars \citep{dt92a}, given
their strong magnetic fields. Observationally, these are either seen
in X-rays - in which case they are called ``anomalous X-ray pulsars''
or AXPs - or in soft $\gamma$-rays as so-called ``soft gamma
repeaters'' or SGRs. Radio emission from magnetars was expected to be
absent because the magnetic fields are too strong to allow standard
scenarios of radio wave creation, but this view was recently
compromised with the discovery of pulsed radio emission from the AXP
XTE J1810$-$197 \citep{crh+06}.

\subsection{Binary and Millisecond Pulsars}\label{ss:BMSP}

Finally, in the bottom left-hand corner of the $P-\dot{P}$ diagram,
there is a second island of pulsars, composed of the millisecond
pulsars (MSPs). These pulsars, with higher spin periods, slower
spindowns and far weaker magnetic fields ($B_{\rm 0} = 10^8$ to
$10^{10}$ Gauss), were first discovered in 1982 \citep{bkh+82}. As
Figure \ref{fig:PPdot} shows, this class of pulsar is found in binary
systems much more often than any other class of pulsar. Current theory
suggests this is a side-effect of the evolutionary cycle of MSPs,
since they are predicted to originate from the heavier star in a
binary system, as outlined below.

At the start of this chapter, we have described the evolution of a
star based on the progression of nuclear fusion in its core: nuclei
get fused into heavier elements until iron is formed. The speed of
this process is proportional to the mass of the star cubed: heavier
stars undergo greater gravitational forces and therefore incite faster
nuclear fusion to provide a greater pressure, resulting in hydrostatic
equilibrium. This implies that the stages of a star's life cycle are
shorter for heavier stars. A consequence of this is that the heaviest
star of a binary system is the first one to evolve and - if it is
sufficiently heavy - become a neutron star.  When subsequently the
companion star evolves into a red giant, its outer shells can grow
beyond the equipotential surface of the two stars and hence matter
will transfer onto the neutron star. Conservation of angular momentum
of this matter (which falls off the outer shells of the giant and onto
the far smaller core of the neutron star), causes the neutron star to
increase its rotational frequency by orders of magnitude
\citep{bv91}. During this spin-up phase, the binary system is observed
as an X-ray binary, as the accreting matter emits strongly in X-rays
\citep{bv91}.

At this point, there are again two distinct possibilities: either the
companion is not very massive, which means its evolution will go
slowly and it will eventually become a white dwarf. In that case the
matter transfer will continue for a long while and the neutron star
will be spun up to periods of milliseconds - as is clearly the case
for the first MSP discovered (PSR B1937+21, $P=1.558\,$ms) and for all
MSPs in the bottom-left corner of the $P-\dot{P}$ diagram. The
alternate possibility is that the companion star is heavy enough to
become a neutron star itself. In that case its evolution will go more
rapidly and the spin-up period will not last as long. This results in
a double neutron star system, of which one star resides in the main
$P-\dot{P}$ island with the normal pulsars and the other on the
``bridge'' linking the normal pulsars with the MSPs - with a pulse
period of the order of tens to hundreds of milliseconds. As derived in
detail by \citet{sb76}, this is how the first binary pulsar to be
discovered evolved: PSR B1913+16 with a pulse period of $P=59\,$ms
\citep{ht75a}.

There are currently 111 MSPs known\footnote{According to the ATNF
Pulsar catalogue: http://www.atnf.csiro.au/research/pulsar/psrcat;
\citet{mhth05} and following the definition of $P<20$\,ms and
$\dot{P}<10^{-16}$.} of which 49 reside in globular clusters
(GCs). There are two main reasons why such a large fraction of known
MSPs reside in GCs. Firstly, the high stellar density of GCs allows
formation of binary star systems through capture in close
encounters. This results in a larger density of X-ray binary sources,
which in turn generates a relatively larger amount of MSPs
\citep{vlv89}. The second reason is that GCs are easy targets for
surveys, while surveys for non-globular MSPs require vastly larger
amounts of observing time due to the inherently larger sky
coverage. The stellar density of GCs has a two-fold effect on pulsar
timing. Firstly, it causes acceleration terms that affect timing
stability over long time spans (years to decades; see Chapter
\ref{chap:20PSRS}). Secondly, it drastically increases the likelihood
of binary systems being disrupted. Because of this, only about half of
the GC MSPs are part of a binary system. For the non-globular (or
field) MSPs, close to 75\% are binaries. The origin of the 17 single
field MSPs has been an item of some debate. While supernovae of either
star in a binary can - and do - increase the spatial velocity and
orbital eccentricity of the system, in the case of a neutron star with
a less massive companion, this effect should not nearly be large
enough to disrupt the system. This has led to various scenarios for
the formation of single MSPs, including neutron star mergers
\citep{hv83a} and the disruption of the low-mass companion star into
either a debris disk or a planet-sized object \citep{rs83}.

\section{Pulsar Timing}
\label{intro:pt}

\subsection{Pulsar Timing Basics}
The lighthouse model suggests that the pulses of radio emission we
receive from pulsars always come from the same phase in the pulsar's
rotation. If this is true and the region from which the emission comes
is stable, then the times-of-arrival (TOAs) of the pulses can be
determined and compared to a timing model. In practice this is indeed
possible, though there are two reasons why pulses are generally not
timed individually. Firstly, it has been shown that the emission is
not perfectly stable: the shape of single pulses varies considerably
\citep{hmt75}. Averaging a train of consecutive pulses does,
generally, result in a stable \emph{average profile}, which can be
timed to high precision. The second reason for adding consecutive
pulses is to reduce the radiometer noise: application of elementary
radio astronomy theory to pulsed emission \citei{as e.g. in}{lk05}
shows that the strength of the emission (\emph{signal}) relates to the
noise introduced by the system temperature (\emph{noise}) as follows:
\begin{equation}
  {\rm SNR} = \sqrt{N_{\rm p} B t} \left(\frac{GS_{\rm peak}}{T_{\rm
  sys}}\right) \sqrt{\frac{P-W}{W}},
\label{eq:Radiometer}
\end{equation}
with SNR the signal-to-noise ratio, $N_{\rm p}$ the number of
polarisations measured, $B$ the bandwidth, $t$ the integration time of
the profile, $S_{\rm peak}$ the peak brightness of the pulsar, $T_{\rm
sys}$ the system noise temperature, $W$ the on-pulse width, $P$ the
pulse period and $G=\eta_{\rm A}A/(2 k_{\rm B})$ is the telescope gain
based on the aperture efficiency $\eta_{\rm A}$, the telescope
aperture $A$ and Boltzmann's constant $k_{\rm B}$.

\begin{figure}
  \centerline{\psfig{figure=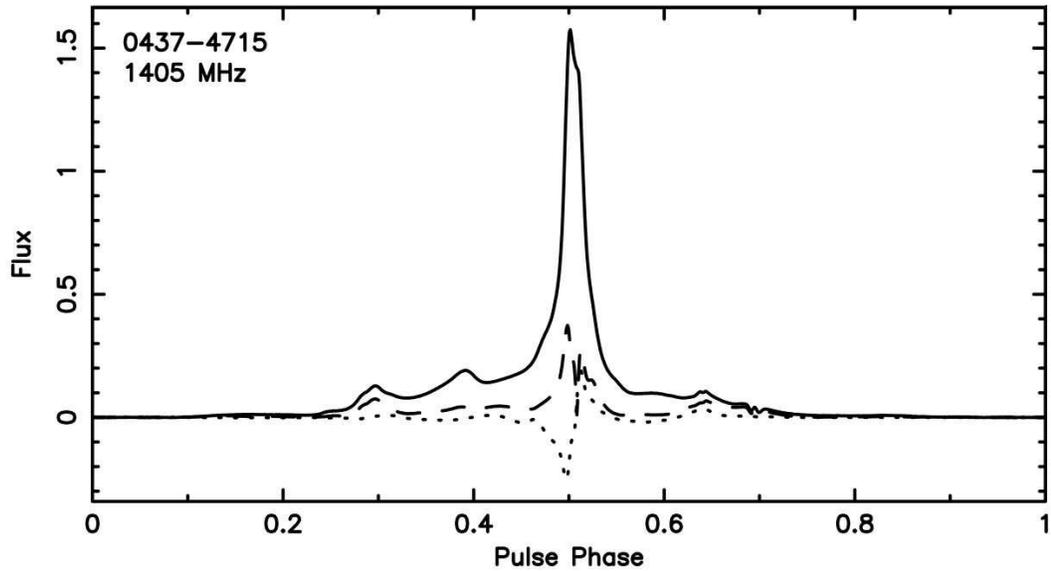,width=14cm,angle=0}}
  \caption[PSR J0437$-$4715 pulse profile]{Pulse profile of PSR
	J0437$-$4715 at an observing frequency of 1405\,MHz, showing the
	linear (dashed) and circular (dotted) Stokes parameters as a
	function of pulse phase.}
  \label{fig:Profile}
\end{figure}

The main complexity of determining TOAs lies in the complicated shape
many pulse profiles have. As an example, the average pulse profile of
PSR J0437$-$4715 (which will be described in much more detail in
Chapter \ref{chap:0437}) is presented in Figure
\ref{fig:Profile}. Clearly, the peak in itself could be used for
timing, but much more precise measurements can be achieved by using
the entire pulse profile.
Practically, therefore, timing works as follows: during observations,
the period of the pulse is derived from a timing model derived from
previous observations. Given that period, the data are \emph{folded}
in real time - i.e. data samples with the same phase (time modulo
pulse period) are averaged. After storing the times when the
observation started and finished, based on the observatory atomic
clock, the folded observation is stored with a timestamp denoting its
centre. At this point the actual position of the pulse is still not
yet defined, so the timestamp in itself only provides the time of the
observation, not the time the ``average'' pulse arrived. For that
information to be derived, the observation is cross-correlated with a
template profile. These templates can fundamentally be any
non-constant function, but in order to take full advantage of the
pulse shape, it is made to resemble the average pulse as closely as
possible. Throughout the analysis done for this thesis, the standard
profiles are constructed through addition of the brightest
observations available. An alternative method that is gaining
popularity, is to construct an analytic template, based on fitting of
standard analytic functions to a high SNR observation. This method has
the advantage of having a noiseless baseline.

From the cross-correlation of the observation and the standard
profile, one can derive the phase offset between the two profiles,
which can be added to the time stamp to provide a site-arrival-time
(SAT). It must be noted that these times are not absolute in the sense
that they depend on the pulse phase of the template profile. However,
if the same template is used for all observations, this constant
offset will be the same in all SATs and is therefore irrelevant, since
absolute phase is practically inachievable.

The next step in the analysis is to transfer the SAT to an inertial
reference frame: the Earth's rotation around the Sun causes the SATs
to be strongly influenced by the position of the Earth throughout the
year. To translate the SATs to barycentric-arrival-times or BATs (in
this context, the barycentre is the centre of mass of the Solar
System), one needs accurate predictions of the masses and positions of
all the major Solar System objects at any point in time. These
predictions - so-called Solar System ephemerides (SSE) - are provided
by several organisations worldwide. For the work presented in this
thesis, the DE200 and DE405 JPL SSE were used \citep{sta04b}.
Alternatives are the INPOP06 from the Observatoire de Paris
\citep{fmlg08} and the EPM2004 from the Russian Academy of Sciences
\citep{pit05}. This variety demonstrates that, while these ephemerides
are all up to very high standards, they do contain uncertainties and
errors, which are not taken into account in the conversion from SAT to
BAT - and are therefore a cause of unaccounted timing irregularities,
as we will show in more detail in \S\ref{SSE}.

Finally, the BATs are subtracted from arrival times predicted by a
timing model of the pulsar. The difference between these two
quantities (the \emph{timing residuals}) are the real tools of pulsar
timing: initially they are used to improve the estimates of timing
model parameters, but since all unmodelled physics is contained within
them, it is the analysis of these timing residuals that will allow
measurement of new effects, determination of new parameters and
provide limits on the influence of predicted effects, such as
gravitational wave backgrounds. In the remainder of this section, we
will provide an insight into the potential components and complexities
of pulsar timing models.

\subsection{Spin and Astrometric Parameters}
\begin{figure}
  \centerline{\psfig{figure=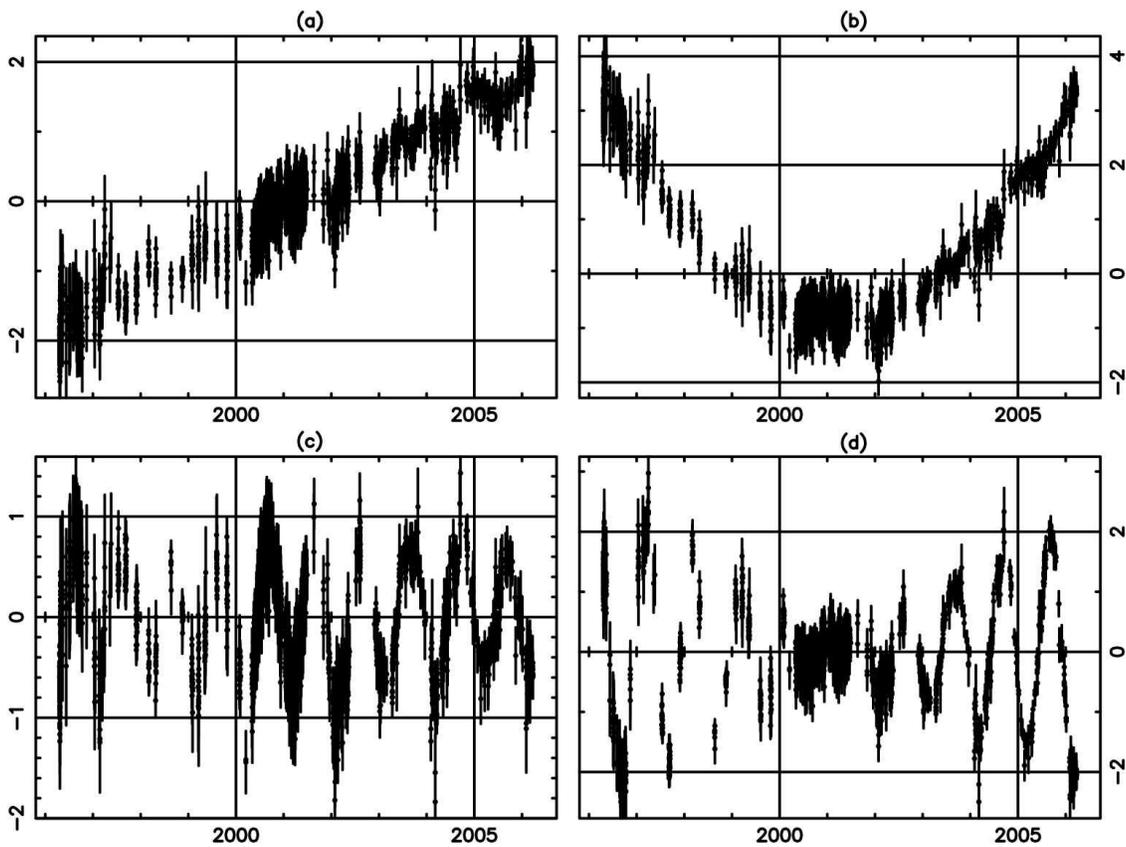,width=15cm,angle=0.0}}
  \caption[Timing residual signatures of errors in astrometric
  parameters]{Residual signatures of basic timing parameters. Year is
  displayed on the x-axis; the residuals in microseconds are displayed
  on the y-axis. (a) Linear trend due to error in $\nu$; (b) quadratic
  signature of $\dot{\nu}$; (c) sine wave with yearly periodicity due
  to erroneous pulsar position; (d) growing sine wave due to wrong
  proper motion. All four examples are based on the PSR J0437$-$4715
  data set which will be fully described and analysed in Chapter
  \ref{chap:0437}.}
  \label{fig:Signatures}
\end{figure}

Every timing model starts out with the most fundamental parameters,
which are determined at the discovery of the pulsar: spin frequency
($\nu$), position in right ascension and declination ($\alpha$ and
$\delta$, respectively) and dispersion measure ($DM$, discussed in the
next section). After about a year, the frequency derivative
($\dot{\nu}$) can also be determined. These five parameters constitute
a simple timing model and their effect on timing residuals can
relatively easily be understood. An incorrect pulse frequency in
the timing model will cause the predicted arrival time of a pulse to
become incrementally wrong with time, resulting in a linear trend in
the residuals (Figure \ref{fig:Signatures} a). A significant error in
$\dot{\nu}$ causes the frequency to be increasingly off and therefore
results in a steepening residual trend - seen as a quadratic signature
in the timing residuals (Figure \ref{fig:Signatures} b). An incorrect
position corrupts the transfer from SAT to BAT and thus introduces a
sine wave with a period of a year (Figure \ref{fig:Signatures}
c). After a while, the proper motion ($\mu$) may also be detected. Its
effect is also a sine wave, but with an amplitude that increases
linearly with time, as shown in Figure \ref{fig:Signatures} d.

\begin{figure}
  \centerline{\psfig{figure=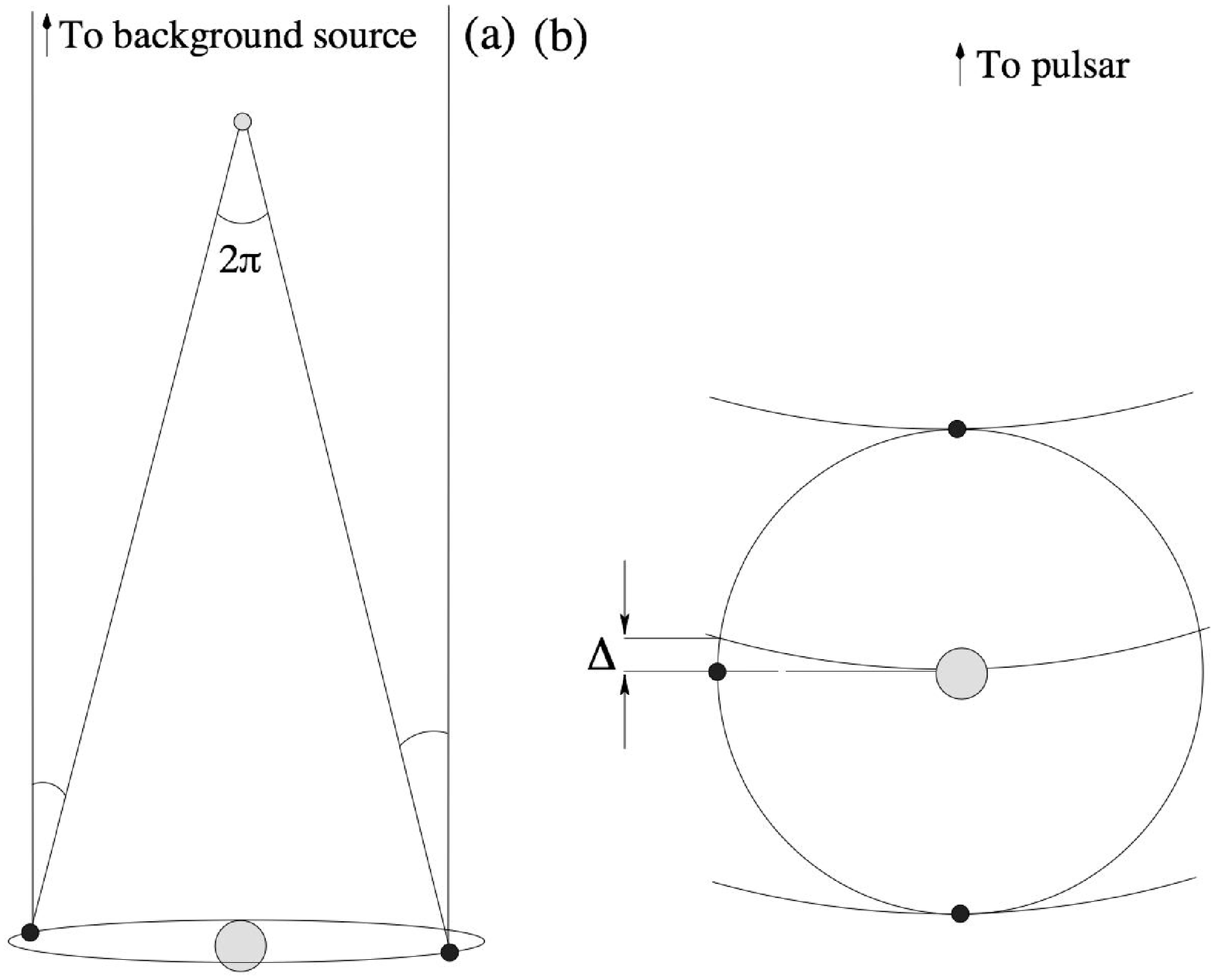,width=15cm,angle=0.0}}
  \caption[Traditional and timing parallax]{(a) Traditionally,
  distances are calculated based on the yearly change in position of a
  nearby source with respect to a background source. (b) In pulsar
  timing, the background source is not observed, but the curvature of
  the wavefront originating at the pulsar is also inversely
  proportional to the distance to the pulsar and this curvature can be
  measured by means of the half-yearly delay $\Delta$ indicated in the
  figure.} 
  \label{fig:Parallax}
\end{figure}

A final astrometric parameter that can be determined in some cases,
though not all, is parallax. In all astronomy short of pulsar timing,
parallax is the apparent yearly wandering of a nearby star with
respect to a background source such as a galaxy (see Figure
\ref{fig:Parallax} a). This changing of the relative position between
the star and background object is caused by the fact that the Earth
moves so that we look at the object from a slightly different
angle. In pulsar timing, the relative position of the pulsar with
respect to a background object is not measurable, since only the
pulsar is timed. However, the closer the pulsar is, the stronger the
wave front is curved - this induces a delay that is maximal when the
pulsar is at a right angle to the Earth-Sun line (see Figure
\ref{fig:Parallax} b). As opposed to the geometric parallax, though,
pulsar timing parallax signatures are practically unmeasurable when
the pulsar is far away from the ecliptic plane - since the same part
of the wavefront hits the Earth at all positions in its orbit. (This
is particularly true because the Earth's orbit is nearly circular.)

Mathematically, these astrometric delays can be derived from the
relative positions of the pulsar and Earth. Let $\vec{p}$ be the
vector pointing from the telescope to the pulsar, $\vec{r}(t)$ the
vector from the telescope to the Solar System barycentre (SSB) and
$\vec{d}$ the vector from the SSB to the pulsar. Now also introduce
the pulsar's velocity vector $\vec{v}$ so that $\vec{p} =
\vec{r}+\vec{d}+(t-t_{\rm 0})\vec{v}$. Following the analysis by
\citet{ehm06}, the travel time between the pulsar and the telescope
becomes (ignoring binary effects and effects due to the interstellar
medium):
\begin{equation}\label{eq:Geometric}
  |\vec{p}| = |\vec{d}| + |\vec{v}_{\parallel}| (t-t_{\rm 0}) +
   |\vec{r}_{\parallel}| + \frac{1}{|\vec{d}|} \left(
  \frac{|\vec{v}_{\perp}|^2}{2} (t-t_{\rm 0})^2 +
  \vec{v}_{\perp}\centerdot\vec{r}_{\perp} (t-t_{\rm 0}) +
  \frac{|\vec{r}_{\perp}|^2}{2}\right),
\end{equation}
where $\parallel$ denotes projections onto the line of sight and
$\perp$ projections perpendicular to the line of sight.

Adopting the notation $d = |\vec{d}|$, introducing the proper motion
$\vec{\mu} = \vec{v}_{\perp}$ and translating into a timing delay, we
obtain:
\begin{equation}\label{eq:TimingModel}
  \Delta_{\rm geom} = (p-d)/c = \frac{v_{\rm r}}{c} (t-t_{\rm 0}) +
  \frac{r_{\parallel}}{c} + \frac{v_{\rm T}^2 (t-t_{\rm 0})^2}{2cd} +
  \frac{\vec{\mu}\centerdot \vec{r}_{\perp}}{cd}(t-t_{\rm 0}) +
  \frac{r_{\perp}^2}{2cd}
\end{equation}

The different terms can be distinguished as follows. The first term is
a secular increase in distance due to the radial velocity of the
pulsar. Since this introduces a linearly time-varying delay, it is
indistinguishable from the spin period of the pulsar and can therefore
not be individually measured. The second term describes the varying
distance between the Earth and the pulsar caused by the orbital motion
of the Earth - which brings it closer or further depending on the time
of year. This type of delay - the varying light-travel time across an
orbit - is called a ``Roemer delay''\footnote{The Roemer delay is
named after the Danish astronomer who first used it to measure the
speed of light based on the orbits of the Galilean moons.}. The third
term, which grows quadratically with time, is the Shklovskii effect,
first identified by \citet{shk70} and is due to the apparent
acceleration away from us as the pulsar travels in a straight line
tangent to the plane of the sky. This effect is easily derived by
considering the distance between the pulsar and Earth ($d$) and the
tangential velocity ($v_{\rm T}$), perpendicular to the line of
sight. Designating the initial distance $d_{\rm 0}$, we get the
relationship: $d = \sqrt{d_{\rm 0}^2 + v_{\rm T}^2
t^2}$. Differentiating twice results in ${\rm d}^2 d/{\rm d}t^2 =
d_{\rm 0}^2 v_{\rm T}^2 / \sqrt{(d_{\rm 0}^2 + v_{\rm T}^2
t^2)^3}$. Approximating $d_{\rm 0} \approx d = \sqrt{d_{\rm 0}^2 +
v_{\rm T}^2 t^2}$ now results in ${\rm d}^2 d/{\rm d}t^2 \approx
v_{\rm T}^2 / d$ - which is the apparent radial acceleration. The
second-last term of Equation \ref{eq:TimingModel} is the actual proper
motion term. It has a yearly signature (as shown by $\vec{r}_{\perp}$)
and the size of the signature grows linearly in time, as described
before and shown in Figure \ref{fig:Signatures} d. Finally, the delay
proportional to $r_{\perp}^2$, is the timing parallax signature which,
due to the square, has a half-yearly signature as shown in Figure
\ref{Fig::Px}.

The above description of the geometric timing delays ignored some of
the more subtle effects, as a careful read of \citet{ehm06} will
show. Most specifically, we have in this analysis ignored several
higher order terms, mainly for purposes of clarity. Also the Einstein
and aberration delays were ignored. The aberration delay is caused by
the relative motion of the pulsar and the observer. The Einstein delay
arises because of the general relativistic time dilation experienced
in fields with different gravitational strength. Through the motion of
the planets in our Solar System, the gravitational strength (and
therefore, the relative speed of time) changes as a function of space
and time. A full treatment \citep{if99} of the integrated effect of
these delays on pulsar timing, shows that correction is needed to
achieve timing at the precisions we have today.

\subsection{Effects of the Interstellar Medium}\label{ssec:ISM}

\begin{figure}
  \centerline{\psfig{figure=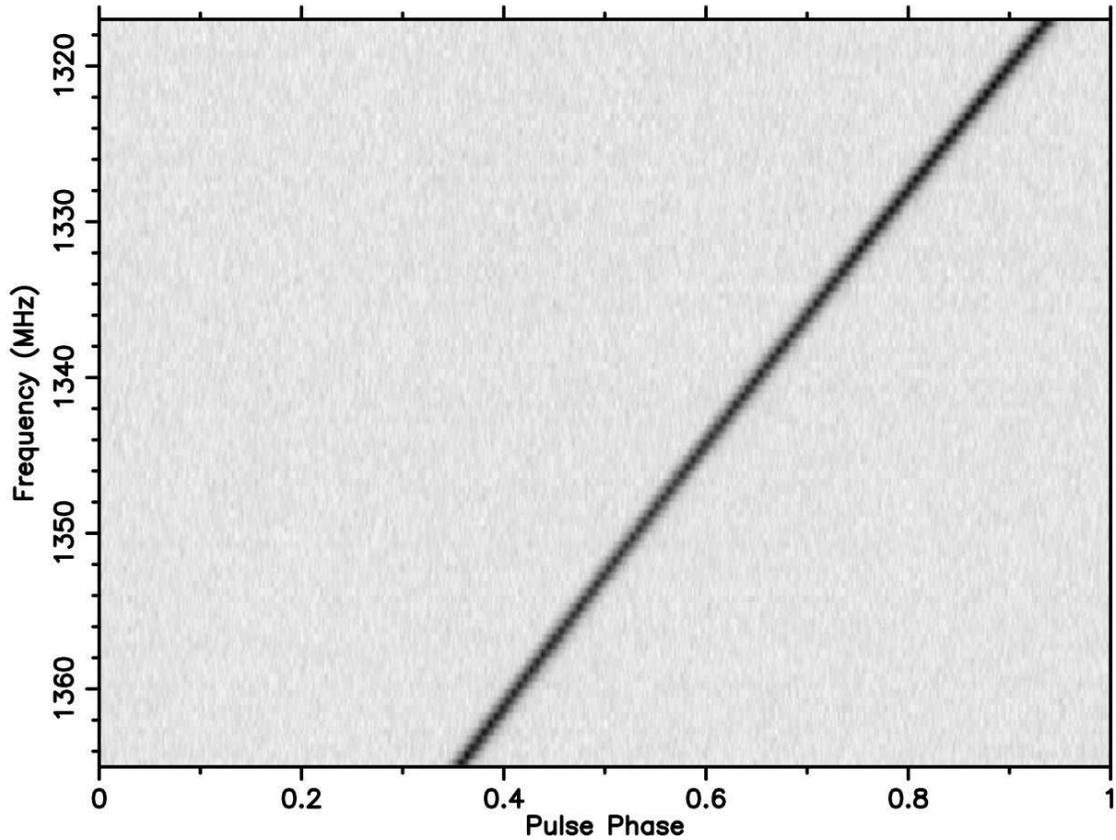,width=15cm,angle=0.0}}
  \caption[Dispersion of PSR J1909$-$3744 emission across the 20\,cm
    observing band]{Frequency-phase plot of a pulse profile of PSR
    J1909$-$3744, uncorrected for interstellar dispersion. The
    pixellation in y-direction is due to the limited number of
    frequency channels in the observational setup (see Chapter
    \ref{chap:Hardware} for more information). While the dispersion
    delay is dependent on the square of the observing frequency, the
    low DM and small bandwidth of this observation allow only a very
    weak quadratic trend to be seen.}
  \label{fig:Dispersion}
\end{figure}

Ever since the exposition of special relativity \citep{ein05}, it has
been understood that the speed of light is constant and independent of
the reference frame of the observer. However, this is only true in a
vacuum: in all other media, the speed of light is determined by the
refractive index $n$ of the medium: $ c = c_{\rm 0}/n$ (with $c_{\rm
0}$ the speed of light in vacuum). Since $n\neq1$ for the ionised
interstellar medium and since $n$ varies strongly with the frequency
of the light, the travel time of a given wave changes with observing
frequency. In the case of pulsar timing, this effect causes the same
pulse to be observed first at higher observational frequencies and
later at lower frequencies. This is clearly illustrated in Figure
\ref{fig:Dispersion}.

This dispersive effect needs to be remedied at two different
points. Firstly, it needs to be countered when integrating over
frequency - otherwise the pulse profiles will be dramatically smeared
out, which will worsen any achievable timing precision. Secondly it
needs to be corrected in the actual timing, in case observations at
different observing frequencies are included in the same data
set. \citet{lk05} derive the time delay as a function of observing
frequency to be:
\begin{equation}\label{eq:DispersionDelay}
  \Delta_{\rm ISM} = \frac{D}{f^2} \int_{\rm 0}^d n_{\rm e} {\rm d}l,
\end{equation}
with $n_{\rm e}$ the electron density per cm$^3$ and the dispersion
constant:
\begin{equation}\label{eq:DispersionConstant}
  D =\frac{e^2}{2\pi m_{\rm e}c} \approx 4.15\times 10^3\,{\rm
  MHz}^2{\rm pc}^{-1}{\rm cm}^3{\rm s}.
\end{equation} 
Notice that the observation of a single pulse across a broad enough
bandwidth can be used to calculate the integrated electron density
between us and the pulsar. This quantity is called the
\emph{dispersion measure}, DM:
\begin{equation}\label{eq:DM}
  DM = \int_{\rm 0}^d n_{\rm e}{\rm d}l.
\end{equation}
Given this definition, a measurement of the $DM$ can be combined with
a measure of distance (either from timing or VLBI) to provide a
precise value for the average electron density towards the
pulsar. This in turn provides an input to models of the Galactic
electron density, as presented by \citet{cl02}. Inversely, such models
can be used to make first-order estimates of the distance to a pulsar,
based on its measured $DM$.

By means of illustration, Figure \ref{fig:Dispersion} shows a delay of
0.58 pulse periods between the frequencies of 1317\,MHz and
1365\,MHz. Given the period of PSR J1909$-$3744 to be 2.947\,ms and
rewriting Equation \ref{eq:DispersionDelay} to calculate the
difference between two bands:
\begin{equation}\label{eq:DMdelay}
  t_{\rm 2}-t_{\rm 1} = D\times DM\times
  \left(f_2^{-2}-f_1^{-2}\right),
\end{equation}
we obtain $DM \approx 10.34$\,cm$^{-3}$pc, which compares well with
the catalogue value of 10.3940\,cm$^{-3}$pc.

It is important to notice that the electron density is not necessarily
constant throughout space. Since the pulsar, the Solar System and the
interstellar medium in between are all in motion, the electron density
along the line of sight is therefore expected to change as a function
of time, which will change the $DM$, too. The density of ionised
particles contained within the Solar wind also varies strongly -
especially as the lines of sight for some pulsars travel closely to
the Sun at some times during the year. Detailed analyses of both of
these effects are presented by You et al. (2007a, 2007b) and
\citet{ojs07}\nocite{yhc+07,yhc+07b}. At a much lower level, the ISM
also causes scattering and scintillation, as reviewed in, for example,
\citet{ric90}. While proper mitigation strategies for these effects
will become increasingly important in high-precision timing, they are
not significant for the work presented in this thesis (see
\S\ref{sec:stab:disc}).
%

\subsection{Geometric Effects of a Binary System}\label{ssec:BinaryEffects}
In standard Newtonian mechanics, the orbit of any body around another
can be determined based on the following parameters:
\begin{description}
  \item[Binary period, $P_{\rm b}$,]{measured in days.}
  \item[Semi-major axis, $a$,]{measured in light-seconds.}
  \item[Orbital eccentricity, $e$,]{between 0 (circular) and 1 (parabola).}
  \item[Inclination angle, $i$,]{defined as the angle between the
  angular momentum vector of the binary orbit and the line of
  sight. Away from the observer (clockwise rotation) is 0$^{\circ}$,
  towards the observer (counterclockwise rotation) is 180$^{\circ}$.}
  \item[Longitude of the ascending node, $\Omega$,]{measured from
  North through East, towards the ascending node. The ascending node
  is the point in the orbit where the pulsar crosses the plane of the
  sky, moving away from the observer.}
  \item[Angle of periastron, $\omega$,]{measured from the
  ascending node along with the binary rotation.}
  \item[Time of periastron passage, $T_{\rm 0}$,]{given as MJD.}
\end{description}
Note that $a$ and $i$ are not generally measured independently, but
rather in combination through the projected semi-major axis, $x = a
\sin{i}$.

Given these definitions, we can expand Equation \ref{eq:Geometric} to
include geometric effects of the binary system. Introducing vector
$\vec{b}$ from the barycentre of the binary system (BB) to the pulsar
and redefining $\vec{d}$ to point from the SSB to the BB, Equation
\ref{eq:TimingModel} can (to first order) be expanded with the
following parameters \citei{as in}{ehm06}:
\begin{equation}\label{eq:BinGeometry}
  \Delta_{\rm Bin} = \frac{b_{\parallel}}{c} +
  \frac{1}{cd}\left((t-t_{\rm 0})\vec{v}_{\perp}\centerdot
  \vec{b}_{\perp}+\vec{r}_{\perp}\centerdot\vec{b}_{\perp} +
  \frac{b_{\perp}^2}{2}\right).
\end{equation}
(Notice various deformations due to general relativity (GR) are also
needed in this treatment, as well as derivatives of parameters such as
$\omega = \omega_{\rm 0}+\dot{\omega}(t-t_{\rm 0})$, for example. For
a full relativistic treatment, see \citet{dd86} and \citet{ehm06}.

In Equation \ref{eq:BinGeometry}, $b_{\parallel}$ is the Roemer delay
of the binary system (equivalent to $r_{\parallel}$ in the single
pulsar case) and can be expanded in terms of the binary parameters as
follows \citei{after}{dd86}:
\begin{equation}
  \Delta_{\rm R} = \frac{b_{\parallel}}{c} = \frac{a \sin{i}}{c}
  \left( \sin{\omega} (\cos{u} - e) + \sqrt{1-e^2} \cos{\omega}
  \sin{u}\right),
\end{equation}
with $u$ the eccentric anomaly defined from $n(t-T_{\rm 0}) = u - e
\sin{u}$ (and $n$ an integer).

The second term, $\vec{v}_{\perp}\centerdot \vec{b}_{\perp}$,
demonstrates the effect the proper motion has on the orbital
parameters. As the binary system moves across the sky, we observe it
from an ever-changing angle, which observationally results in a
secular change in the projected semi-major axis $x$ and angle of
periastron $\omega$, as first described by \citet{kop96}. The third
term, $\vec{r}_{\perp}\centerdot\vec{b}_{\perp}$ is a combined effect
of the orbital motions of the Earth and pulsar, which causes
additional delays of a similar type to those of the timing
parallax. This effect, which was first described by \citet{kop95} is
therefore named the ``annual-orbital parallax'' (AOP). Both the proper
motion effect and the AOP were first measured in the J0437$-$4715
binary system (respectively by \citet{sbm+97} and \citet{vbb+01}). The
final term, $b_{\perp}^2$ is comparable to the earlier $r_{\perp}^2$
and is effectively the parallax effect due to the binary motion of the
pulsar - therefore named ``orbital parallax''. It was first derived in
parallel with the AOP effect by \citet{kop95} but has not been
measured to date.

From \citet{kop96}, we can obtain the full expansion of the proper
motion effect in terms of the binary parameters:
\begin{eqnarray}\label{eq:KopPM}
  \Delta_{\rm Bin, PM}& =& \frac{(t-t_{\rm 0})
  \vec{\mu}\centerdot\vec{b}_{\perp}}{c} \notag \\ 
  & = & \frac{x (1-e \cos{u}) \cos{(\omega+A_{\rm e})}}{\sin{i}}
  (\mu_{\rm \alpha} \cos{\Omega} + \mu_{\rm \delta} \sin{\Omega})\notag \\ 
  & & + x \cot{i} (1-e \cos{u}) \sin{\left(\omega + A_{\rm e}\right)}
  (-\mu_{\rm \alpha} \sin{\Omega} + \mu_{\rm \delta} \cos{\Omega}),
\end{eqnarray}
where the true anomaly, $A_{\rm e}$, is defined as: 
\begin{equation*}
  A_{\rm e} = 2 \arctan{\left(\sqrt{\frac{1+e}{1-e}}
  \tan{\frac{u}{2}}\right)}
\end{equation*}
Likewise, the AOP effect can be written out as follows, from \citet{kop95}:
\begin{multline}\label{eq:KopAOP}
  \Delta_{\rm Bin, AOP} =
  \frac{\vec{r}_{\perp}\centerdot\vec{b}_{\perp}}{cd}
  = \frac{x}{d}
   \Big[
	 \left(\Delta_{I_{\rm 0}} \sin{\Omega} -
	 \delta_{J_{\rm 0}} \cos{\Omega}\right) 
	 R \cot{i}\\
	 -\left(\Delta_{I_{\rm 0}} \cos{\Omega} + 
	 \Delta_{J_{\rm 0}} \sin{\Omega}\right)
	 Q \csc{i} \Big],
\end{multline}
with $\Delta_{I_{\rm 0}} = -\vec{r}\centerdot\vec{I_{\rm 0}}$ and
$\Delta_{J_{\rm 0}} = -\vec{r}\centerdot\vec{J_{\rm 0}}$ the
X and Y components of the Sun-Earth vector on the plane of the
sky. ($\vec{I_{\rm 0}}$ and $\vec{J_{\rm 0}}$ are unit vectors
connected to the BB and pointing North and East respectively.) $R$
and $Q$ are functions defined as follows:
\begin{eqnarray}
  R = \sin{\omega} (\cos{u} - e) + \sqrt{1-e^2} \cos{\omega} \sin{u}\\
  Q = \cos{\omega} (\cos{u} - e) - \sqrt{1-e^2} \sin{\omega} \sin{u}.
\end{eqnarray}

The important point to note about these terms, is their dependence on
the orbital inclination, $i$, and the longitude of the ascending node,
$\Omega$. Without these so-called ``Kopeikin terms'', the measurement
of $\Omega$ would be impossible. In the next section we will show that
$i$ can be measured due to general relativistic effects, so the
independent determination of the inclination angle through the
Kopeikin terms provides a test of these general relativistic
predictions - as described in \citet{vbb+01}. Alternatively, the GR
and Kopeikin effects can be combined to provide a more precise
measurement - this approach will be used in Chapter \ref{chap:0437}.

\subsection{Relativistic Effects in Binary Systems}\label{sec:GR}

The timing formulae described in the preceding sections have ignored
any effects due to general relativity (GR). Now, we will highlight the
most important general relativistic additions to that timing
model. While the focus will be on the binary system of the pulsar, it
must be noted that these effects play in the Solar System as well,
albeit at a lower level.

One of the problems of Keplerian dynamics that was solved by the
introduction of GR, was the perihelion advance of Mercury. This same
effect has been readily observed in several binary pulsar systems and
is predicted to be \citei{as in}{tw82}:
\begin{eqnarray}\label{eq:Omdot}
  \dot{\omega} & = & 3 \left(\frac{2 \pi}{P_{\rm b}}\right)^{5/3} 
  \left(\frac{G M}{c^3}\right)^{2/3}\frac{1}{1-e^2}\notag\\
  & \approx & 0.19738 \left(\frac{M}{M_{\odot}}\right)^{2/3}
  \left(\frac{P_{\rm b}}{1\,{\rm day}}\right)^{-5/3}\frac{1}{1-e^2},
\end{eqnarray}
with $M = M_{\rm psr} + M_{\rm c}$ the total system mass, $M_{\odot}$
the mass of the Sun and $\dot{\omega}$ in units of degrees per year.

The next two post-Keplerian parameters are measured through what is
now known as the \emph{Shapiro Delay}, after the scientist who first
proposed this test of GR \citep{sha64}. The effect in question is the
time delay introduced by a gravitational potential along the line of
sight. While Shapiro originally envisaged this test to take place in
the Solar System through transmission of radio waves past the edge of
the Sun, it can be readily observed in binary pulsar systems that have
a nearly edge-on orbit ($i\approx 90^{\circ}$). Clearly, the amplitude
of the effect (the \emph{range}, $r$) is dependent on the companion
mass, while its evolution as a function of binary phase (the
\emph{shape}, $s$) is determined by how closely the rays pass by the
companion star - and therefore depends on the inclination angle of the
system. More precisely, the two relativistic parameters are predicted
to be \citei{see, e.g.}{sta03}:
\begin{eqnarray}
  r& =& \frac{G M_2}{c^3} = 4.9255\times 10^{-6} s
  \left(\frac{M_{\rm c}}{M_{\odot}}\right)\\
  s& =& \frac{cxG^{-1/3}}{M_{\rm c}}\left(\frac{2\pi M}{P_{\rm
  b}}\right)^{2/3}\notag\\
  & =& 0.1024 \left(\frac{P_{\rm b}}{1\,{\rm day}}\right)^{-2/3}
  \left(\frac{M}{M_{\odot}}\right)^{2/3} 
  \left(\frac{M_{\rm c}}{M_{\odot}}\right)^{-1} 
  \left(\frac{a \sin{i}}{c}\right)
\end{eqnarray}
and in GR, $s = \sin{i}$. 

The fourth relativistic effect is the orbital decay due to
gravitational wave emission. First measured in the original binary
pulsar system PSR B1913+16 \citep{tw82}, this parameter has provided
the first indirect detection of gravitational waves. The predicted
size of this effect is, as presented by \citet{tw82}:
\begin{eqnarray}\label{eq:PbdotGW}
  \dot{P}_{\rm b} & = & - \frac{192 \pi}{5 c^5} 
  \left(\frac{2 \pi G}{P_{\rm b}}\right)^{5/3}
  \frac{1+\frac{73}{24}e^2+\frac{37}{96}e^4}{(1-e^2)^{7/2}} 
  \frac{M_{\rm psr}M_{\rm c}}{M^{1/3}}\notag\\
  & = & -2.1719\times 10^{-14}
  \frac{1+\frac{73}{24}e^2+\frac{37}{96}e^4}{(1-e^2)^{7/2}} 
  \left(\frac{P_{\rm b}}{1 {\rm day}}\right)^{-5/3}
  \left(\frac{M_{\rm psr} M_{\rm c}}{M_{\odot}^2}\right)
  \left(\frac{M}{M_{\odot}}\right)^{-1/3}
\end{eqnarray}
(where $\dot{P}_{\rm b}$ is unitless). Notice, however, that this is
not the only contribution to the observed orbital period
derivative. As will be explained in more detail in \S\ref{Dist}, the
Shklovskii effect discussed earlier, as well as some accelerations
caused by the mass distribution in the Galaxy, both influence the
effective value of $\dot{P}_{\rm b}$.

The fifth and final relativistic effect that can be measured through
pulsar timing, is the transverse doppler and gravitational redshift
parameter, $\gamma$. It is caused by the fact that the progress of
time is strongly affected by the strength of the gravitational
potential. As a pulsar in an eccentric orbit moves closer or further
away from its companion star, the gravitational potential varies and,
consequentially, so does the clock rate. The theoretical prediction
for this parameter is \citei{from}{tw82}:
\begin{eqnarray}\label{eq:Gamma}
  \gamma & = & \frac{G^{2/3}e}{c^2}
  \left(\frac{P_{\rm b}}{2\pi}\right)^{1/3} \frac{M_{\rm c}(M+M_{\rm
	  c})}{M^{4/3}}\notag\\
  & = & 6.926\times 10^{-3}s 
  \left(\frac{P_{\rm b}}{1 {\rm day}}\right)^{1/3}e 
  \frac{M_{\rm c}(M+M_{\rm c})}{M_{\odot}^2}
  \left(\frac{M}{M_{\odot}}\right)^{-4/3}
\end{eqnarray}

One point of note is that all of the relativistic effects presented
here (equations \ref{eq:Omdot} through to \ref{eq:Gamma}) are only
dependent on the Keplerian parameters presented in the previous
section and the masses of the binary system: $M_{\rm psr}$ and $M_{\rm
c}$. This implies that, as soon as two GR effects are measured in
addition to the Keplerian parameters, the magnitude of all other
effects can be predicted. The overdetermined character of these
equations has allowed the most stringent tests of GR to date
\citep{ksm+06}.

\section{Gravitational Waves and Pulsar Timing Arrays}
\label{intro:pta}

The interaction of the gravitational force with matter is described by
the Einstein field equations which are, like Maxwell's equations for
the electromagnetic force, wave equations \citei{see, e.g.}{sch93b}.
The concept that acceleration of masses creates a gravitational wave
(GW) like the acceleration of electric charge generates an
electromagnetic wave, was one of the main relativistic predictions
that remained untested for more than half a century. As mentioned in
the previous section, \citet{tw82} finally proved the veracity of this
prediction by accurately demonstrating that the energy loss from the
binary pulsar system B1913+16 equated the predicted energy loss due to
gravitational wave emission. Until today, however, no direct detection
of gravitational waves has been made so all characteristics of these
waves (such as, for example, polarisation and velocity) remain
untested.

In this section, the case for direct detection of GWs through pulsar
timing will be outlined. First some initial experiments and concepts
will be described in \S\ref{GWStart}. Next, \S\ref{ssec:Limits} will
discuss the recent history of limits on the gravitational wave
background (GWB). In \S\ref{sec:GWBSources} we will outline the
potential sources of GWs that might be detected by pulsar timing
arrays (PTAs) and \S\ref{ssec:PTAs} will provide predictions for GW
sensitivity of those PTAs.

\subsection{Initial Pulsar - GW Experiments}\label{GWStart}
In \S\ref{sec:GR}, we saw that the gravitational field of a companion
star delays pulsar radiation when it passes close to the star. When
GWs travel past the line of sight, the changing gravitational
potential this entails has a similar effect, even though the details
differ. This idea was first explored by \citet{saz78}, who analysed
the effect GWs from stellar binaries would have on timing, if the
binary was situated close to the line of sight between the pulsar and
the Earth. His work showed that a close enough alignment is rather
unlikely and that, even if such an alignment existed, it could prove
impossible to distinguish the GW-induced sinusoid from a planet
orbiting the pulsar, since the effect would only be seen in a single
pulsar. However, the potential to detect the influence of GWs from a
supermassive black hole (SMBH) binary system, proved more
promising. Because of the extreme gravitational character of these
systems, their effect is predicted to be significant over cosmological
distances - which implies it would affect the timing of all pulsars,
not just one. This idea was further explored by \citet{det79} who
first used pulsar timing residuals to place a limit on the energy
density of a stochastic background of such cosmological GW sources. A
few years later, \citet{hd83} first used the fact that the effect of
this gravitational wave background (GWB) must be correlated between
different pulsars. To understand why, considering a single
gravitational wave is a helpful stepping stone. As with
electromagnetic waves, gravitational waves act perpendicularly to
their direction of propagation. However, unlike electromagnetic waves,
they have a quadrupolar signature. This means that, perpendicular to
the GW's direction of travel, pulsars in opposite parts of the sky
undergo identical effects (positive correlation) and pulsars offset by
90$^{\circ}$ undergo opposed effects (negative correlation), while
there is no impact on pulsars along the direction of propagation.
\begin{figure}
  \centerline{\psfig{figure=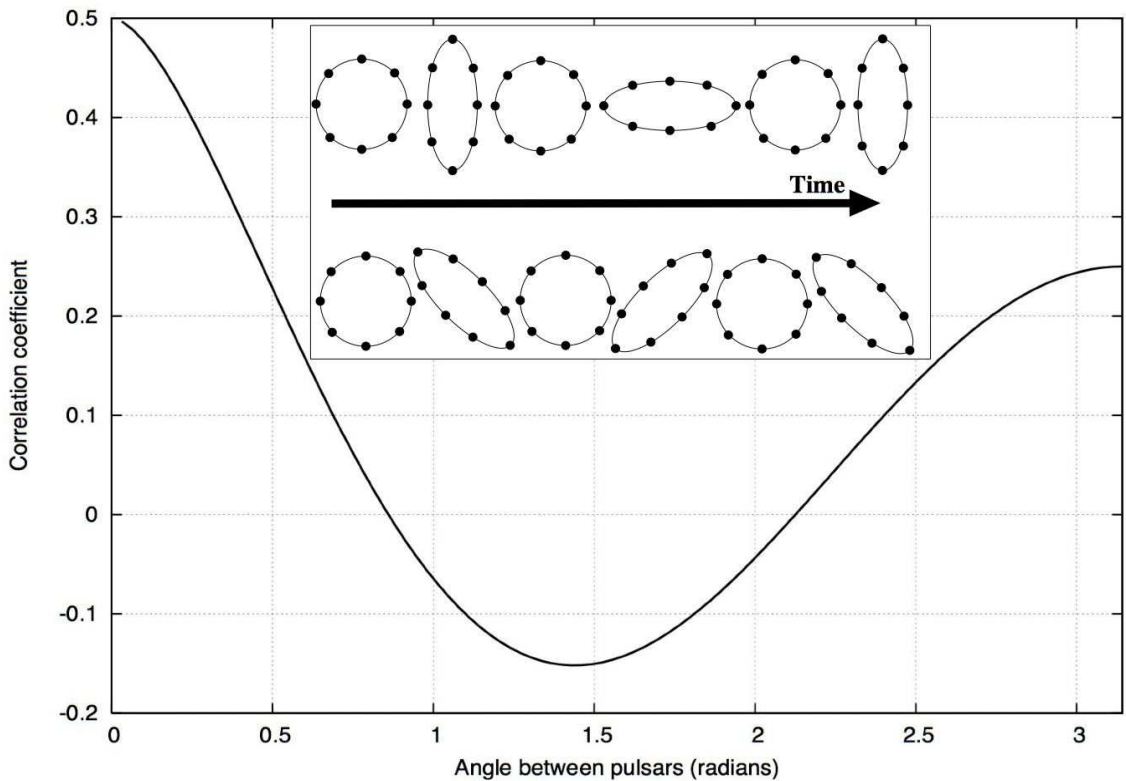,width=15cm}}
  \caption[Hellings \& Downs curve and effect of a gravitational wave
	on a ring of test particles]{Hellings \& Downs curve as a function
	of angular separation between pulsars. Notice the correlation only
	rises up to 0.5. This is because the GW effect on the Earth is
	correlated between pulsars, but the equally large effect on the
	pulsar itself is uncorrelated. The inset shows the effect of a
	gravitational wave on a ring of test particles. The direction of
	propagation of the wave is perpendicular to the page and the
	images show the evolution as a function of time, progressing
	horizontally. Top row: +-polarised GW; bottom row:
	$\times$-polarised GW.}
  \label{fig:GWEffect}
\end{figure}
This quadrupolar effect - as shown in Figure \ref{fig:GWEffect} -
turns out to be characteristic for a gravitational wave background
(GWB) as well, as demonstrated by \citet{hd83}. Mathematically, it is
described as follows \citei{as given by}{jhlm05}:
\begin{equation}\label{eq:HDCurve}
  \zeta(\theta) = \frac{3}{2}x \log{x} - \frac{x}{4} + \frac{1}{2}
\end{equation}
with $x = 0.5 \left( 1 - \cos{\theta}\right)$ and $\theta$ the angular
separation of the pulsars on the sky. This expected correlation as a
function of angle between the pulsars is commonly referred to as the
``Hellings \& Downs curve''.

With the first MSP discovered in 1982, by the end of the 1980s the
superior timing stability of these more rapidly rotating neutron stars
was acknowledged and the concept of a pulsar timing array (PTA) was
proposed (in short succession by \pcite{rom89} and \pcite{fb90}). The
fundamental idea behind a PTA is to time a group of highly stable MSPs
and analyse the correlations between their timing residuals in order
to optimise sensitivity to several corrupting effects. Three sources
of correlations are expected to exist in the timing residuals of
MSPs. The first source consists of errors in the observatory
clocks. This would affect all pulsars in the same way at the same
time: the correlation coefficient will be positive and equal for all
pulsar pairs - this is a monopole correlation. The second source of
correlations are inaccuracies in the SSE. At any point in time, an
error in the SSE will correspond to an artificial offset in the
calculation of the SSB. This will imply that pulses from pulsars in
the direction of the artificial offset will be seen to arrive late;
pulsars in the opposing hemisphere will be considered early and
pulsars at right angles with the offset will be unperturbed. This
correlation signature is therefore a dipole defined by the error in
the SSE. Finally, correlations due to a GWB would induce correlations
as outlined in the previous paragraph and shown in Figure
\ref{fig:GWEffect}. This effect is quadrupolar and therefore
fundamentally different - and easily distinguishable - from both clock
errors and errors in the SSE.

\subsection{Limits on the Power in the GWB}\label{ssec:Limits}
In the decade that followed the proposal to construct timing arrays,
most efforts focussed on using pulsar timing to limit the potential
power in the GWB. The main reason for this was that too few stable
pulsars to attempt a detection were known. Following the summary of
\citet{jhv+06}, the spectral power induced by a GWB in pulsar timing
residuals, is:
\begin{equation}\label{eq:GWBResPower}
  P(f) = \frac{h_{\rm c}(f)^2}{12 \pi^2 f^3} =
  \frac{A^2}{12\pi^2}\frac{f^{2\alpha-3}}{f_{\rm 0}^{2\alpha}}, 
\end{equation}
where $h_{\rm c}(f)$ is the characteristic strain spectrum, defined
through:
\begin{equation}\label{eq:h_cintro}
  h_{\rm c}(f) = A \left(\frac{f}{f_{\rm 0}}\right)^{\alpha}
\end{equation}
with $A$ the amplitude of the GWB, $f$ the frequency in units of
yr$^{-1}$ and $f_{\rm 0} = 1$\,yr$^{-1}$. $\alpha$ is the spectral
index of the GWB, which depends on the type of GWB considered, as
detailed in \S\ref{sec:GWBSources}. Since $\alpha < 0$ for all
predicted backgrounds (see \S\ref{sec:GWBSources}), the spectral index
of the effect on the residuals is highly negative and therefore
strongly dominated by low frequencies - implying a strong correlation
between GWB sensitivity and data span. Note the effect on the timing
residuals is not only dependent on the GWB amplitude $A$, but also on
the spectral index of the GWB, $\alpha$. This implies that any limit
on $A$ derived from pulsar timing residuals will have to specify the
spectral index under consideration as well, resulting in different
limits on $A$ for different GWBs, as will follow.

The first limit on the strength of the GWB derived from MSP timing,
was presented by \citet{srtr90}, who used seven years of Arecibo data
on PSRs J1939+2134 and J1857+0943\footnote{The original names for
these pulsars are PSRs B1937+21 and B1855+09. Throughout this thesis
the more recent J2000 names are used.} and assumed a spectral index
$\alpha = -1$. At $95\%$ confidence, they limited the energy density
of the GWB per unit logarithmic frequency interval to $ \Omega_{\rm
g}h^2 < 4\times 10^{-7}.$ This can be converted into the amplitude
used above, through Equation 3 of \citet{jhv+06}:
\begin{equation}
  \Omega_{\rm gw}(f) = \frac{2}{3}\frac{\pi^2}{H_{\rm 0}^2} f^2
  h_{\rm c}(f)^2
\end{equation}
and hence:
\begin{equation}
  \Omega_{\rm gw}(f) h^2 = \frac{2 \pi^2}{3\times
  10^4}\left(\frac{F_{\rm 1}}{F_{\rm 2}}\right)^2
  \frac{f^{2\alpha+2}}{f_{\rm 0}^{2\alpha}} A^2, 
\end{equation}
where we defined the Hubble constant as $H_{\rm 0} = 100 h$\,km/s/Mpc
and introduced the normalisation factors $F_{\rm 1} = 3.0856\times
10^{19}$\,km/Mpc and $F_{\rm 2} = 3.15576\times 10^7$\,s/yr to convert
units properly. Numerically, this approximates as:
\begin{equation}\label{eq:AOmConversion}
  \Omega_{\rm gw}(f) h^2 \approx 6.29\times 10^{20} A^2 f^{2\alpha + 2}.
\end{equation}
The limit of $\Omega_{\rm g} h^2 < 4\times 10^{-7}$ for $\alpha = -1$
can therefore be converted to $A < 3\times 10^{-14}$.

The analysis performed by \citet{srtr90} was based on a type of power
spectrum constructed from Gram-Schmidt orthonormal polynomials that
were fitted to increasingly short subsets of the timing residuals to
provide a measure of power at increasingly high frequencies. While
these ``Gram-Schmidt spectra'' are very powerful tools for
investigations of very steep power spectra \citep{db82,dee84}, their
translation into traditional power spectra and Fourier transforms is
unclear and the interpretation of any feature of these spectra is
therefore difficult. They can, however, be used to determine limits by
comparing the power levels obtained from actual data to levels
calculated for a supposed GWB amplitude.

This analysis, which was reproduced with longer data sets by
\citet{ktr94} and \citet{lom02}, was fundamentally sound, though the
final calculation of certainty levels was flawed with the potential to
obtain probabilities in excess of unity, as pointed out by
\citet{td96}. These last authors proposed an alternative approach to
the problem and applied the Neyman-Pearson statistical test, resulting
in a $95\%$ confidence limit of $ \Omega_{\rm g}h^2 < 1.0\times
10^{-8}$ (equivalent to $A < 4\times 10^{-15}$). Their approach
compares the likelihood of a ``zero hypothesis'', $H_{\rm 0}$, which
states that all timing residuals are purely due to statistically white
noise (i.e. radiometer noise) to the likelihood of an alternative
hypothesis, $H_{\rm 1}$, which states the timing residuals are a
combination of white noise and GWB. This analysis was invalidated in
turn by \citet{mzvl96}, who pointed out it didn't properly account for
errors of the first and second kinds, which quantify the probabilities
of incorrect assessment of hypotheses. \citet{mzvl96} subsequently
proposed a Bayesian approach, which resulted in the weaker limit of
$\Omega_{\rm g}h^2 < 9.3\times 10^{-8}$ at $95\%$ confidence ($A <
1\times 10^{-14}$).

More recently, \citet{jhv+06} developed a Monte-Carlo based method of
limiting the GWB amplitude. Their method uses newly developed software
that enables simulation of GWB effects on timing residuals in a way
that allows a direct comparison of real data with GW-affected,
simulated, data \citep{hjl+09}. This has the advantage of allowing a
more rigorous statistical analysis, though the \citet{jhv+06} method
implicitly requires the pulsar timing data to be $100\%$ statistically
white, which is a problematic requirement, especially for data sets
with long time spans. Notwithstanding this restriction, they obtained
the most stringent limits to date: $\Omega_{\rm g}h^2 < 2.0\times
10^{-8}$ (or $A < 6\times 10^{-15}$) at $95\%$ confidence for a
background with $\alpha = -1 $ and $A < 1.1\times 10^{-14}$ at 95\%
confidence for a background with $\alpha = -2/3$. (Note the limit on
$\Omega_{\rm g}h^2$ is dependent on the gravitational wave frequency,
as shown in Equation \ref{eq:AOmConversion}. So unless $\alpha = -1$,
one should always specify the GW frequency connected to the
$\Omega_{\rm g}h^2$ limit. For ease of use we provide limits on $A$
instead, except when quoting from literature. These limits can easily
be converted using Equation \ref{eq:AOmConversion}.) Based on the same
simulation software, \citet{vlml08} have proposed a Bayesian technique
to both detect and limit the GWB in PTA data sets, but this technique
has not yet been applied to actual data.

\subsection{PTA-detectable GW sources}\label{sec:GWBSources}
There is a large variety of sources to which pulsar timing arrays
could be sensitive, including both single sources and
backgrounds. Single sources, such as binary SMBHs, are possibly
non-existent in the Milky Way. \citet{lb01} analysed the effect of a
potential black hole binary with total mass $5\times 10^6\,{\rm
M}_{\odot}$ in the centre of the Milky Way. They concluded that the
induced timing residuals would be at or below the 10\,ns level - which
is far below current timing sensitivity. In the case of gravitational
wave \emph{backgrounds} (GWBs), however, stronger signals might be
expected.

As shown in the overview by \citet{mag00}, there are several predicted
origins for GWBs that would be detectable through PTA research. The
most important one of these is a background of binary SMBHs in the
relatively nearby (redshift $z \approx$1-2) Universe. Based on the
premise of hierarchical galaxy formation, simulations such as those of
\citet{rr95a}, \citet{jb03}, \citet{wl03a} and \citet{eins04}, have
shown that as galaxies merge, the black holes at their centres
initially become a binary pair and eventually merge. This would give
rise to both SMBHs and a large number of black hole (and SMBH)
binaries and mergers - which generate gravitational waves as they
spiral in as well as during and after the merging event. The ensemble
of a large number of these creates a background with a spectral index
of $-2/3$ and amplitudes predicted to lie between $10^{-15}$ and
$10^{-14}$. As a point of comparison, the most stringent limit from
pulsar timing to date places a bound on this background of
$A\le1.1\times 10^{-14}$ \citep{jhv+06}.

There are, however, some potential problems with these
models. Firstly, it is unclear whether there is sufficient orbital
momentum loss for the SMBH binary to merge within a Hubble time. At
the initial stages the black holes are surrounded by accretion disks
which cause orbital energy loss through friction and AGN
activity. However, it is possible that all the stellar material and
dust surrounding the black holes will be discarded long before a
merger event.  This would leave gravitational radiation as the only
means to lose energy, but this radiation is far less powerful than
that generated in a merger event and it releases too little energy to
cause a collapse within a Hubble time. Another major issue lies in
simplifications within the models and large uncertainties in input
parameters to the simulations. A recent analysis by \citet{svc08}
shows that the spectral index of $\alpha = -2/3$ strongly depends on
the merging history of galaxies and is probably
underestimated. Furthermore, they demonstrate \citei{as did}{rr95a}
that at higher frequencies, the GWB would be dominated by a few bright
sources at smaller distance. While this means that the results of the
simulations - and the predictions following from them - are far less
secure than one would hope for, it also provides additional value to a
potential detection, as this will uncover significant (and currently
inaccessible) information about galaxy formation history.

A second potential GWB in the PTA sensitivity range originates from
cosmic strings \citep{cbs96,dv05}. The background generated by these
would have a steeper power spectrum (spectral index $\alpha = -7/6$)
than the GWB from SMBH mergers and is therefore more easily detectable
over longer lengths of time. The main problems with these backgrounds
are the very limited knowledge of input parameters to the models and
the potential non-existence of the cosmic strings
altogether. Currently predicted amplitudes for this background lie
between $10^{-16}$ and $10^{-14}$, but current limits already place a
bound at $3.9\times 10^{-15}$ \citep{jhv+06}.

The third potential background is the gravitational wave equivalent to
the cosmic microwave background: it is composed of gravitational waves
created in the Big Bang \citep{gri05,bb08}. Spectrally speaking this
background lies between the previous two, with a spectral index
expected around $-0.8$ or $-1.0$. The amplitudes predicted by
\citet{gri05} for this background are between $10^{-17}$ and
$10^{-15}$, making it the weakest of the three backgrounds and
therefore the least likely one to be detected any time soon. Other
authors \citei{such as }{bb08} predict even lower amplitudes for this
GWB.

Given the amplitudes, spectral indices and caveats concerning the
backgrounds discussed above, the remainder of this thesis will focus
on the GWB due to SMBH coalescence, assuming for ease of use a
spectral index of $\alpha = -2/3$ and amplitude range of $10^{-15} -
10^{-14}$. We notice that the analysis by \citet{svc08} warrants a
more complex analysis in order to make detailed assessment of actual
detection or exclusion of astrophysical models based on limits, but
given the large uncertainties in any of these models, we take these
values to provide a decent first order approximation.

\subsection{Pulsar Timing Arrays}\label{ssec:PTAs}
With the large number of pulsar discoveries in surveys after 1990
\citep{mld+96,cnst96,ebvb01,mlc+01}, the feasibility of PTA-type
projects has improved, inspiring a new assessment of the PTA concept
by \citet{jhlm05}. They presented for the first time a thorough
analysis of the timing precision required for detection of a GWB. In
the case of a homogeneous timing array (i.e. a timing array in which
every pulsar has an identical timing residual RMS), they derived the
following sensitivity curve:
\begin{equation}\label{eq:PTASensitivity}
  S =
  \sqrt{\frac{M\left(M-1\right)/2}
	{1+\left[\chi\left(1+\bar{\zeta}^2\right)+2\left(\sigma_{\rm
		n}/\sigma_{\rm g}\right)^2+\left( \sigma_{\rm n}/ \sigma_{\rm
  g}\right)^4 \right] / \left(N \sigma_{\zeta}^2\right)}},
\end{equation}
where $M$ is the number of pulsars in the PTA, $N$ is the number of
observations for each of these pulsars, $\zeta$ is the Hellings \&
Downs correlation between two pulsars, $\sigma_{\rm n}$ is the RMS of
the non-GW noise, $\sigma_{\rm g}$ is the residual RMS caused by the
GWB and
\[
\chi = \frac{1}{N\sigma_{\rm g}^4} \sum_{i=0}^{N-1} \sum_{j=0}^{N-1}
c_{\rm ij}^2
\]
with $c$ the GW-induced correlation between the pulsar residuals. One
consequence of Equation \ref{eq:PTASensitivity} is that the
sensitivity saturates: for very strong GWBs, $\sigma_{\rm g}\gg
\sigma_{\rm n}$ and therefore the equation reduces to $S \approx 0.16
\sqrt{M\left(M-1\right)}$, only dependent on the number of
pulsars. However, as also described by \citet{jhlm05}, prewhitening
schemes could reduce this self-noise and assure increased sensitivity
to well beyond this threshold. 

Another way of rewriting Equation \ref{eq:PTASensitivity} is to
evaluate the GWB amplitude at which such saturation becomes
significant - effectively the lowest amplitude to which the PTA is
sensitive. This amplitude is approached as $N \sigma_{\rm g}^2$
becomes much larger than $\sigma_{\rm n}^2$. Using $13 \sigma_{\rm
n}^2 = N \sigma_{\rm g}^2$ as a threshold (the factor of 13 was
derived to achieve the $3 \sigma$ detection level in the case of a
timing array with 20 pulsars) and using Equation 8 from
\citet{jhlm05}:
\begin{equation}
  N \sigma_{\rm g}^2 = \frac{N A^2}{12\pi^2\left(2-2\alpha\right)}
 \frac{\left( f_{\rm l}^{2\alpha-2}-f_{\rm h}^{2\alpha-2}\right)}
 {f_{\rm 0}^{2 \alpha}}
\end{equation}
where $A$ is the amplitude of the GWB, $\alpha$ the spectral index of
the GWB, $f_{\rm l}$ is the lowest spectral frequency the pulsar data
set is sensitive to and $f_{\rm h}$ is the high-frequency
cutoff. \citei{Notice the factor of 12 based on the more recent
definition of $A$ as given in}{jhv+06}. Following \citet{jhlm05}, we
use $f_{\rm l} = T^{-1}$ and $f_{\rm h} = 4T^{-1}$ where $T$ is the
length of the data set. Rewriting Equation \ref{eq:PTASensitivity}, we
derive the lowest amplitude at which a $3\sigma$ detection can be
made (for an array with 20 MSPs):
\begin{equation}\label{eq:PTASScaling}
  A_{\rm S = 3} \approx 2.3\times 10^{-12}\frac{\sigma_{\rm
  n}}{T^{5/3}\sqrt{N}}
\end{equation}
for a background with spectral index $\alpha = -2/3$ (see
\S\ref{sec:GWBSources}). Notice the units for Equation
\ref{eq:PTASScaling} are $\mu$s for $\sigma_{\rm n}$ and years for
$T$. This relation determines the fundamental trade-off any PTA will
be determined by, showing the strong dependence on the length of the
observational campaign, $T$, as well as on the timing precision,
$\sigma_{\rm n}$. As a baseline scenario for future PTA efforts,
\citet{jhlm05} proposed a PTA based on weekly observations of 20 MSPs,
timed at 100\,ns residual RMS for 5 years (i.e. $N = 250$, $M = 20$,
$\sigma_{\rm n} = 0.1\,\mu$s, $T = 5$\,years). Such an array would be
sensitive at the $3\sigma$ level to backgrounds with amplitudes larger
than $10^{-15}$ (see \S\ref{sec:GWBSources}).
 
\section{Thesis Structure}
\label{intro:struct}

The remainder of this thesis is structured as follows. In Chapter
\ref{chap:Hardware}, the hardware used in our observations will be
described, along with some fundamentals of radio astronomy required
for this description. As the baseline scenario for PTAs presented
above shows, pulsar timing needs to achieve high levels of timing
precision ($\sim 100\,$ns) and maintain this timing precision over
many years ($T > 5$\,yrs). Chapters \ref{chap:0437} and
\ref{chap:20PSRS} address these requirements. Specifically, Chapter
\ref{chap:0437} presents the highest-precision pulsar timing data set
that has a time span of a decade. Such data sets enable several
interesting investigations into the pulsar astrometric and binary
parameters, which are also presented. Chapter \ref{chap:20PSRS}
contains the first large sample of MSPs to be timed over substantial
timescales (12\,yrs on average) and presents an analysis of MSP
stability and the predictions for PTA sensitivity following from
that. The sensitivity analysis we use goes beyond the simple
homogeneous PTA presented in \S\ref{ssec:PTAs} and uses the analysis
presented in Appendix \ref{app:Sens}. Having presented some of the
longest MSP timing data sets at high timing precision, Chapter
\ref{chap:GWBLimit} presents a straightforward method of limiting the
strength of the GWB based on these data sets and derives a new limit
on that background, from our data. The results are interpreted and
summarised in Chapter \ref{chap:conclusion}. The various abbreviations
and symbols used throughout this thesis are listed for easy reference
in Appendix \ref{app:nomen}.

\chapter[Observing Systems]{Radio Astronomy Fundamentals and Observing
  Hardware}
\label{chap:Hardware}
\noindent \textsf{It is a riddle wrapped in a mystery inside an enigma\\}
\vspace{0.25cm}
\textit{Winston Churchill, 1939}
\vspace{1.5cm}

\section{Abstract} 
In this chapter a brief overview is given of the hardware with which
the data analysed in subsequent chapters were acquired. Since this
thesis reports on ten years of pulsar timing, the variety of backends
is large and almost provides a full overview of historic pulsar timing
observing systems. Given the increasing complexity of these systems
over the years, we have chosen to adopt a chronological approach in
our discussion (Section \ref{sec:Hardware}), describing the oldest
backends - the analogue filterbanks - first in \S\ref{sec:AFB},
followed by the autocorrelation spectrometers in \S\ref{sec:FPTM} and,
finally, the coherent dedispersion baseband systems in
\S\ref{sec:CPSR}. The actual instruments used, are listed in
\S\ref{sec:Inst}. First, some fundamental principles of radio
astronomy will be outlined in Section \ref{sec:RA}.

\section{Fundamentals of Radio Astronomy}\label{sec:RA}
\subsection{Blackbody Radiation and Brightness Temperature}
\begin{figure}
  \centerline{\psfig{figure=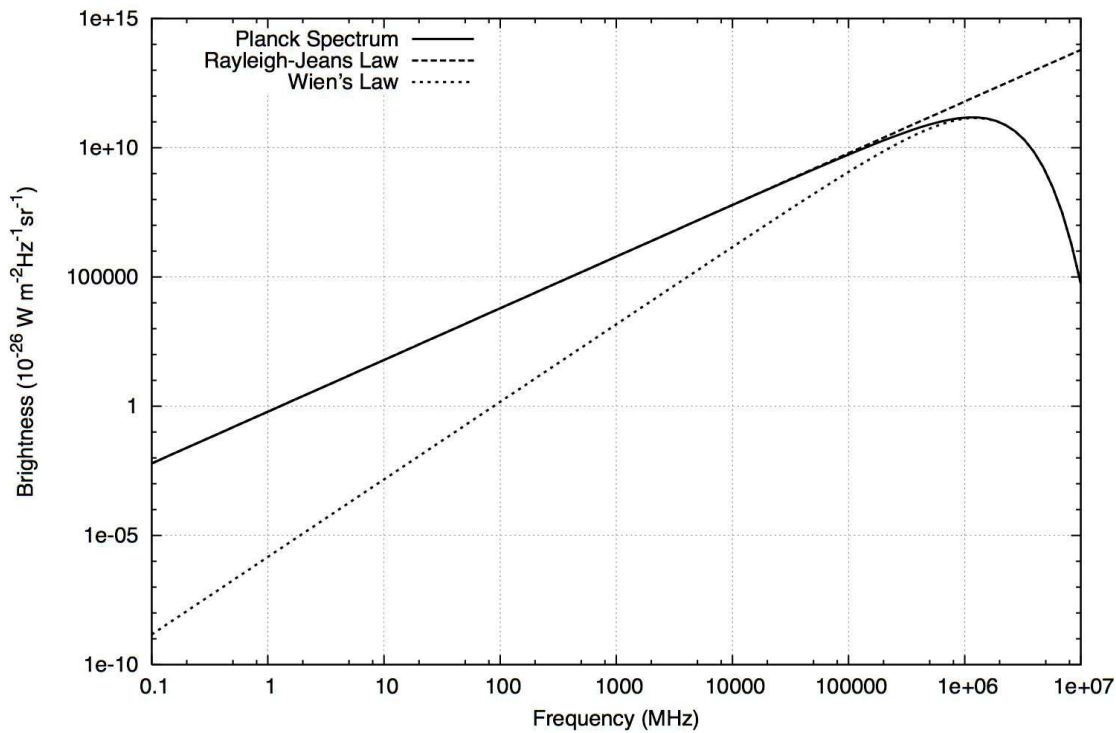,width=15cm,angle=0}}
  \caption[20\,K blackbody spectrum]{Planck spectrum for a blackbody
    at 20\,K (full line). Also shown are the Rayleigh-Jeans
    approximation (dashed line) and Wien's law (dotted line),
    demonstrating the different regimes in which these laws
    approximate the Planck spectrum.}
  \label{fig:BlackBody}
\end{figure}

Macroscopic bodies of finite temperature emit radiation. In most
cases, the spectrum of this radiation is reasonably well approximated
by that of blackbody radiation. A blackbody is a hypothetical object
with perfect absorption and emission properties. The spectrum of its
heat-induced emission is defined by Planck's law:
\begin{equation}\label{eq:Planck}
  B(\nu,T) = \frac{2 h \nu^3}{c^2} \frac{1}{e^{h\nu/kT}-1},
\end{equation}
with $B$ the brightness in W m$^{-2}$ Hz$^{-1}$ sr$^{-1}$, $\nu$ the
frequency of the emission, $T$ the temperature of the body, $h$
Planck's constant, $c$ the speed of light and $k$ Boltzmann's
constant. Figure \ref{fig:BlackBody} shows the Planck spectrum for a
blackbody of temperature 20\,K, along with two approximations. The
first one of these is the Rayleigh-Jeans law, which approximates a
Planck spectrum at the low frequency end:
\begin{equation}\label{eq:RayleighJeans}
  B(\nu,T) = \frac{2\nu^2}{c^2}kT.
\end{equation}
The second approximation, Wien's law, approximates at the
high-frequency end:
\begin{equation}\label{eq:Wien}
  B(\nu,T) = \frac{2h\nu^3}{c^2}e^{-h\nu/kT}.
\end{equation}
From Figure \ref{fig:BlackBody}, it is quite clear that, at radio
frequencies, the Rayleigh-Jeans law is a simple but precise
approximation that can easily be used, at least for the 20\,K body
shown in the Figure. Equation \ref{eq:Planck} shows that the peak of
a blackbody spectrum increases to higher frequencies as the
temperature of the body increases. More precisely, Wien's displacement
law predicts the peak frequency to be:
\begin{equation}\label{eq:WienDisplacement}
\frac{\nu_{\rm max}}{\rm GHz} = 48.8\frac{T}{\rm K}
\end{equation}
\citei{after}{rw00}. This implies that the Rayleigh-Jeans law will be
a useful approximation for all practical radio astronomy purposes. 

Based on these formulae, it is possible to prove that pulsar radiation
is not thermal in character. Consider the Crab pulsar, B0531+21, as an
example. Its average flux is 14\,mJy ($= 14\times 10^{-29}\,$W
m$^{-2}$Hz$^{-1}$) at an observing frequency of 1.4\,GHz
\citep{lylg95}. Also, based on its $DM$ of 56.791\,cm$^{-3}$pc
\citep{cr71}, its distance is estimated to be 2.49\,kpc \citei{as
  described in \S\ref{ssec:ISM}, based on the Galactic ISM model
  of}{tc93}. Assuming an emission region of 10\,km radius, the solid
angle of this emission would be: $\pi($10\,km/2490\,pc$)^2 = 5.3\times
10^{-32}$\,sr and the pulsar therefore has a brightness of $2.6\times
10^3$\,W m$^{-2}$Hz$^{-1}$sr$^{-1}$. Inserting this into Equation
\ref{eq:RayleighJeans} results in a blackbody temperature of $T_{\rm
  b} = 4\times10^{24}\,$K. Given Equation \ref{eq:WienDisplacement},
such a brightness temperature would be expected to have the peak of
its emission at frequencies around $20\times 10^{25}\,$GHz, which is
well beyond the gamma ray part of the spectrum. If pulsar emission
were thermal in origin, therefore, it should be easily visible across
all bands\footnote{While some pulsars - like the Crab pulsar B0531+21
  are indeed seen across the spectrum, their spectral index at radio
  frequencies is still negative - implying an inversion with respect
  to the Planck spectrum}. This argument is taken one step further by
\citet{mt77}, in relating the particle energy required for incoherent
radiation. In the case of thermal emission, the required energy of the
particles involved is bounded as $k T_{\rm b} < \epsilon$, with $k$
the Boltzmann constant and $T_{\rm b}$ the brightness
temperature. This would imply particle energies of $\epsilon >
3.4\times 10^{20} {\rm eV}$. Such high particle energies cannot be
produced by any known processes and as a consequence coherent emission
must lie at the basis of the pulsar emission at radio
wavelengths. Finally, as seen in Figure \ref{fig:BlackBody}, the
spectral index of thermal radiation at radio wavelengths would be
expected to be positive. For pulsars the spectral index is generally
negative \citep{lylg95}.

\subsection{Noise and Amplification in the Signal Chain}\label{ssec:Amplify}
A simple consequence of the theory of blackbody radiation is that all
systems in the signal chain radiate energy and hence contribute to the
noise in the signal. Now consider a system chain with $N$ elements,
each with noise temperature $T_{\rm i}$ and gain $G_{\rm i}$ ($G > 1$
for amplifiers, $G < 1$ for all other elements). The input power or
astronomical signal is: $P_{\rm 0} = k T_{\rm A}$. After the first
element, the power becomes:
\[
P_1 = k(T_{\rm A}+T_1)G_1.
\]
For a system chain with $N$ elements, the power at the end will
therefore be:
\[
P_N = k(T_{\rm A}+T_{\rm X}) \prod_{i=1}^N G_{\rm i}
\]
with the additional noise temperature: 
\begin{equation}
T_{\rm X} = T_1 + \frac{T_2}{G_1}+\frac{T_3}{G_1G_2}+\ldots +
\frac{T_N}{G_1G_2\ldots G_{N-1}} = T_1+\sum_{i=2}^N
\frac{T_i}{\prod_{j=1}^{i-1}G_j}.
\end{equation}
This clearly demonstrates the importance of the first element in the
system: its noise temperature is the most prominent addition to the
system-induced noise and its gain reduces all other
contributions. This is the reason why the receiving systems are
generally cooled to temperatures of several tens of Kelvin and why
the first element after the receiver is a low-noise amplifier (as
depicted in Figure \ref{fig:SignalChain}).

\subsection{Polarisation}
Electromagnetic radiation manifests itself as a transverse wave with
perpendicular magnetic and electric fields. A monochromatic electric
wave can therefore be represented as a vector perpendicular to the
direction of propagation:
\begin{eqnarray}\label{eq:EField}
  e_{\rm x} &=& a_{\rm x} \cos \Big(2\pi(z/\lambda-\nu t)+\phi_{\rm
  1}\Big)\notag\\
  e_{\rm y} &=& a_{\rm y} \cos \Big(2\pi(z/\lambda-\nu t)+\phi_{\rm
  2}\Big)\\
  e_{\rm z} &=& 0,\notag
\end{eqnarray}
in which $a_{\rm x}$ and $a_{\rm y}$ are the amplitudes in x and y
directions and $\phi_{\rm 1}-\phi_{\rm 2}$ is the phase offset between
the two wave components. Traditionally, the characteristics of an
electric wave are expressed through the Stokes parameters, which are
defined as follows:
\begin{eqnarray}
  S_{\rm 0} & = I & = a_{\rm x}^2+a_{\rm y}^2\notag\\
  S_{\rm 1} & = Q & = a_{\rm x}^2-a_{\rm y}^2\notag\\
  S_{\rm 2} & = U & = 2a_{\rm x}a_{\rm y}\cos(\phi_{\rm 1}-\phi_{\rm 2})\\
  S_{\rm 3} & = V & = 2a_{\rm x}a_{\rm y}\sin(\phi_{\rm 1}-\phi_{\rm
  2}).\notag
\end{eqnarray}
In simple terms, $I$ can be though of as the total power of the wave,
$Q$ as a measure of ellipticity along the axes, $U$ is a measure of
ellipticity at 45$^{\circ}$ to the axes and $V$ is a measure of
circular polarisation. However, this image only applies to
waves with an infinitessimal bandwidth. For practical applications, we
will now rederive these relationships for a signal with finite
bandwidth. For such a signal, Equation \ref{eq:EField} becomes:
\begin{eqnarray}
  e_{\rm x}^{\rm r} = \int_0^{\infty} b_{\rm x}(\nu)
  \cos\Big(\phi(\nu)-2\pi\nu t\Big) {\rm d}\nu\notag\\
  e_{\rm y}^{\rm r} = \int_0^{\infty} b_{\rm y}(\nu)
  \cos\Big(\phi(\nu)-2\pi\nu t\Big) {\rm d}\nu
\end{eqnarray}
Where $b(\nu)$ is the amplitude as a function of observing frequency,
otherwise known as the bandpass of the signal. This definition can
logically be expanded into a Fourier transform:
\begin{equation}
  e_{\rm x} = e_{\rm x}^{\rm r} + i \int_0^{\infty} b_{\rm x}(\nu)
  \sin\Big(\phi(\nu)-2\pi\nu t\Big) {\rm d}\nu,
\end{equation}
and similarly for the y-component. This extended definition of the
electric field is known as the \emph{analytic signal}. An alternative
but equivalent way to consider this is as the expansion of the
components of the electric vector as follows:
\begin{equation}
  e(t) = e^{\rm r}(t)+i e^{\rm r}(t)*h(t),
\end{equation}
where $*$ denotes convolution and $h(t) = (\pi t)^{-1}$. The
convolution $x(t)*h(t)$ is the Hilbert transform and it can be more
readily analysed in the Fourier domain, since the convolution theorem
states that convolution in the time domain is equivalent to
multiplication in the Fourier domain. Hence, given the Fourier
transform of $h(t)$ to be:
\begin{equation}
  H(\nu) = 
  \begin{cases}
	-i & {\rm if\ } \nu > 0 \\
	 i & {\rm if\ } \nu < 0 
  \end{cases}
\end{equation}
and therefore:
\begin{eqnarray}
  E(\nu) & = & FT(e(t)) = E(\nu) + i E(\nu) H(\nu)\notag\\
   & = & 
  \begin{cases}
	2 E(\nu) & {\rm if\ } \nu > 0 \\
	0 & {\rm if\ } \nu < 0 
  \end{cases}
\end{eqnarray}
with $E^{\rm r}(\nu)$ the Fourier transform of the signal $e^{\rm
  r}(t)$ and $E(\nu)$ the Fourier transform of the analytic signal. 

We have defined the analytic signal as:
\begin{equation}
  \vec{e}(t) = 
	 \binom{e_{\rm x}(t)}{e_{\rm y}(t)} =
	 \binom{a_{\rm x}e^{i(\phi_{\rm x}(t)-2\pi\nu_{\rm 0}t)}}
	 {a_{\rm y}e^{i(\phi_{\rm y}(t)-2\pi\nu_{\rm 0}t)}}.
\end{equation}
The coherency matrix $\bar{\rho}$ of this vector can be related to the
Stokes parameters as outlined by \citet{bri00}:
\begin{equation}\label{eq:CoherencyMatrix}
  \bar{\rho} = \left( 
  \begin{array}{cc}
	\langle|e_{\rm x}|^2\rangle & \langle e_{\rm x} e_{\rm
	y}^*\rangle\\
	\langle e_{\rm y}e_{\rm x}^*\rangle & \langle | e_{\rm
	y}|^2\rangle
  \end{array}\right)
  = \frac{1}{2}\left(
  \begin{array}{cc}
	S_{\rm 0}+S_{\rm 1} & S_{\rm 2}-i S_{\rm 3} \\
	S_{\rm 2}+i S_{\rm 3} & S_{\rm 0}-S_{\rm 1}
  \end{array}
  \right)
\end{equation}
It is easily derived from this that the Stokes parameters can be
derived from the coherency products as follows:
\begin{equation}
  \left(
  \begin{array}{c}
	S_{\rm 0}\\
	S_{\rm 1}\\
	S_{\rm 2}\\
	S_{\rm 3}
  \end{array}
  \right)
  = 
  \left(
  \begin{array}{c}
	\langle|e_{\rm x}|^2\rangle + \langle|e_{\rm y}|^2\rangle \\
	\langle|e_{\rm x}|^2\rangle - \langle|e_{\rm y}|^2\rangle \\
	2\Re(\langle e_{\rm y}e_{\rm x}^*\rangle)\\
	2\Im(\langle e_{\rm y}e_{\rm x}^*\rangle) 
  \end{array}
  \right),
\end{equation}
with $\Re(\langle e_{\rm y}e_{\rm x}^*\rangle)$ and $\Im(\langle
e_{\rm y}e_{\rm x}^*\rangle)$ the real and imaginary parts of $\langle
e_{\rm y}e_{\rm x}^*\rangle$, respectively. It is important to note
that $e_{\rm x}^2 = a_{\rm x}^2$ and $e_{\rm y}^2 = a_{\rm y}^2$ are
simply the amplitudes of the signals in the two orthogonal directions
we commenced with (Equation \ref{eq:EField}) and are therefore readily
measured. Also, $2\Re(\langle e_{\rm y}e_{\rm x}^*\rangle) = 2\langle
a_{\rm x}a_{\rm y} \cos(\phi_{\rm x}-\phi_{\rm y})\rangle$ is the
average power of the product of those two original signals and
therefore also easily measured. Finally, $2\Im(\langle e_{\rm y}e_{\rm
x}^*\rangle) = 2 a_{\rm x}a_{\rm y} \sin(\phi_{\rm x}-\phi_{\rm
y})\rangle = 2 a_{\rm x}a_{\rm y} \cos(\phi_{\rm x}-\phi_{\rm
y}-\pi/2)\rangle$ is the same averaged power, but now with one signal
shifted by 90$^{\circ}$. All four of these numbers - and therefore all
the values of the coherency matrix $\bar{\rho}$ (Equation
\ref{eq:CoherencyMatrix}), are readily determined from two orthogonal
probes, both in hardware and in software. Most of the receivers used
for data acquisition in this thesis measure two orthogonal and linear
components to the electric field as described above. On a more general
note, however, it is possible to rewrite Equation \ref{eq:EField} in
terms of two circular probes with opposite handedness - in which case
the same results ensue. Such a derivation is beyond the scope of this
introduction, but can be found in \citet{rw00}.

%
%
\section{Observing Hardware}\label{sec:Hardware}
\subsection{Basic Signal Chain}\label{sec:Chain}
\begin{figure}
  \centerline{\psfig{figure=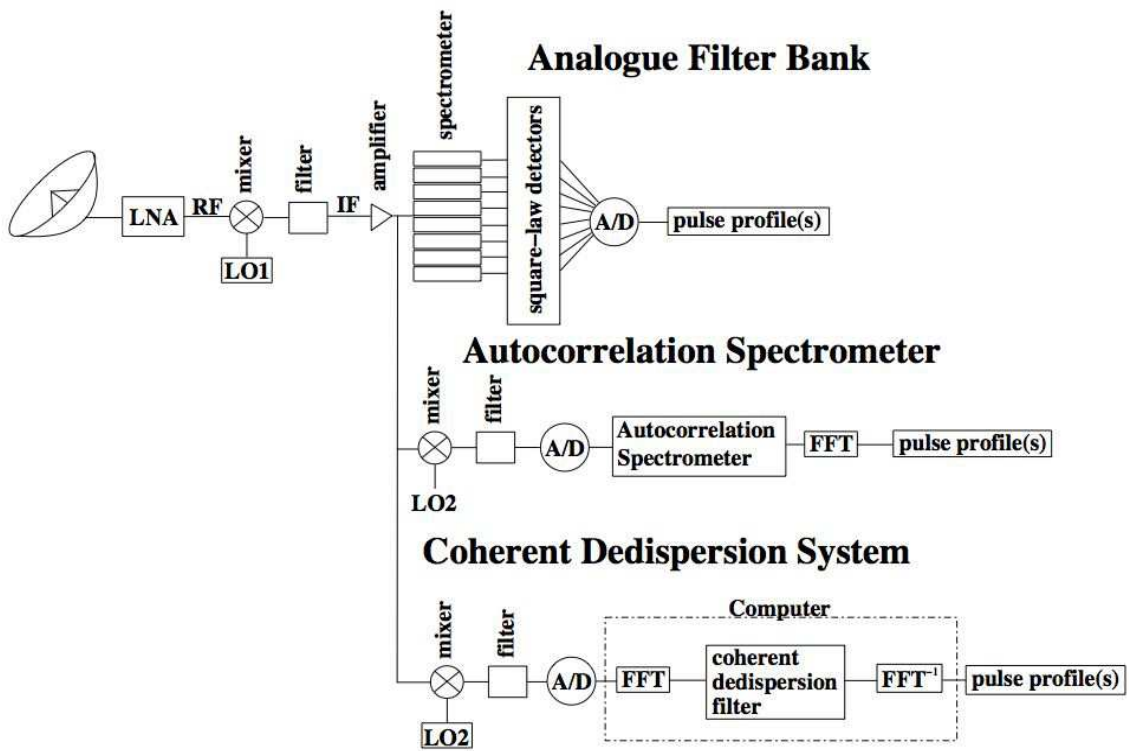,width=15cm,angle=0}}
  \caption[Signal chain for different backend systems]{Fundamental
  system chain for different backends used for the radio pulsar
  observations analysed in this thesis.}
  \label{fig:SignalChain}
\end{figure}

%
Figure \ref{fig:SignalChain} shows the components of a signal chain
for pulsar timing observations. The first elements of the system are
standard in all radio astronomy observations and will be discussed
here. Starting at the left end, we have the radio antenna, which
focusses the radio waves originating at the astronomical source. To
this end, the telescope surface is parabolic, with a receiver (which
converts the electromagnetic radiation into voltages) in the
focus\footnote{Note that in the case of a Cassegrain system there is
only a reflective surface near the focus of the paraboloid, while the
receivers are placed at the secondary focus, either to the side of the
antenna (Green Bank Telescope, e.g.) or, more commonly, in the dish
surface (Australia Telescope Compact Array, e.g.). For ease of
discussion, we will consider the set-up of the Parkes Radio Telescope,
which has the receivers and focus cabin at the focus of the
paraboloid.}. For reasons clarified in \S\ref{ssec:Amplify}, the
receivers are cooled and the signal is passed through a cryogenically
cooled low noise amplifier (LNA). Next the signal will be transfered
to the ground or control room, where all other hardware resides. This
involves data transfer via cables that normally attenuate
high-frequency ($\nu \approx 1\,$GHz) signals. In order to transfer
the radio signal with minimal loss, its frequency is therefore first
downconverted. This is accomplished by mixing the signal with that
from a local oscillator (LO) at a precisely defined
frequency. Mathematically speaking, this mixing is a multiplication of
two waves, which equates to an upconverted signal at the summed
frequency and a downconverted signal at the difference of the
frequencies:
\[
\cos{\nu_{\rm obs}t}\times\cos{\nu_{\rm LO}t} =
0.5\big(\cos{(\nu_{\rm obs}-\nu_{\rm LO})t}+\cos{(\nu_{\rm
	obs}+\nu_{\rm LO})t}\big)
\]
The downconverted signal is selected by running the signal through a
low-pass filter subsequently. Notice that, while the LO frequency
$\nu_{\rm LO}$ is precisely defined, the observed signal has a finite
bandwidth $B$ that depends on the receiver response. The downconverted
signal therefore has a bandwidth as well, with a frequency range of
$\nu_{\rm obs}-\nu_{\rm LO}-B/2$ to $\nu_{\rm obs}-\nu_{\rm
LO}+B/2$. This means that if the LO frequency is smaller than the
observed centre frequency ($\nu_{\rm LO} < \nu_{\rm obs}$), the
resulting downconverted signal will be an exact copy of the original
bandpass, simply translated in frequency - this is called \emph{upper
sideband} downconversion. If the reverse is true: $\nu_{\rm LO} >
\nu_{\rm obs}$, then the bandpass is mirrored in addition to being
translated. This is called \emph{lower sideband}
downconversion. Through downconversion, the radio frequency (RF)
signal is reduced to an intermediate frequency (IF) signal. The
low-pass filter used in this downconversion process can also be used
as a bandpass filter to select the frequency range required by the
backend. Alternatively, another stage of filtering may need to be
applied. Following another step of amplification, the signal is passed
on to the different pulsar instruments or backends, which are
described below.

\subsection{Analogue Filter Banks}\label{sec:AFB}
There are two types of resolution a pulsar backend attempts to
achieve. Firstly, time resolution is required at a level well below
the pulse period. Secondly, frequency resolution is required in order
to enable mitigation of the dispersive effects illustrated in Figure
\ref{fig:Dispersion} and described by Equation \ref{eq:DMdelay}. The
easiest means to achieve frequency resolution is to pass the signal
through a series of parallel bandpass filters with adjoining frequency
responses. Each of these filters creates a single frequency channel,
the width of which determines the frequency resolution of our
data. The analogue power is subsequently sampled in each of these
channels, at a given sampling periodicity that determines the time
resolution. Finally, the signals of the different channels are shifted
in time relative to each other, according to the expected DM delay
calculated through Equation \ref{eq:DMdelay}. While this corrects for
the DM delay between the channels, it is incapable of correcting the
smearing within each channel. 

\begin{figure}
  \centerline{\psfig{figure=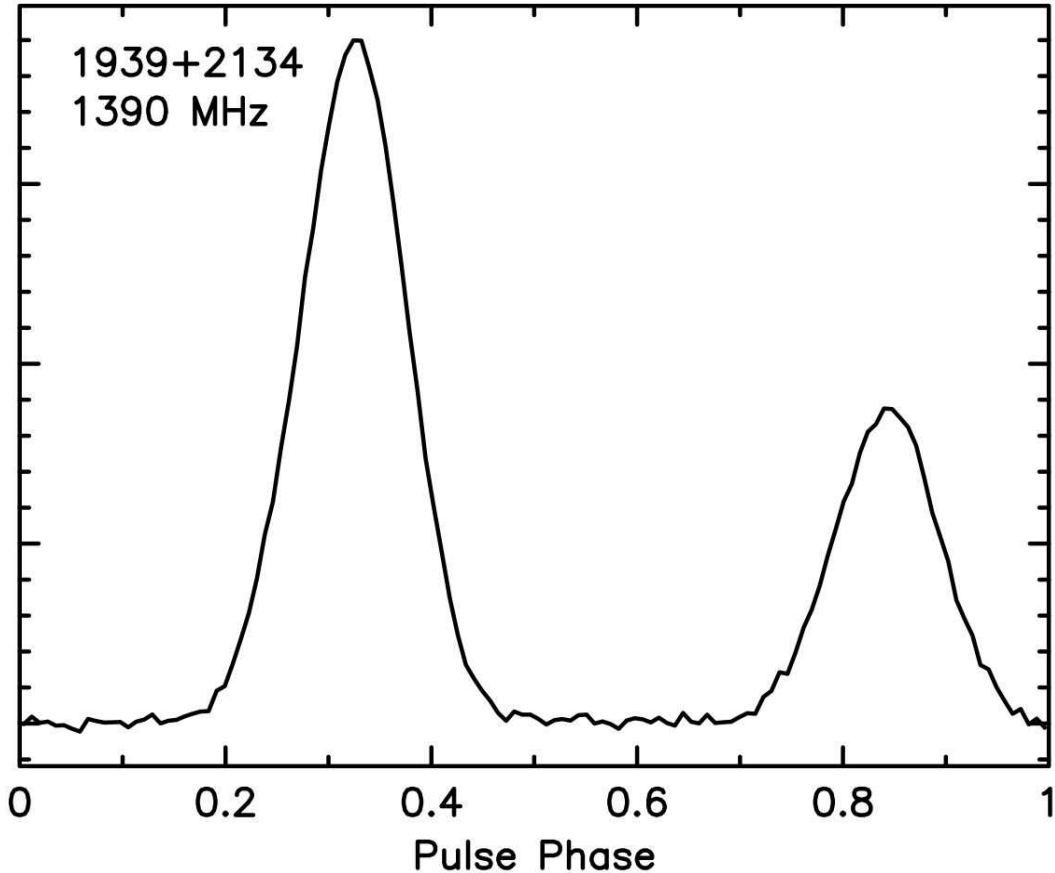,width=14cm,angle=0}}
  \caption[Analogue filter bank profile of PSR J1939+2134]{Pulse
  profile of PSR J1939+2134, taken with an analogue filter bank
  backend. The smearing of the pulse peak by 9 bins is clearly visible
  when compared to Figure \ref{fig:CPSR2profile}.}
  \label{fig:AFBprofile}
\end{figure}

The analogue filter bank system used in this thesis, provides 512
frequency channels over a total bandwidth of 256\,MHz - resulting in a
500\,kHz channel bandwidth at a centre frequency of around
1400\,MHz. Equation \ref{eq:DMdelay} can be used to calculate the DM
smearing within a channel - for PSR J1939+2134, for example:
$$
\Delta t = 4.15\times 10^3 DM \left(1399.75^{-2}-1400.25^{-2}\right)
= 107\,\mu{\rm s}
$$ with DM $= 71.0226\,$cm$^{-3}$pc. With 128 time bins across a
profile for a pulsar with $P = 1.5578\,$ms, this results in nearly
nine bins of smearing, which can be seen in Figure
\ref{fig:AFBprofile}, when compared to \ref{fig:CPSR2profile}.

There are a few disadvantages to these systems. Firstly, the
spectrometer (bandpass filters) is implemented in hardware and is thus
inflexible: the channel number and width cannot easily be changed for
pulsars with different dispersion measures. This leads to large
smearing for pulsars with high DM (like the example of Figure
\ref{fig:AFBprofile}, PSR J1939+2134). Secondly and more importantly,
there is a trade-off between channel number and time resolution. On
the one hand one desires a large number of very narrow channels so
that the DM smearing can be optimally corrected. Whereas on the other,
the narrower a single channel is, the less time resolution it will
have, owing to the finite rise time of the filter. Finally, the cost
of a large number of filters can be high and the response may differ
between filters, leading to systematic errors.

\subsection{Autocorrelation Spectrometers}\label{sec:FPTM}
\begin{figure}
  \centerline{\psfig{figure=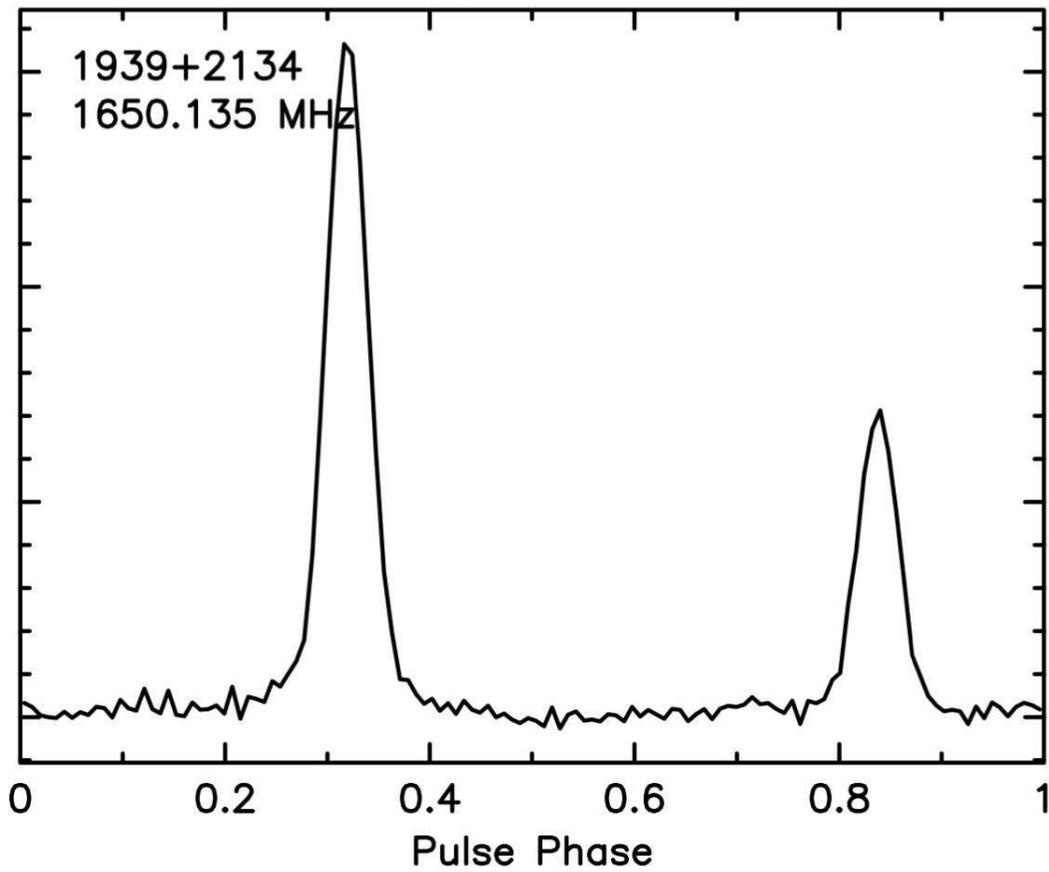,width=14cm,angle=0}}
  \caption[Autocorrelation spectrometer profile of PSR
  J1939+2134]{Pulse profile of PSR J1939+2134, taken with the FPTM
  autocorrelation spectrometer backend. The smearing is considerably
  reduced when compared to Figure \ref{fig:AFBprofile}, but sharp
  features visible with coherent dedispersion backends (Figure
  \ref{fig:CPSR2profile}), remain unresolved.}
  \label{fig:FPTMprofile}
\end{figure}

As rederived in \citet{rw00}, the Wiener-Khinchin theorem states that
the autocorrelation function of a signal is the Fourier transform of
the power spectrum of that same signal. This fact is used by
autocorrelation spectrometers to facilitate obtaining frequency
resolution in configurable (flexible) hardware. In practise such a
system works in the following steps:

\begin{description}
  \item[Digitisation of the analogue signal:]{The analogue signal,
    $S(t)$, is sampled at intervals of $t_{\rm samp}$. Notice this
    digitisation step comes at the start of the process, while it came
    last in the case of analogue filterbanks. We call the digitised
    signal $S_{\rm d}(t)$ and it consists of discrete
    voltages. Sampling is performed at the Nyquist sampling rate:
    $t_{\rm samp} = 1/(2B)$, where $B$ is the bandwidth of the
    signal. For a $256\,$MHz bandwidth system, this implies $t_{\rm
      samp} \approx 2\,$ns. }
  \item[Addition of delays to copies of the signal stream:]{The digitised
    signal, $S_{\rm d}(t)$, is delayed by $2N_{\rm chan}$ lags of size
    $\Delta\tau$, resulting in delayed signals $S_{\rm
      d}(t+i\Delta\tau)$ with $i = 1$ to $N_{\rm chan}$. The number of
    lags used will eventually determine the number of frequency
    channels in our data, hence its naming.}
  \item[Autocorrelation of the signal:]{Given the definition of
    autocorrelation:
  \[
  R(i\Delta\tau) = \langle S_{\rm d}(t)\times S_{\rm
  d}(t+i\Delta\tau)\rangle
  \]
  the next step is to multiply the original and delayed signals and to
  average them, resulting in the autocorrelation of the signal, as a
  function of lag: $R(\tau)$. This multiplying and averaging continues
  during the ``dump time'', $t_{\rm dump}$, after which $R(\tau)$ is
  saved. The dump time defines the time resolution of the final
  observation and is therefore required to be much smaller than the
  pulse period: $t_{\rm dump}\ll P$. As an example, consider a 3\,ms
  pulsar and a desired 256 time bins across its profile. This would
  require: $t_{\rm dump} \approx 12\,\mu$s. Assuming we desired 512
  frequency channels, the longest lag in the autocorrelation would be
  $\Delta\tau \times 512 = 1\,\mu$s, which is much less than the dump
  time.}
  \item[Folding at the pulse period:]{In order to increase the SNR and
    decrease the required disk space for data storage, the
    autocorrelation functions at equal pulse phases can be averaged
    for an arbitrary amount of time. This is the first step that can
    be performed in software, while all previous steps usually happen
    in hardware. The autocorrelation spectrometer used to gather some
    of the data for this thesis (the ``fast pulsar timing machine'' or
    FPTM) however, used a numerically clocked oscillator to perform
    the folding in hardware.}
  \item[Fourier transform to obtain a pulse profile:]{At this stage
    we have the autocorrelation values as a function of pulse phase
    and correlation delay. In order to convert this to power as a
    function of phase and frequency (as seen in Figure
    \ref{fig:Dispersion}), we perform a Fourier transform for each
    pulsar phase bin. This is traditionally done off-line, although
    current computing power would be able to do this in real
    time. This finally results in a pulse profile with frequency
    resolution $\Delta\nu\approx B/N_{\rm chan}$ and time resolution
    $t_{\rm samp}$.}
  \item[Dedispersion of the pulse profile:]{As with the analogue
    filterbank systems, dispersion effects are removed by shifting the
    frequency channels with respect to each other. Within the
    frequency channels, the effects remain, so the dispersion smearing
    is still determined by the width of the frequency channels.}
\end{description}
Even though almost all of the above is performed in hardware, this
hardware is much more easily modified and configured than the bandpass
filters of the analogue filter bank system. This implies there is much
larger flexibility in both frequency and time resolution. While the
fundamental trade-off of time versus frequency resolution still holds,
these numbers can more easily be optimised depending on the observed
pulsar, choosing higher frequency resolution for high-DM pulsars and
higher time resolution for low-DM pulsars with narrow pulses. The
observing set-up for the high-DM pulsar PSR J1939+2134 can therefore
provide higher frequency resolution, resulting in less smearing than
was the case for analogue filter bank systems (see figures
\ref{fig:AFBprofile} and \ref{fig:FPTMprofile}).

Another important difference between autocorrelation spectrometers and
analogue filter banks is that the former need a baseband input
signal. Baseband implies that the frequency range of the signal lies
between $0$ and the bandwidth $B$. The main reason for this
requirement is that the Nyquist sampling rate is far reduced, from
$1/(2f_{\rm h})$ to $1/(2B)$, as used above ($f_{\rm h} = f_{\rm
0}+B/2$ being the highest frequency of the IF signal and $B$ the
bandwidth of the signal). The transformation to baseband is
accomplished through a second stage of down-conversion, as depicted in
Figure \ref{fig:SignalChain}.

\subsection{Coherent Dedispersion Systems}\label{sec:CPSR}
In Section \ref{ssec:ISM}, we have commented on the dispersive effects
of the ISM due to the varying group velocity of electromagnetic waves
travelling through an ionised plasma. We also derived the relative
time delay this induced between two frequency channels. This relative
time delay is - as described in \S\ref{sec:AFB} and \S\ref{sec:FPTM} -
corrected in pulsar backend systems after the signal has been detected
and recorded. However, to correct this dispersion in a continuous and
absolute way, i.e. to \emph{phase-coherently} dedisperse a pulsar
signal, requires a somewhat more involved treatment.

The basic signal chain for coherent dedispersion backend systems is
identical to that for autocorrelation spectrometers: the IF signal is
downconverted to baseband and subsequently digitised. Next, the signal
is Fourier transformed to the frequency domain.

The reason for Fourier transforming is that dispersion is
mathematically a convolution process. Using the convolution theorem
which states that convolution in time is multiplication in frequency,
deconvolving in Fourier space can be done through division, which is
computationally easily achieved. The frequency response function
characterising the dispersion effects of the ISM was derived by
\citet{hr75} and shown to have the following analytic form:
\begin{equation}
  H(f_0+\Delta f) = \exp{\left(\frac{i 2\pi D (\Delta f)^2}{f_0^2
  \left(f_0+\Delta f\right)}\right)}
\end{equation}
with $f_0$ the centre frequency, $\Delta f$ (for which holds $-B/2 <
\Delta f < B/2$) the offset from the centre frequency and $D$ the
dispersion constant as defined in Equation
\ref{eq:DispersionConstant}. Removal of the dispersive effects can
therefore easily be accomplished by dividing the Fourier transform of
the signal by this function. At present, computing power is sufficient
to perform this coherent dedispersion in real time, followed by an
inverse Fourier transform, providing the observer with an almost
instantaneous view of the pulsar, if it is sufficiently
bright. However, during the first half of this decade, the raw data
had to be stored for off-line deconvolution.

\begin{figure}
  \centerline{\psfig{figure=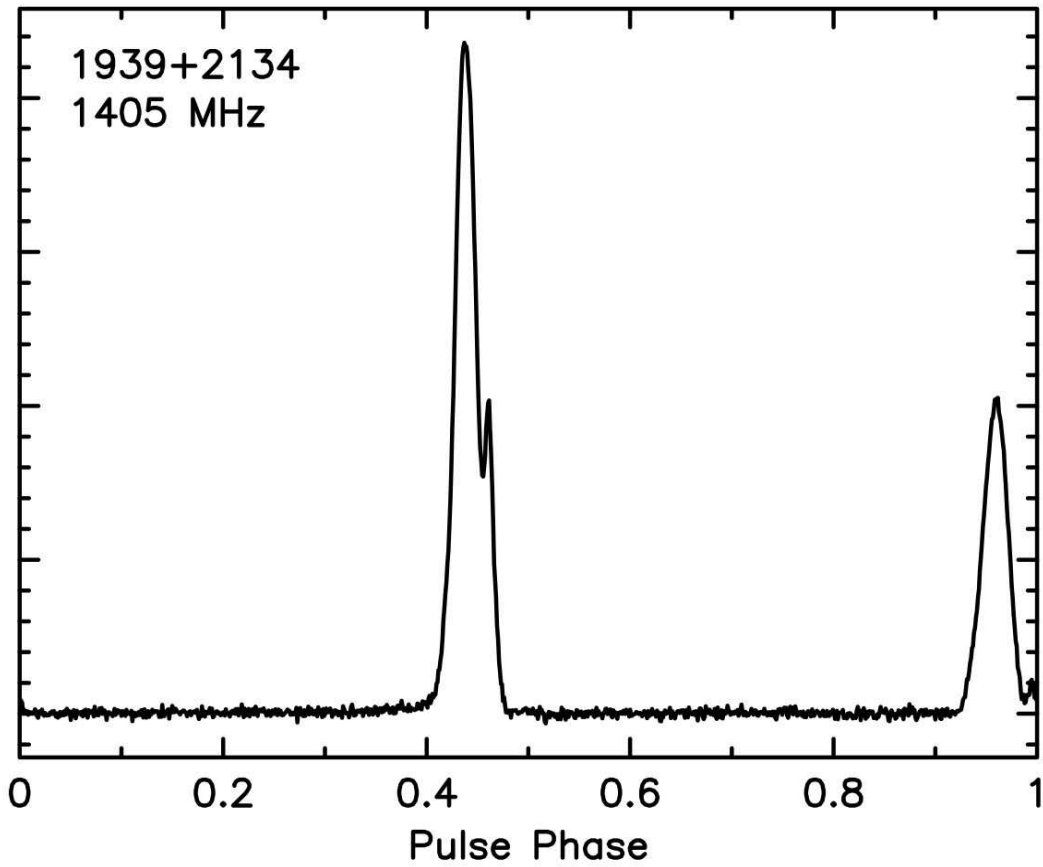,width=14cm,angle=0}}
  \caption[Coherently dedispersed profile of PSR J1939+2134]{Pulse
  profile of PSR J1939+2134, taken with the CPSR2 coherent
  dedispersion backend system. Sharp features are now fully resolved,
  in contrast to observations made with other systems (figures
  \ref{fig:AFBprofile} and \ref{fig:FPTMprofile}).}
  \label{fig:CPSR2profile}
\end{figure}

Since the interstellar dispersion effects are now removed within the
frequency channels, a small channel bandwidth is not urgently required
anymore. Given the fundamental limitation that frequency and time
resolution are inversely proportional in Fourier analysis, this
reduced need for small bandwidths of frequency channels, allows higher
time resolution. An example for a coherently dedispersed pulse is
shown in Figure \ref{fig:CPSR2profile}, which shows a narrow spike on
the trailing edge of the main pulse of PSR J1939+2134. This narrow
spike was unresolved with the older backends, mostly due to DM
smearing.

\subsection{Overview of Instruments}\label{sec:Inst}
Five different instruments were used in the data collection for this
thesis. Two of these were only used on PSR J0437$-$4715, as described
in Chapter \ref{chap:0437}. These were:
\begin{description}
  \item[S2:] The S2 VLBI recorder is a 16\,MHz bandwidth recorder that
  stores raw data with 31\,ns time resolution onto eight SVHS tapes
  for offline coherent dedispersion. More details are provided in
  \citet{wvd+98}.
  \item[CPSR:] The first generation Caltech-Parkes-Swinburne Recorder,
  CPSR, recorded a 20\,MHz bandpass at dual polarisation onto DLT
  tapes. The data was coherently dedispersed offline and generated
  pulse profiles with a resolution of 4096 bins per pulse period. More
  information can be found in \citet{vbb+00} and references therein.
\end{description}
The three back ends that were used for both PSR J0437$-$4715 and for
all pulsars described in Chapter \ref{chap:20PSRS}, were:
\begin{description}
  \item[FB:] The analogue filter bank had a 256\,MHz bandwidth with
  512 frequency channels. It generates profiles after offline
  processing in software. Its time resolution of $t_{\rm s} =
  80\,\mu$s limited the effective number of bins to $P/t_{\rm s}$
  where $P$ is the pulsar's pulse period. This system was upgraded in
  the early 2000s to provide higher time resolution and higher
  bandwidth.
  \item[FPTM:] The fast pulsar timing machine is an autocorrelation
  spectrometer with a maximum bandwidth of 256\,MHz and up to 1024
  bins across a profile. Details are provided in \citet{sbm+97} and
  \citet{san01}. 
  \item[CPSR2:] The second generation Caltech-Parkes-Swinburne
  recorder, CPSR2, is a coherent dedispersion baseband system that
  samples two independent 64\,MHz-wide observing bands. For
  observations around an observing frequency of 1400\,MHz, these two
  bands were placed adjacent to each other, with centre frequencies at
  1341 and 1405\,MHz. In case of observations in the 50\,cm
  ($\approx$685\,MHz) band, one observing band was either ignored or
  centred around 3\,GHz wavelength, using the coaxial 10/50\,cm
  receiver at Parkes. More details are provided in \citet{hbo06}.
\end{description}

\chapter{High-Precision Timing of PSR J0437$-$4715}
\label{chap:0437}
\noindent \textsf{Twenty-five years ago, general relativity was often
  thought of more as a branch of mathematics than of physics.\\}
\vspace{0.25cm}
\textit{Backer \& Hellings, ``Pulsar Timing and General Relativity'',
  ARA\&A, 1986}
\vspace{1.5cm}

\newcommand{\pb}{\mbox{$P_{\rm b}$}}
\newcommand{\pbdot}{\mbox{$\dot{\pb}$}}

%

This chapter was previously published as Verbiest et al., ``Precision
Timing of PSR J0437$-$4715: An Accurate Pulsar Distance, a High Pulsar
Mass, and a Limit on the Variation of Newton's Gravitational
Constant'', published in the Astrophysical Journal, volume 679,
pp. 675--680, 2008 May 20. Minor updates and alterations have been
made for the purpose of inclusion in this thesis.

\section{Abstract}
Analysis of ten years of high-precision timing data on the millisecond
pulsar PSR J0437$-$4715 has resulted in a model-independent kinematic
distance based on an apparent orbital period derivative, \pbdot,
determined at the $1.5\%$ level of precision ($D_{\rm k} = 157.0\pm
2.4$\,pc), making it one of the most accurate stellar distance
estimates published to date.  The discrepancy between this measurement
and a previously published parallax distance estimate is attributed to
errors in the DE200 Solar System ephemerides. The precise measurement
of \pbdot\ allows a limit on the variation of Newton's gravitational
constant, $|\dot{G}/G| \leq 2.3 \times 10^{-11}$\,yr$^{-1}$.  We also
constrain any anomalous acceleration along the line of sight to the
pulsar to $|a_{\odot}/c| \leq 1.5\times 10^{-18}$\,s$^{-1}$ at $95\%$
confidence and derive a pulsar mass, $m_{\rm psr} = 1.76 \pm
0.20\,M_{\odot}$, one of the highest estimates so far obtained.

\section{Introduction}\label{Introduction}

\citet{jlh+93} reported the discovery of PSR J0437$-$4714, the nearest
and brightest millisecond pulsar known. Within a year, the white dwarf
companion and pulsar wind bow shock were observed \citep{bbb93} and
pulsed X-rays were detected \citep{bt93a}. The proper motion and an
initial estimate of the parallax were later presented along with
evidence for secular change in the inclination angle of the orbit due
to proper motion \citep{sbm+97}. Using high time resolution
instrumentation, the three-dimensional orbital geometry of the binary
system was determined, enabling a new test of general relativity \citei{GR;}
{vbb+01}. Most recently, multi-frequency observations were used
to compute the dispersion measure structure function \citep{yhc+07},
quantifying the turbulent character of the interstellar medium towards
this pulsar.

The high proper motion and proximity of PSR J0437$-$4715 led to the
prediction \citep{bb96} that a distance measurement independent of
parallax would be available within a decade, when the orbital period
derivative (\pbdot) would be determined to high accuracy. Even if the
predicted precision of $\approx$ 1\% would not be achieved, such a
measurement would be significant given the strong dependence of most
methods of distance determination on relatively poorly constrained
models and the typically large errors on parallax measurements. Even
for nearby stars, both the Hubble Space Telescope and the Hipparcos
satellite give typical distance errors of $3\%$ \citep{val07} and so
far only two distances beyond $100$\,pc have been determined at
$\approx 1\%$ uncertainty \citep{tlmr07}. This kinematic distance is one
of the few model-independent methods that does not rely upon the
motion of the Earth around the Sun.

As demonstrated by \citet{dt91}, \pbdot\ can also be used to constrain
the variation of Newton's gravitational constant. The best such limit
from pulsar timing to date \citei{$|\dot{G}/G| = (4\pm5)\times
10^{-12}$\,yr$^{-1}$ from PSR B1913+16;}{tay93} is compromised due to
the poorly constrained equation of state (EOS) for the neutron star
companion \citep{nor90}. The slightly weaker but more reliable limit
of $|\dot{G}/G| = (-9\pm 18) \times 10^{-12}$\,yr$^{-1}$ \citei{from
PSR B1855+09, which has a white dwarf companion;}{ktr94} should
therefore be considered instead. However, neither of these limits come
close to that put by lunar laser ranging \citei{LLR;}{wtb04}:
$\dot{G}/G = (4\pm 9) \times 10^{-13}$\,yr$^{-1}$. Besides limiting
alternative theories of gravity, bounds on $\dot{G}$ can also be used
to constrain variations of the Astronomical Unit ($AU$). Current
planetary radar experiments \citep{kb04} have measured a significant
linear increase of $d AU/dt =0.15\pm 0.04$\,m yr$^{-1}$, which may
imply $\dot{G}/G = (-1.0 \pm 0.3)\times 10^{-12}$\,yr$^{-1}$, just beyond
the sensitivity of the limits listed above.

As mentioned before, the EOS for dense neutron star matter is very
poorly constrained. Specifically, it is generally accepted that
nuclear matter would degenerate into quark matter as pressure and
density increase, but the critical pressure and density at which this
would happen are as yet mostly unknown \citep{lp07}. Alternative
scenarios of further degeneration and state changes into hyperons or
Bose-Einstein condensates of pions and/or kaons are also not ruled
out, leading to uncertainty about what the fundamental ground state of
matter is. In order to probe matter at such high densities and
constrain potential EOSs for dense nuclear matter, two avenues are
currently open. One is provided by particle accelerators such as the
large hadron collider (LHC) at CERN and the RHIC at the Brookhaven
national laboratory. The other possibility is provided by probing the
masses and radii (and therefore densities) of neutron stars, the
densest known objects without an event horizon \citep{wei08}.

While no accurate measurement of a neutron star radius has been made
to date, the combination of the requirement for hydrostatic
equilibrium with the pressure expected by a given EOS, provides an
EOS-dependent upper limit on neutron star masses \citei{for a more
  detailed derivation, see}{lp07}. While most measured pulsar masses
fall within a narrow range close to $1.4\,M_{\odot}$, recent results
on the pulsars NGC 6440B, Terzan 5\,I and Terzan 5\,J indicate the
potential for substantially heavier pulsars \citep{frb+08,rhs+05};
however, as discussed in more detail in Section \ref{Mass}, these
predictions do not represent objective mass estimates.

The remainder of this chapter is structured as follows: Section
\ref{Obs} describes the observations, data analysis and general timing
solution for PSR J0437$-$4715. Section \ref{Dist} describes how the
measurement of \pbdot\ leads to a new and highly precise distance. In
Section \ref{GandA}, this measurement is combined with the parallax
distance to derive limits on $\dot{G}$ and the Solar System
acceleration. Section \ref{Mass} presents the newly revised pulsar
mass and our conclusions are summarised in Section \ref{Conc}.

\section{Observations and Data Reduction}\label{Obs}

Observations of PSR J0437$-$4715 were made over a time span of ten
years (see Figure \ref{Fig::Res}), using the Parkes 64-m radio
telescope. Two 20\,cm receiving systems (the central beam of the
Parkes multi-beam receiver \citep{swb96} and the H-OH receiver) were
used and four generations of digital instrumentation (see Table
\ref{tbl::Obs}): the Fast Pulsar Timing Machine (FPTM), the S2 VLBI
recorder and the Caltech-Parkes-Swinburne Recorders (CPSR and CPSR2),
all described in Chapter \ref{chap:Hardware}.

\subsection{Arrival Time Estimation}
For the FPTM, S2 and CPSR backends, the uncalibrated polarisation
data were combined to form the polarimetric invariant interval
\citep{bri00} and each observation was integrated in time and
frequency before pulse arrival times were calculated through standard
cross-correlation with an instrument-dependent template profile.  For
the CPSR2 data, the technique described by \citet{van04a} was used to
calibrate 5 days of intensive PSR J0437$-$4715 observations made on
2003 July 19 to 21, 2003 August 29 and 2005 July 24.  The calibrated
data were integrated to form a polarimetric template profile with an
integration length of approximately 40 hours and frequency resolution
of 500 kHz.  This template profile and Matrix Template Matching
\citei{MTM,}{van06} were used to calibrate the three years of CPSR2
data.  An independent MTM fit was performed on each five-minute
integration, producing a unique solution in each frequency channel, as
shown in Figure 2 of \citet{van06}. The calibrated data were then
integrated in frequency to produce a single full-polarisation profile
at each epoch.  MTM was then used to derive time-of-arrival (TOA)
estimates from each calibrated, five-minute integration.  The
application of MTM during the calibration and timing stages reduced
the weighted RMS of the CPSR2 post-fit timing residuals by a factor of
two.  All the data reduction described above was performed using the
\textsc{psrchive} software package \citep{hvm04}.

\begin{threeparttable}
	\caption[Backends used in the PSR J0437$-$4715 timing
	  analysis]{Characteristics of the timing data from the four
	  instruments used.}
	\begin{tabular}{ccrrr}
	  \hline
	  Backend & Date range  & Bandwidth & RMS\\
	  &  &  &  Residual \\
	  \hline
	  FPTM & 1996 Apr -- 1997 May  & $256\,$MHz & $368\,$ns \\
	  S2 & 1997 Jul -- 1998 Apr & $16\,$MHz & $210\,$ns \\
	  CPSR & 1998 Aug -- 2002 Aug & $20\,$MHz & $218\,$ns \\
	  CPSR2 & 2002 Nov -- 2006 Mar & $2\times 64\,$MHz\tnote{a} &
	  $164\,$ns\\ 
	  \hline
	  \hline
	  Backend & Observation & Number of & TOA \\
	          & length\tnote{b}& TOAs & error\tnote{b} \\
	  \hline
	  FPTM & $10\,$min & 207 & $500\,$ns \\
	  S2 & $120\,$min & 117 & 160\,ns \\
	  CPSR & 15\,min & 1782 & 250\,ns \\
	  CPSR2 & 60\,min & 741 & 140\,ns \\
	  \hline
	\end{tabular}
	\begin{tablenotes}
	\item[a] CPSR2 records two adjacent $64\,$MHz bands simultaneously at
	  20\,cm.
	\item[b] Displayed are typical values only.
	\end{tablenotes}
	\label{tbl::Obs}
\end{threeparttable}

\subsection{Timing Analysis}
Most data were recorded at a wavelength of 20\,cm; however, in the
final three years, simultaneous observations at 10 and 50\,cm were
used to measure temporal variations of the interstellar dispersion
delay \citei{corrections for these variations were implemented in a
way similar to that of}{yhc+07}. A linear trend of these delays was
also obtained for the year of FPTM data, using data at slightly
different frequencies close to $1400$\,MHz.

\begin{threeparttable}
\caption[PSR J0437$-$4715 timing model parameters]{PSR J0437$-$4715
	timing model parameters\tnote{a}}
\begin{tabular}{llrcr}
\hline
Parameter Name & Parameter &\textsc{T2} & M-C & Error\\ 
and Units & Value &Error\tnote{b} &Error\tnote{b} & Ratio \\
\hline
\multicolumn{5}{c}{Fit and Data Set}\\
\hline
MJD range \dotfill & 50191.0--53819.2 & & & \\
Number of TOAs \dotfill & 2847 & & & \\
Rms timing residual ($\mu$s) \dotfill & 0.199 & & & \\
\hline
\multicolumn{5}{c}{Measured Quantities}\\
\hline
 Right ascension, $\alpha$ (J2000)       \dotfill &
 04$^{\mathrm h}$37$^{\mathrm m}$15\fs8147635 & 3 & 29 & 9.8\\
 Declination, $\delta$ (J2000)           \dotfill & 
 $-$47\degr15\arcmin08\farcs624170 & 3 & 34 & 11\\
 Proper motion in $\alpha$, \\
 $\mu_\alpha \cos{\delta}$ (mas yr$^{-1}$)
 \dotfill &  121.453 & 1 & 10 & 8.7 \\ 
 Proper motion in $\delta$, $\mu_\delta$ (mas yr$^{-1}$) \dotfill &
 $-$71.457 & 1 & 12 & 9.0\\ 
 Annual parallax, $\pi$ (mas)  \dotfill & 6.65 & 7 & 51 & 7.9 \\
 Dispersion measure, $DM$ (cm$^{-3}$ pc) \dotfill & 2.64476 & 7 &
 \tnote{d} & \tnote{d} \\
 Pulse period, $P$ (ms)                    \dotfill &
 5.757451924362137 & 2 & 99 & 47\\ 
 Pulse period derivative, $\dot{P}$ (10$^{-20}$) \dotfill &
 5.729370 & 2 & 9 & 4.8 \\
 Orbital period, $P_{\rm b}$ (days)        \dotfill &
 5.74104646\tnote{c} & 108 & 200 & 1.9 \\
 Orbital period derivative, $\dot{P_{\rm b}}$ (10$^{-12}$) \dotfill &
 3.73 & 2 & 6 & 2.5\\
 Epoch of periastron passage,\\
 $T_{0}$ (MJD) \dotfill &
 52009.852429\tnote{c} & 582 & 780 & 1.3\\
 Projected semi-major axis, $x$ (s)      \dotfill &
 3.36669708\tnote{c} & 11 & 14 & 1.4\\
 Longitude of periastron, $\omega_0$ (\degr) \dotfill &
 1.2224\tnote{c} & 365 & 490 & 1.3 \\
 Orbital eccentricity, $e$ (10$^{-5}$) \dotfill & 1.9180 & 3 &
 7 & 2.1\\ 
 Periastron advance, $\dot \omega$ (\degr\ yr$^{-1}$) \dotfill &
 0.01600\tnote{c} & 430 & 800 & 1.8\\ 
 Companion mass, $m_{\rm 2}$ (M$_{\odot})$       \dotfill &
 0.254\tnote{c} & 14 & 18 & 1.3 \\
 Longitude of ascension, $\Omega$ (\degr)  \dotfill &
 207.8\tnote{c} & 23 & 69 & 3.0 \\
 Orbital inclination, $i$ (\degr)          \dotfill & 137.58 & 6
 & 21 & 3.7\\
\hline
\multicolumn{5}{c}{Set Quantities}\\
\hline
Reference epoch for $P$, $\alpha$ & & & & \\
 and $\delta$ determination (MJD)
 \dotfill & 52005 & & & \\
  Reference epoch for DM &  &  & & \\
  determination (MJD) \dotfill & 53211 & & & \\
\hline
\end{tabular}
\begin{tablenotes}
  \item[a] These parameters are determined using \textsc{Tempo2} which
	uses the International Celestial Reference System and Barycentric
	Coordinate Time. As a result this timing model must be modified
	before being used with an observing system that inputs
	\textsc{Tempo} format parameters. See \citet{hem06} for more
	information.
  \item[b] Given uncertainties are $1\sigma$ values in the last digits
	of the parameter values. ``T2'' refers to the formal uncertainties
	provided by the \textsc{Tempo2} software package, ``M-C'' refers
	to the uncertainties resulting from the Monte-Carlo simulations.
  \item[c] Because of large covariances, extra precision is given for
	selected parameters.
  \item[d] Dispersion measure was determined through alignment of
	simultaneous CPSR2 observations centred at 1341\,MHz and
	1405\,MHz. The effect of red noise is therefore not applicable.
\end{tablenotes}
\label{Model}
\end{threeparttable}

The arrival times were analysed using the \textsc{Tempo2} pulsar
timing software package \citep{hem06,ehm06} and consistency with the
earlier program, \textsc{Tempo}, was verified.  The timing model (see
Table \ref{Model}) is based on the relativistic binary model first
derived by \citet{dd86} and expanded to contain the geometric orbital
terms described by \citet{kop95} and \citet{kop96} - see also
\S\ref{ssec:BinaryEffects}. The model is optimised through a standard
weighted least-squares fit in which all parameters are allowed to
vary, including all parameters presented in Table \ref{Model}, as well as
the unknown time delays between data from different instruments, but
excluding the mean value of dispersion measure, which is determined
from the simultaneous CPSR2, 64\,MHz-wide bands centred at 1341 and
1405\,MHz.

A major difference between our implementation of solutions for the
orbital angles $\Omega$ and $i$ and previous efforts
\citep{vbb+01,hbo06} is that they were implemented as part of the
standard fitting routine.  This ensures any covariances between these
and other parameters (most importantly the periastron advance and
companion mass, see Table \ref{Model} and Section \ref{Mass}) are
properly accounted for, thereby yielding a more reliable measurement
error. The previous works 
mentioned above derived these effects from an independent mapping of
$\chi^2$ space, leaving the errors of other parameters unaffected.

\begin{figure}
  \centerline{\psfig{figure=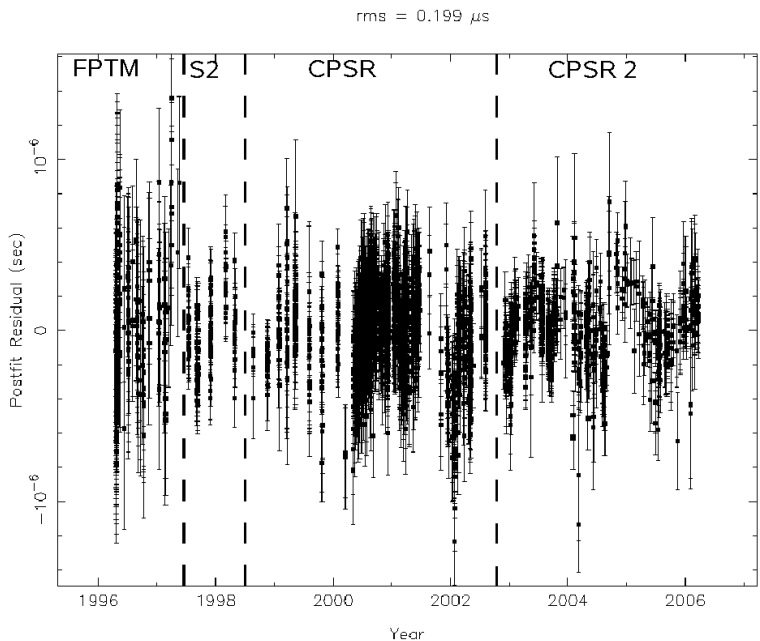,width=10cm,angle=0}}
  \caption[PSR J0437$-$4715 timing residuals]{Combined 20\,cm post-fit
  timing residuals for new and archival PSR J0437$-$4715 timing data. Vertical
  dashed lines separate the different instruments.}
  \label{Fig::Res}
\end{figure}

As can be seen from Figure \ref{Fig::Res}, there are significant
low-frequency structures present in the timing residual data. Since
the standard least-squares fitting routine used in \textsc{Tempo2}
does not account for the effect of such correlations on parameter
estimation, we performed a Monte-Carlo simulation where data sets with
a post-fit power spectrum statistically consistent with that of the
PSR J0437$-$4715 data were used to determine the parameter estimations
uncertainties in the presence of realistic low frequency noise. These
errors, as well as the factors by which the original errors were
underestimated, are shown in Table \ref{Model}. As an example, the
distribution of derived pulsar masses from the Monte-Carlo simulation
is given in Figure \ref{PSRMass}. Because of the dispersion measure
corrections implemented in the final three years of data, one can
expect the spectrum of these most precise data points to contain less
low-frequency noise than the ten year data set as a whole. We
therefore expect the errors resulting from this analysis to be
slightly overestimated. Ongoing research into extending the fitting
routine with reliable whitening schemes to avoid spectral leakage and
hence improve the reliability of the measured parameters, is expected
to reduce these errors by factors of around two. All errors given in
this paper are those resulting from the Monte-Carlo simulations,
unless otherwise stated. The simulations also showed that any biases
resulting from the red noise are statistically negligible for the
reported parameters. (A full description of this Monte-Carlo technique
and the whitening schemes mentioned will be detailed in a future
publication.)

\begin{threeparttable}
  \caption[Comparison of Solar System ephemeris models for PSR
	J0437$-$4715]{Comparison of DE200 and DE405 results for PSR
	J0437$-$4715\tnote{a}} 
  \begin{tabular}{lll}
	\hline
	Parameter name & DE200 result & DE405 result \\
	\hline
	Rms residual (ns) \dotfill & 281 & 199 \\
	Relative $\chi^2$ \dotfill & 2.01  & 1.0 \\
	Parallax, $\pi$ (mas) \dotfill & 7.84(7)  & 6.65 (7) \\
	Parallax distance, $D_{\pi}$ (pc) \dotfill & 127.6(11) & 150.4(16) \\
	Previously published $\pi$ (mas) \dotfill 
	& 7.19(14)\tnote{c} & 6.3(2)\tnote{d} \\
	Kinematic distance, $D_k$ (pc) \dotfill & 154.5 (10) & 156.0 (10) \\  
	$D_k$ corrected for Galactic effects (pc) \dotfill & 
	155.5 (10) & 157.0 (10) \\
	Variation of Newton's gravitational & & \\
	constant, $| {\dot{G} / G} |$ (10$^{-12}$ yr$^{-1}$)\dotfill &
	$-21.2(22)$\tnote{b}  & $-5.0(26)$\tnote{b} \\
	Total proper motion, $\mu_{tot}$ (mas yr$^{-1}$) \dotfill & 140.852(1)
	& 140.915(1) \\ 
	Companion mass, $m_{\rm c}$ ($M_{\odot}$) \dotfill & 0.263(14)
	& 0.254(14) \\ 
	Pulsar mass, $m_{\rm psr}$ ($M_{\odot}$) \dotfill & 1.85(15)
	& 1.76(15) \\ 
	Periastron advance, $\dot{\omega}$ (\degr\ yr$^{-1}$) \dotfill &
	0.020(4)  & 0.016(4) \\ 
	GR prediction of $\dot{\omega}$ (\degr\ yr$^{-1}$)
	\dotfill & 0.0178(9)  & 0.0172(9) \\
	\hline
  \end{tabular}
  \begin{tablenotes}
  \item[a] Numbers in parentheses represent the formal \textsc{Tempo2}
	$1\,\sigma$ uncertainty in the last digits quoted, unless otherwise
	stated.
  \item[b] Given are $2\,\sigma$ errors, i.e. $95\%$ confidence levels.
  \item[c] \citet{vbb+01}
  \item[d] \citet{hbo06}
  \end{tablenotes}
  \label{DEModels}
\end{threeparttable}

\subsection{Solar System Ephemerides}\label{SSE}

Pulsar timing results are dependent on accurate ephemerides for the
Solar System bodies. The results presented in this paper were obtained
using the DE405 model \citep{sta04b} and, for comparison, selected
parameters obtained with the earlier DE200 model are shown in Table
\ref{DEModels}. The greatly reduced $\chi^2$ indicates that the newer
Solar System ephemerides are superior to the earlier DE200,
reinforcing similar conclusions of other authors
\citep{sns+05,hbo06}. We notice the parallax value changes by more
than $10\,\sigma$ and that the different derived values are closely
correlated with the ephemeris used. Although the effect is not as
dramatic as it appears because of the under-estimation of the
\textsc{Tempo2} errors, the fact that the DE405 results agree much
better with the more accurate kinematic distance (discussed in the
next section), strongly suggests that the differences are due to the
ephemeris used and confirms that the DE405 ephemeris is superior.
Finally, we note that the DE405 measurement of $\dot{\omega}$ ($0.016
\pm 0.008\,^{\circ}{\rm yr}^{-1}$) is consistent with the GR
prediction for this system ($0.0172 \pm 0.0009\,^{\circ}{\rm
yr}^{-1}$).

\section{Kinematic Distance}\label{Dist}

\begin{figure}
  \centerline{\psfig{figure=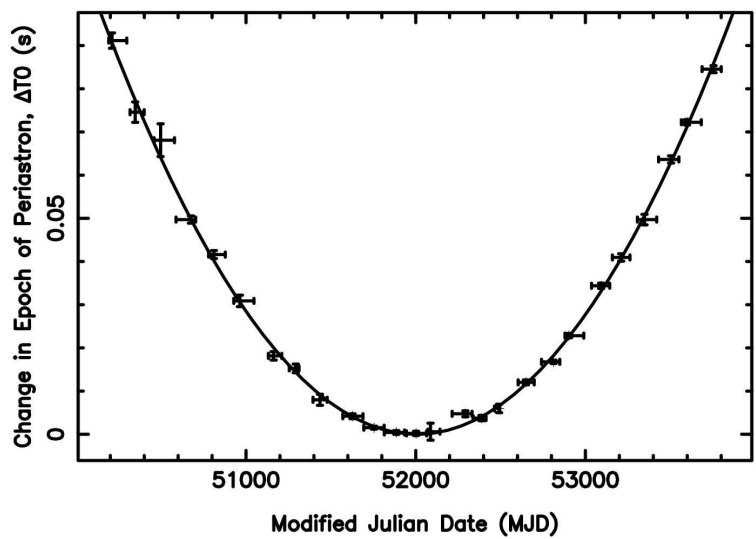,width=10cm}}
  \caption[$T_{\rm 0}$ variations due to orbital period
  increase]{Variations in epoch of periastron passage ($T_{\rm 0}$)
  due to apparent orbital period increase. A steady increase in
  orbital period is equivalent to a quadratic increase in $T_{\rm 0}$
  relative to periastron times for a constant orbital period.  For
  this plot, $T_{\rm 0}$ was measured on data spans of up to 120 days
  with a model having no orbital period derivative. The formal
  one-$\sigma$ measurement errors reported by \textsc{Tempo2} are
  shown by vertical error bars and the epochs over which the
  measurements were made are shown by horizontal bars. As the mean
  measurement time was determined through a weighted average of the
  data contained in the fit, these horizontal bars need not be centred
  at the mid time associated with the measurement.  The parabola shows
  the effect of the \pbdot\ value obtained from a fit to the data
  shown in Figure \ref{Fig::Res}.}
  \label{Fig::Pbdot}
\end{figure}

As shown in Figure \ref{Fig::Pbdot}, the long-term timing history
enables precise measurement of the orbital period derivative, \pbdot $
= (3.73\pm0.06)\times 10^{-12}$.  This observed value represents a
combination of phenomena that are intrinsic to the binary system and
dynamical effects that result in both real and apparent accelerations
of the binary system along the line of sight \citep{bb96}; i.e.
\begin{equation}\label{PbdotEq::Basis}
\dot{P}_{\rm b}^{\rm obs} =
\dot{P}_{\rm b}^{\rm int} +
\dot{P}_{\rm b}^{\rm Gal} +
\dot{P}_{\rm b}^{\rm kin}
\end{equation}
where ``obs'' and ``int'' refer to the observed and intrinsic values;
``Gal'' and ``kin'' are the Galactic and kinematic contributions.

Intrinsic orbital decay is a result of energy loss typically due to
effects such as atmospheric drag and tidal dissipation; however, in a
neutron star--white dwarf binary system like PSR J0437$-$4715, energy loss is
dominated by quadru\-po\-lar gravitational wave emission. For this
system, GR predicts \citep{tw82} $\dot P_{\rm b}^{\rm
GR}=-4.2\times10^{-16}$, two orders of magnitude smaller than the
uncertainty in the measured value of \pbdot.

Galactic contributions to the observed orbital period derivative
include differential rotation and gravitational acceleration
\citep{dt91}.  The differential rotation in the plane of the Galaxy is
estimated from the Galactic longitude of the pulsar and the
Galactocentric distance and circular velocity of the Sun. Acceleration
in the Galactic gravitational potential varies as a function of height
above the Galactic plane \citep{hf04b}, which may be estimated using
the parallax distance and the Galactic latitude of the
pulsar. Combining these terms gives $\dot{P}_{\rm b}^{\rm Gal} = (-1.8
-0.5) \times 10^{-14} = -2.3\times10^{-14}$, which is of the same
order as the current measurement error.

\begin{figure}
  \centerline{\psfig{figure=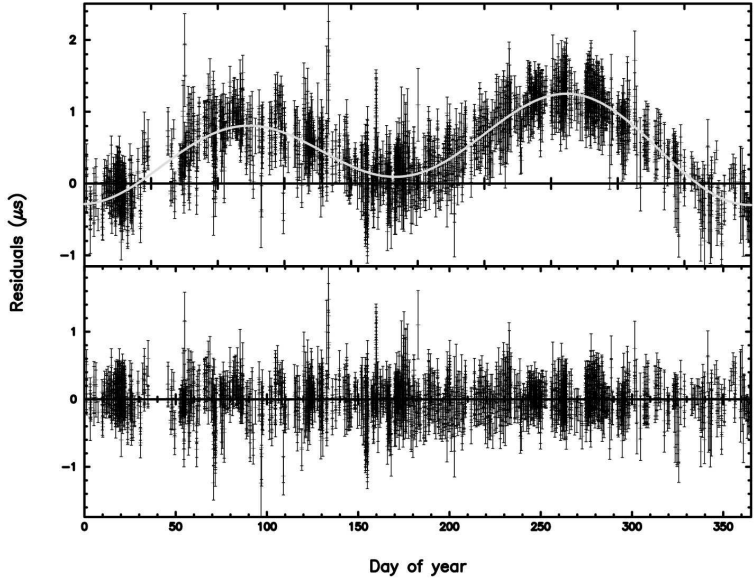,width=10cm}}
  \caption[Parallax signature of PSR J0437$-$4715]{Parallax signature
  of PSR J0437$-$4715. Top: Timing residuals for PSR J0437$-$4715 as a
  function of day of year (starting on 18 November), without parallax
  but with all remaining parameters at their best-fit values. The
  smooth curve represents the model fit of a parallax of 6.65
  mas. Bottom: The same timing residuals with parallax included in the
  model. The overall RMS for the top and bottom plots is $524$ and
  $199$\,ns respectively.  The double-humped signature specific to
  parallax originates from the delay in pulse time-of-arrival (TOA) as
  the Earth orbits the Sun and samples different parts of the curved
  wave-front originating at the pulsar.}
  \label{Fig::Px}
\end{figure}

Given the negligible intrinsic contribution, Equation
\ref{PbdotEq::Basis} can be simplified and rewritten in terms of the
dominant kinematic contribution known as the Shklovskii effect
\citep{shk70}, an apparent acceleration resulting from the non-linear
increase in radial distance as the pulsar moves across the plane
perpendicular to the line of sight; quantified by the proper motion,
$\mu$ and distance $D$ from the Earth:
\begin{equation}\label{PbdotEq::Shk}
\dot{P}_{\rm b}^{\rm obs} -
\dot{P}_{\rm b}^{\rm Gal} \simeq
\dot{P}_{\rm b}^{\rm kin} = \frac{\mu^2D}{c} P_{\rm b},
\end{equation}
where $c$ is the vacuum speed of light. Using the measured values of
$\mu$, \pb and \pbdot, Equation \ref{PbdotEq::Shk} is used to derive
the kinematic distance \citep{bb96}: $D_k = 157.0 \pm 2.4$\,pc.  This
distance is consistent with the one derived from the timing parallax
($D_{\rm \pi} = 150\pm12$\,pc -- see also Figure \ref{Fig::Px}) and
with the VLBI parallax derived for this system: $D_{\rm VLBI} = 156.3
\pm 1.3\,$pc \citep{dvtb08}. Our measurement is, with a relative error
of $1.5\%$, comparable in precision to the best parallax measurements
from VLBI \citep{tlmr07,dvtb08} and better than typical relative
errors provided by the Hipparcos and Hubble space telescopes
\citep{val07}.

Given the dependence of parallax distances on ephemerides, as
described in Section \ref{SSE}, it is interesting to note the
robustness of $D_{\rm k}$.  Also, Table \ref{Model} shows that the
presence of red noise corrupts the parallax error by a factor of 7.9,
whereas \pbdot\ is only affected by a factor of 2.5.  These facts
clearly indicate the higher reliability of $D_{\rm k}$ as compared to
$D_{\rm \pi}$.

\section[Limits on \pbdot\ Anomalies]{Limits on \pbdot\ Anomalies:
  $\dot{G}$ and the Acceleration of the Solar System}\label{GandA} 

Any anomalous orbital period derivative can be constrained by
substituting the parallax distance into Equation 2, yielding
\begin{eqnarray}
  \Big(\frac{\dot{P}_{\rm b}}{P_{\rm b}}\Big)^{\rm excess} & = &
  \big(\dot{P}_{\rm b}^{\rm obs} - \dot{P}_{\rm b}^{\rm Gal} -
  \dot{P}_{\rm b}^{\rm kin}\big)/P_{\rm b}\nonumber\\ 
  & = & (3.2 \pm 5.7) \times 10^{-19}\, {\rm s}^{-1}.
\end{eqnarray}
in which the error is almost exclusively due to the parallax
uncertainty. Following \citet{dt91}, this can be translated into a
limit on the time derivative of Newton's gravitational constant (given
are $95\%$ confidence levels):
\begin{equation}\label{eq:GdotLimitJoris}
\frac{\dot{G}}{G} = -\frac{1}{2}\Big(\frac{\dot{P_{\rm b}}}{P_{\rm
    b}}\Big)^{\rm excess} = (-5\pm 18) \times 10^{-12}\,{\rm yr}^{-1}
\end{equation}

This limit is of the same order as those previously derived from
pulsar timing (see Section \ref{Introduction}) but have been further
improved by \citet{dvtb08} who used the VLBI parallax distance in
combination with our \pbdot measurement to achieve a better limit
still:
\begin{equation}
\frac{\dot{G}}{G} = \left(-5 \pm 26\right)\times 10^{-13}\,{\rm
  yr}^{-1}
\end{equation}
at 94\,\% certainty. This limit is close to that put by LLR:
\mbox{$(4\pm 9) \times 10^{-13}$\, yr$^{-1}$}\citep{wtb04}. The LLR
experiment is based on a complex $n$--body relativistic model of the
planets that incorporates over 140 estimated parameters, such as
elastic deformation, rotational dissipation and two tidal dissipation
parameters. In contrast, the PSR J0437$-$4715 timing and VLBI results
are dependent on a different set of models and assumptions and
therefore provide a useful independent confirmation of the LLR
result.

A recent investigation into the possible causes of a measured
variability of the Astronomical Unit \citei{$AU$;}{kb04} has refuted
all but two sources of the measured value of $d AU/dt = 0.15 \pm
0.04\, {\rm m/yr}$. \citet{kb04} state that the measured linear
increase in the $AU$ would be due to either systematic effects or to a
time-variation of $G$ at the level of $\dot{G}/G = (-1.0\pm 0.3) \times
10^{-12}\, {\rm yr}^{-1}$, comparable to, but inconsistent with, the
LLR limit.

The anomalous \pbdot\ measurements of a number of millisecond pulsars
have also been used to place limits on the acceleration of the Solar
System due to any nearby stars or undetected massive planets
\citep{zt05}. The PSR J0437$-$4715 data set limits any anomalous Solar
System acceleration to $ | a_{\odot}/c | \leq 1.5\times 10^{-18}\,
{\rm s}^{-1}$ in the direction of the pulsar with $95$\% certainty.
This rules out any Jupiter-mass objects at distances less than
$117\,$AU along the line of sight, corresponding to orbital periods of
up to 1270\,years. Similarly, this analysis excludes any Jupiter-mass
planets orbiting PSR J0437$-$4715 between $\sim$5 and $117\,$AU along
the line of sight. \citet{zt05} also compared the sensitivity of this
limit to that of optical and infra-red searches for trans-Neptunian
objects (TNOs) and concluded that beyond $\sim 300$\,AU the
acceleration limit becomes more sensitive than the alternative
searches. At a distance of $300\,$AU from the Sun, the $95\%$
confidence upper limit on the mass of a possible TNO (in the direction
of the pulsar) is $6.8$ Jupiter masses. The precise VLBI measurement
of parallax mentioned above improves these limits somewhat, as
reported in \citet{dvtb08}.

\section{Pulsar Mass}\label{Mass}

\begin{figure}
  \centerline{\psfig{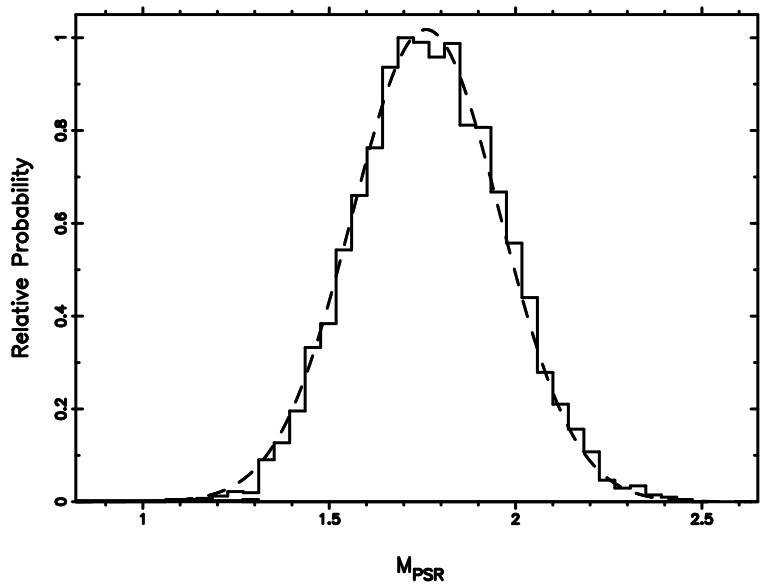}}
  \caption[Probability distribution for mass of PSR
  J0437$-$4715]{Pulsar mass probability distribution. The solid line
  shows the histogram of 5000 pulsar masses derived from a Monte-Carlo
  simulation with power spectrum and sampling equal to that of the
  PSR J0437$-$4715 data set. The dashed line is a Gaussian distribution with a
  mean value of $m_{\rm psr} = 1.76\,M_{\odot}$ and standard deviation
  of $0.20\,M_{\odot}$.}
  \label{PSRMass}
\end{figure}

A combination of the mass function and a measurement of the Shapiro
delay range can be used to obtain a measurement of the pulsar
mass. Using this method, \citet{vbb+01} derived a mass for PSR
J0437$-$4715 of $1.58 \pm 0.18\,M_{\odot}$ whereas \citet{hbo06}
obtained $1.3 \pm 0.2\,M_{\odot}$. It should be noted, however, that
these values resulted from a model that incorporated geometric
parameters first described by \citet{kop95} and \citet{kop96}, but
covariances between these and other timing parameters (most
importantly the companion mass or Shapiro delay range) were not taken
into account. Whilst the length of the data sets used by these authors
was only a few years, it can also be expected that some spectral
leakage from low-frequency noise was unaccounted in the errors of
these previously published values. As described in Section \ref{Obs},
the Monte-Carlo simulations and extended fitting routines implemented
for the results reported in this paper do include these covariances
and spectral leakage; it can therefore be claimed that the current
estimates (at $68\%$ confidence) of $m_{\rm c} = 0.254 \pm
0.018\,M_{\odot}$ and $m_{\rm psr} = 1.76 \pm 0.20\,M_{\odot}$, for
the white dwarf companion and pulsar respectively, reflect the
measurement uncertainty more realistically than any previous estimate.
The distribution of $m_{\rm psr}$ that follows from the 5000
Monte-Carlo realisations is shown in Figure \ref{PSRMass}, together
with a Gaussian with mean $1.76$ and standard deviation $0.20$. This
demonstrates the symmetric distribution of the pulsar mass likelihood
distribution, induced by the precise determination of the orbital
inclination angle.

We also note that the new mass measurement of PSR J0437$-$4715 is the
highest obtained for any pulsar to date. Distinction needs to be made
between the objective mass estimate presented in this paper and the
subjective mass predictions presented in \citet{rhs+05} and
\citet{frb+08}. The pulsar mass confidence interval presented in this
paper is derived from the measurement uncertainties of all relevant
model parameters, including the well-determined orbital inclination
angle, $i$. In contrast, $i$ is unknown in the Terzan 5\,I and J
\citep{rhs+05} and PSR J1748$-$2021B \citep{frb+08} binary systems and
the posterior probability intervals for the pulsar masses presented in
these works are based upon the prior assumption of a uniform
distribution of $\cos i$. These fundamental differences must be
accounted for in any subsequent hypothesis testing.  Consequently, PSR
J0437$-$4715 is currently the only pulsar to provide reliable
constraints on EOSs based on hyperons and Bose-Einstein condensates as
described by \citet{lp07}. Simulations with \textsc{Tempo2} indicate
that a forthcoming observational campaign with a new generation of
backend systems can be expected to increase the significance of this
measurement by another factor of about two in the next few years.
 
\section{Conclusions}\label{Conc}

We have presented results from the highest-precision long-term timing
campaign to date. With an overall residual RMS of 199\,ns, the
10\,years of timing data on PSR J0437$-$4715 have provided a precise
measurement of the orbital period derivative, \pbdot, leading to the
first accurate kinematic distance to a millisecond pulsar: $D_{\rm k}
= 157.0\pm 2.4$\,pc. Application of this method to other pulsars in
the future can be expected to improve distance estimates to other
binary pulsar systems \citep{bb96}.

Another analysis based on the \pbdot\ measurement places a limit on
the temporal variation of Newton's gravitational constant. We find a
bound comparable to the best so far derived from pulsar timing:
$\dot{G}/G = (-5\pm18)\times 10^{-12}\,{\rm yr}^{-1}$.  A VLBI
parallax measurement for this pulsar has further improved this limit,
enabling an independent confirmation of the LLR limit.

Previous estimates of the mass of PSR J0437$-$4715 have been revised
upwards to $m_{\rm psr} = 1.76 \pm 0.20\, M_{\odot}$, which now makes
it one of the few pulsars with such a heavy mass measurement.  A new
generation of backend instruments, dedicated observing campaigns and
data prewhitening techniques currently under development should
decrease the error in this measurement enough to significantly rule
out various EOSs for dense nuclear matter.

\chapter[MSP Stability and GW Detection]{Stability of
  Millisecond Pulsars and Prospects for Gravitational Wave Detection}
\label{chap:20PSRS}
\noindent \textsf{[...]regular timing observations of 40 pulsars each
  with a timing accuracy of 100\,ns will be able to make a direct
  detection of the predicted stochastic background from coalescing
  black holes within 5 years.\\}
\vspace{0.25cm}
\textit{Jenet et al., ``Detecting the Stochastic Gravitational Wave
  Background using Pulsar Timing'', The Astrophysical Journal, 2005}
\vspace{1.5cm}

This chapter will be submitted to Monthly Notices of the Royal
Astronomical Society for publication as Verbiest et al.,
``On the stability of millisecond pulsars and prospects for
gravitational wave detection'', in 2009. Minor alterations have been
made for the purpose of inclusion in this thesis.

\section{Abstract}
Analysis of high-precision timing observations of an array of $\sim$20
millisecond pulsars (a so-called ``timing array'') may ultimately
result in the detection of a stochastic gravitational wave background
(see also \S\ref{intro:pta}). The overall timing precision achievable
using a given telescope and the stability of the pulsars themselves
determine the duration of an experiment required to detect a given
stochastic background level. We present the first long-term,
high-precision timing and stability analysis of a large sample of
millisecond pulsars used in gravitational wave detection projects. The
resulting pulsar ephemerides are provided for use in future
observations. Intrinsic instabilities of the pulsar or the observing
system are shown to contribute to timing irregularities only at or
below the 100\,ns level for our most precisely timed pulsars. Based on
this stability analysis, realistic sensitivity curves for planned and
ongoing timing array efforts are determined. We conclude that, given
the stability of the investigated millisecond pulsars, prospects for
gravitational wave detection within five years to a decade are good
for current timing array projects in Australia, Europe and North
America and for the South African SKA pathfinder telescope, MeerKAT.

\section{Introduction}\label{sec:intro}
The rotational behaviour of pulsars has long been known to be
predictable, especially in the case of MSPs. Current models suggest
that such pulsars have been spun up by accretion from their binary
companion star to periods of several milliseconds, making them much
faster than the more numerous younger pulsars, which typically have
periods of seconds (as outlined in \S\ref{ss:BMSP}). MSPs are
generally timed 3-4 orders of magnitude better than normal pulsars and
on timescales of several years, it has been shown that some MSPs have
a stability comparable to the most precise atomic clocks
\citep{mte97}. This intrinsic stability is most clearly quantified
through the technique of pulsar timing, which compares arrival times
of pulses to a model describing the pulsar, its binary orbit and the
ISM between the pulsar and Earth \citei{as described in
\S\ref{intro:pt} and by}{ehm06}. This technique has enabled precise
determination of physical parameters at outstanding levels of
precision, such as the orbital characteristics of binary star systems
\citei{e.g.}{vbb+01}, the masses of pulsars and their companions
\citei{e.g.}{jhb+05,nic06} and the turbulent character of the ISM
\citei{e.g.}{yhc+07}. The strong gravitational fields of pulsars in
binary systems have also enabled outstanding tests of GR and
alternative theories of gravity, as described by, e.g., \citet{ksm+06}
and \citet{bbv08}. Finally, pulsars have provided the first evidence
that gravitational waves exist at levels predicted by GR \citep{tw82}
and have placed the strongest limit yet on the existence of a
background of gravitational waves \citep{jhv+06}. It is predicted
\citei{most recently by}{jhlm05} that pulsar timing will also
enable a direct detection of such a GWB, as fully discussed in
\S\ref{intro:pta}. 

A main result that follows from the work of Jenet et al. (2005) is
Equation 1.34, replicated below: 
\begin{equation}
  A_{\rm S=3} \approx 2.3\times 10^{-12} \frac{\sigma_{\rm
      n}}{T^{5/3}\sqrt{N}},
\end{equation}
where $A_{\rm S=3}$ is the lowest GWB amplitude at which a given
PTA achieves a 3$\sigma$ sensitivity, $T$ is the data span,
$\sigma$ is the typical RMS and $N$ is the number of TOAs. As
described in \S1.3.4, this results in the standard PTA scenario
proposed by Jenet et al. (2005): $N = 250$, $\sigma_{\rm n} =
0.1\,\mu$s, $T = 5\,$years and $M = 20$. However, depending on
achievable timing precision of MSPs, an alternative PTA could
achieve the same results through timing of 20\,MSPs on a biweekly
basis for ten years with an RMS of close to 300\,ns. This raises
two questions related to the potential of PTAs to detect a
GWB. First, down to which precision can MSPs be timed
($\sigma_{\rm min}$) and second, can high timing precision be
maintained over long campaigns (i.e. does $\sigma/T^{5/3}$
decrease with time)?

It has been shown for a few pulsars that timing at a precision of a
few hundred nanoseconds is possible for campaigns lasting a few
years. Specifically, \citet{hbo06} presented a timing RMS of 200\,ns
over two years of timing on PSRs J1713+0747 and J1939+2134 and 300\,ns
over two years of timing on PSR J1909$-$3744; \citet{sns+05} reported
an RMS of 180\,ns on six years of timing on PSR J1713+0747 and
Verbiest et al. (2008, also Chapter 3) timed PSR J0437$-$4715 at
200\,ns over ten years. It has, however, not been demonstrated thus
far that MSPs can be timed with an RMS residual of $\le 100$\,ns over
five years or more.

The second question - whether high timing precision can be maintained
over ten years or longer, also remains unanswered. \citet{ktr94}
detected excessive low-frequency noise in PSR J1939+2134;
\citet{sns+05} presented apparent instabilities in long-term timing of
PSR J1713+0747 and in Chapter 3, we noted a low-frequency structure in
the timing residuals of PSR J0437$-$4715, but apart from these, no
long-term timing of MSPs has been presented to date. Given the high
timing precision reported on all three sources, it is unclear how
strongly the reported low-frequency noise would affect the use of
these pulsars in a GWB detection effort.
In this chapter we present the first high-precision stability analysis
for a sample of 20 MSPs, which have been timed for ten years on
average. \S\ref{sec:obs} describes the source selection, observing
systems and data analysis methods used. \S\ref{sec:timing} provides
the timing models and residual plots for all pulsars in our sample. In
accordance with previous publications, we present twice the formal
1\,$\sigma$ uncertainties on our parameters, though we defer a full
discussion of these timing model parameters and their uncertainties to
a later paper. In \S\ref{sec:stability} we analyse the stability of
two of the most precisely timed MSPs and quantify the different noise
sources present in our timing residuals. Specifically, we separate the
levels of low-frequency noise, radiometer noise and effects dependent
on observing frequency. In \S\ref{sec:PTA}, we use this stability
information to calculate sensitivity curves for the ongoing Parkes
pulsar timing array \citei{PPTA;}{man08}, European pulsar timing array
\citei{EPTA;}{jsk+08} and the North American nanohertz observatory for
gravitational waves (NANOGrav\footnote{http://www.nanograv.org})
projects. We also assess the usefulness of the two square kilometre
array (SKA) pathfinder telescopes currently being built (the
Australian SKA pathfinder - ASKAP - and the extended Karoo array
telescope - MeerKAT) for PTA-type projects. In \S\ref{sec:conclusions}
we summarise our results.

\section{Observations and Data Reduction}\label{sec:obs}
\subsection{Sample Selection}
The data presented in this chapter have been collated from two pulsar
timing programmes at the Parkes radio telescope. The oldest of these
commenced during the Parkes 70\,cm millisecond pulsar survey
\citep{bhl+94}, aiming to characterise properly the astrometric and
binary parameters of the MSPs found in the survey. Initial timing
results from this campaign were published by \citet{bbm+97} and
\citet{tsb+99}. The bright millisecond pulsars PSRs J1713+0747 and
B1937+21 (both discovered earlier at Arecibo) were also included in
this programme. A few years later, as new discoveries were made in the
Swinburne intermediate latitude survey \citep{ebvb01}, these pulsars
were also added, resulting in a total of 16 MSPs that were regularly
timed by 2006. Improved timing solutions for these 16 pulsars were
presented by \citet{hbo06} and \citet{ojhb06}.

Besides the projects described above, the PPTA project commenced more
regular timing observations of these pulsars in late 2004, expanding
the number of MSPs to 20 (listed in Table \ref{tab:psrs}) and adding
regular monitoring at low observing frequencies ($685$\,MHz) in order
to allow correction for variations of the ISM electron density. A
detailed analysis of these low frequency observations and ISM effects
was recently presented by \citet{yhc+07} and an analysis of the
combined data on PSR J0437$-$4715 was presented in Chapter
\ref{chap:0437}. For this pulsar we will use the timing results
presented in that chapter; for all other pulsars we will present our
improved timing models in \S\ref{sec:timing}.

\subsection{Observing Systems}\label{sec:obsSys}
Unless otherwise stated, the data presented were obtained at the
Parkes 64\,m radio telescope, at a wavelength of 20\,cm. Two receivers
were used: the H-OH receiver and the 20\,cm multibeam receiver
\citep{swb+96}. Over the last five years, observations at 685\,MHz
were taken with the 10/50\,cm coaxial receiver for all pulsars;
however, they were only used directly in the final timing analysis of
PSR J0613$-$0200, whose profile displays a sharp spike at this
frequency, which can be resolved with coherent dedispersion. For PSRs
J1045$-$4509, J1909$-$3744 and J1939+2134, the 685\,MHz observations
were used to model and remove the effects of temporal variations in
interstellar dispersion delays and hence included indirectly in the
timing analysis.

Three different observing backend systems were used. Firstly, the FPTM
\citei{as described in \S\ref{sec:Inst} and by}{sbm+97,san01}, between
1994 and November 2001. Secondly, the 256\,MHz bandwidth analogue
filterbank (FB) was used in 2002 and 2003. Finally, the CPSR2 back end
\citei{see \S\ref{sec:Inst} and}{hbo06} was used from November 2002
onwards.
\clearpage
\begin{threeparttable}
  \caption[Pulsars in our sample]{Pulsars in our sample. Column 2
  gives the reference for the discovery paper, while column 3 provides
  references to recent or important publications on timing of the
  sources. For the three pulsars with original B1950 names, these
  names are given as footnotes to the J2000.0 names.}
  \begin{tabular}{llrrrr}
	  \hline
	  Pulsar & Discovery & \multicolumn{1}{l}{Previous} &
	  \multicolumn{1}{l}{Pulse} & \multicolumn{1}{l}{Orbital} &
	  \multicolumn{1}{l}{Dispersion} \\ 
	  name &  &  \multicolumn{1}{l}{timing}
	  & \multicolumn{1}{l}{period} & 
	  \multicolumn{1}{l}{period} & \multicolumn{1}{l}{measure}\\
	  & & \multicolumn{1}{l}{solution\tnote{a}} &
	  \multicolumn{1}{l}{(ms)} &
	  \multicolumn{1}{l}{(d)}&
	  \multicolumn{1}{l}{(cm$^{-3}$ pc)} \\ 
	  \hline
	  J0437--4715 & \citet{jlh+93} & 1, 2 &  5.8 &   5.7 &  2.6\\
	  J0613--0200 & \citet{lnl+95} & 3  &  3.1 &   1.2 & 38.8\\
	  J0711--6830 & \citet{bjb+97} & 3, 4  &  5.5 &   --  & 18.4\\ 
	  J1022+1001  & \citet{cnst96} & 3  & 16.5 &   7.8 & 10.3\\
	  J1024--0719 & \citet{bjb+97} & 3  &  5.2 &   --  &  6.5\\
	  \\
	  J1045--4509 & \citet{bhl+94} & 3  &  7.5 &   4.1 & 58.2\\
	  J1600--3053 & \citet{ojhb06} & 5  &  3.6 &  14.3 & 52.3\\
	  J1603--7202 & \citet{llb+96} & 3  & 14.8 &   6.3 & 38.0\\
	  J1643--1224 & \citet{lnl+95} & 4  &  4.6 & 147.0 & 62.4\\
	  J1713+0747  & Foster et al. (1993)\nocite{fwc93}  & 3, 6  &  4.6
	  &  67.8 & 16.0\\ 
	  \\
	  J1730--2304 & \citet{lnl+95} & 4 &  8.1 &   --  &  9.6\\
 	  J1732--5049 & \citet{eb01b}  & 7  &  5.3 &   5.3 & 56.8\\
	  J1744--1134 & \citet{bjb+97} & 3  &  4.1 &   --  &  3.1\\
	  J1824$-$2452\tnote{b} & \citet{lbm+87} & 8, 10 &  3.1 &   --  &120.5\\
	  J1857+0943\tnote{b} & \citet{srs+86} & 3, 9  &  5.4 &  12.3 & 13.3\\
	  \\
	  J1909--3744 & \citet{jbv+03} & 3, 11  &  2.9 &   1.5 & 10.4\\
	  J1939+2134\tnote{b} & \citet{bkh+82} & 3, 9  &  1.6 &   --  & 71.0\\
	  J2124--3358 & \citet{bjb+97} & 3  &  4.9 &   --  &  4.6\\
	  J2129--5721 & \citet{llb+96} & 3  &  3.7 &   6.6 & 31.9\\
	  J2145--0750 & \citet{bhl+94} & 3, 12  & 16.1 &   6.8 &  9.0\\
	  \hline
	\end{tabular}
  \begin{tablenotes}
	\item[a] References: (1) Chapter \ref{chap:0437} and
	\citet{vbv+08}; (2) \citet{vbb+01}; (3) \citet{hbo06}; (4)
	\citet{tsb+99}; (5) \citet{ojhb06}; (6) \citet{sns+05}; (7)
	\citet{eb01b}; (8) \citet{hlk+04}; (9) \citet{ktr94}; (10)
	\citet{cb04}; (11) \citet{jhb+05}; (12) \citet{lkd+04}
	\item[b] PSRs J1824$-$2452, J1857+0943 and J1939+2134 are also
 	  known under their B-names: PSRs B1821$-$24, B1855+09 and
 	  B1937+21, respectively.
  \end{tablenotes}
  \label{tab:psrs}
\end{threeparttable}
\begin{landscape}
\oddsidemargin=-0.3cm
\begin{figure}
  \centerline{\psfig{angle=-90.0,width=22cm,figure=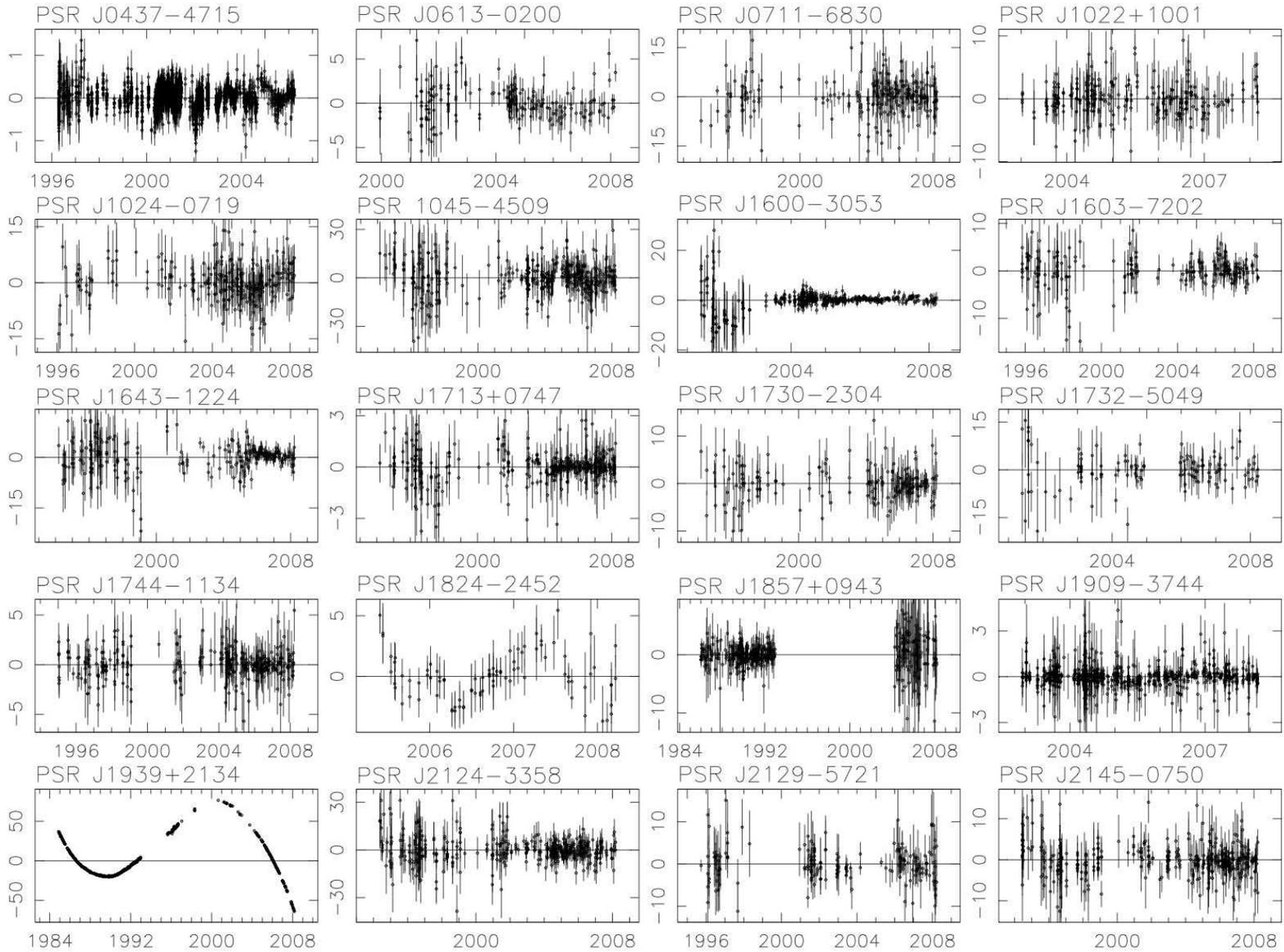}}
	\caption[Timing residuals of the 20 MSPs in our sample]{Timing
    residuals of the 20 pulsars in our sample. Scaling on the x-axis
    is in years and on the y-axis in $\mu$s. For PSRs J1857+0943 and
    J1939+2134, these plots include the Arecibo data made publically
    available by \citet{ktr94}; all other data are from Parkes, as
    described in \S\ref{sec:obs}. Sudden changes in white noise levels
    are due to changes in pulsar backend set-up - see \S\ref{sec:obs}
    for more details.}
  \label{fig:Residuals}
\end{figure}
\end{landscape}

\subsection{Arrival Time Estimation}
The processing applied differs for data from different observing
backends. The FPTM data were calibrated using a real-time system to
produce either two or four Stokes parameters which were later combined
into Stokes I. The FB data were produced from a search system with no
polarimetric calibration possible. This system produced Stokes I
profiles after folding 1-bit data. Data from both of these systems
were integrated in frequency and time to produce a single profile for
each observation. These observations were $\sim$25\,minutes in
duration. For CPSR2 data, in order to minimise the effects of aliasing
and spectral leakage, 12.5\% of each edge of the bandpass was
removed. To remove the worst radio frequency interference, any
frequency channel with power more than $4\sigma$ in excess of the
local median was also removed (``local'' was defined as the nearest 21
channels and the standard deviation $\sigma$ was determined
iteratively). CPSR2 also operated a total power monitor on microsecond
timescales, which removed most impulsive interference.

The CPSR2 data were next integrated for five minutes and calibrated
for differential gain and phase to correct for possible asymmetries in
the receiver hardware. If calibrator observations were available
(especially in the years directly following the CPSR2 commissioning,
observations of a pulsating noise source, needed for polarimetric
calibration, were not part of the standard observing
schedule). Subsequently the data were integrated for the duration of
the observation, which was typically 32\,minutes for PSRs
J2124$-$3358, J1939+2134 and J1857+0943 and 64\,minutes for all other
pulsars. In the case of PSR J1643$-$1224, the integration time was
32\,minutes until December 2005 and 64\,minutes from 2006
onwards. Finally, the CPSR2 data were integrated in frequency and the
Stokes parameters were combined into total power. CPSR2 data that did
not have calibrator observations available were processed identically,
except for the calibration step. While for some pulsars (like PSR
J0437$-$4715) these uncalibrated data are provably of inferior quality
\citei{see, e.g.}{van06}, in our case this is largely outweighed by
the improved statistics of the larger number of TOAs and by the
extended timing baseline these observations provided. We therefore
include both calibrated and uncalibrated observations in our data sets.

To obtain pulse TOAs, the total intensity profiles thus obtained were
cross-correlated with pulsar and frequency-dependent template
profiles. These template profiles were created through addition of a
large number of observations and were phase-aligned for both CPSR2
observing bands. As there were only few high signal-to-noise
observations obtained with the FPTM and FB backends for most pulsars,
these data were timed against standards created with the CPSR2
backend. This may affect the reliability of their derived TOA
errors. For this reason we have evaluated the underestimation of TOA
errors for each backend separately. We note that these factors do not
vary much with backend, which indicates that the application of the
CPSR2 templates to the FB and FPTM data does not affect the timing
significantly. While the TOA errors were generally determined through
the standard Fourier phase gradient method, the Gaussian interpolation
method produced more precise estimates for pulsars with low
signal-to-noise ratios \citep{hbo05a} - specifically for PSRs
J0613$-$0200, J2129$-$5721, J1732$-$5049, J2124$-$3358 and
J1045$-$4509. The PSRCHIVE software package \citep{hvm04} was used to
perform all of the processing described above.

\section{Timing Results}\label{sec:timing}
The \textsc{tempo2} software package \citep{hem06} was used to
calculate the residuals from the TOAs and initial timing solutions
(Table \ref{tab:psrs}). In order to account for the unknown
instrumental delays and pulsar-dependent differences in observing
setup, arbitrary phase-offsets between the backends were
introduced. Where available, data at an observing frequency of
685\,MHz were included in an initial fit to inspect visually the
presence of DM variations. In the case of PSRs J1045$-$4509,
J1909$-$3744, J1939+2134 and J0437$-$4715, such variations were
obvious and dealt with in the timing software through a method similar
to that presented by \citet{yhc+07}. We updated all the pulsar
ephemerides to use International Atomic Time (implemented as TT(TAI)
in \textsc{tempo2}) and the DE405 Solar System ephemerides
\citep{sta04b}. In order to achieve a reduced $\chi^2$ value of close
to unity, the TOA errors were multiplied by backend and
pulsar-dependent error factors which were generally close to unity. A
summary of the lengths of the data sets and the achieved timing
precision can be found in Table \ref{tab:summary}, highlighting the
superior timing precision of PSRs J1909$-$3744, J0437$-$4715 and
J1713+0747 when compared to other pulsars. While the residual RMS has
been an oft-quoted measure of data quality, it does not take the
density of observations into account and can therefore be
misleading. We hence introduce the concept of the ``normalised RMS''
(column 5 of Table \ref{tab:summary}), which is the theoretical RMS
one would get by averaging all TOAs within a year:
\begin{equation}
\sigma_{\rm Norm} = \frac{\sigma}{\sqrt{N T/T_{\rm 0}}},
\end{equation}
with $\sigma$ the RMS of the timing residuals, $N$ the number of TOAs
and $T$ the time span, normalised by $T_{\rm 0} = 1$\,yr. For
comparison of data sets from different projects or telescopes the
normalised RMS provides a more objective measure of data quality than
the residual RMS. We therefore encourage future authors to use this
statistic instead.

The timing residuals for our data sets are presented in Figure
\ref{fig:Residuals} and the timing models are presented in tables
\ref{tab:SinglePsrs1}, \ref{tab:SinglePsrs2}, \ref{tab:binpsrs1},
\ref{tab:binpsrs2}, \ref{tab:binpsrs3} and \ref{tab:binpsrs4}. While
several parameters are listed that have not previously been published,
the analysis by \citet{vbv+08} has demonstrated the unreliability of
parameter uncertainties resulting from standard pulsar timing
techniques, especially in the presence of (even small amounts of)
low-frequency noise. The focus of this present paper is on the overall
pulsar stability and implications for pulsar timing array science, we
therefore defer the discussion of these new astrometric parameters -
along with a reliable analysis of the parameter uncertainties - to a
later paper. However, we encourage observers to use the improved
models when observing. We also note that all but a few of the
parameters in our timing models are consistent with those published
previously.

\begin{threeparttable}
  \caption[Summary of the timing results]{Summary of the timing
  results, sorted in order of decreasing timing precision. The columns
  present the pulsar name, the RMS of the timing residuals, the length
  of the data set, the number of TOAs, the normalised RMS, the second
  period derivative (which is not contained in our timing models, but
  determined independently as a measure of stability) and the
  stability parameter $\Delta_{\rm 8}$. For $\ddot{\nu}$, the numbers
  in brackets represent twice the formal 1\,$\sigma$ errors in the
  last digit quoted. See \S\ref{sec:timing} and \S\ref{sec:stability}
  for details.}
  \begin{tabular}{cr@{. }lr@{.}lrrr@{.}lc}
	  \hline
	  Pulsar & \multicolumn{2}{c}{rms} & 
	  \multicolumn{2}{c}{T} & \multicolumn{1}{c}{N$_{\rm pts}$} & 
	  \multicolumn{1}{c}{Norm.} & \multicolumn{2}{c}{$\ddot{\nu}$} 
	  & $\Delta_{\rm 8}$\\
	  name & \multicolumn{2}{c}{($\mu$s)} &
	  \multicolumn{2}{c}{(yr)}&  & 
	  \multicolumn{1}{c}{rms (ns)} &
	  \multicolumn{2}{c}{($10^{-27}$\,s$^{-1}$)} & \\
	  \hline
	  J1909$-$3744 & 0&166 &  5&2 & 893 &  13 &    1&1(4) & $-5.51$ \\
	  J0437$-$4715 & 0&199 &  9&9 &2847 &  12 & $-$0&23(4)& $-5.79$ \\
	  J1713+0747   & 0&204 & 14&0 & 392 &  39 & $-$0&01(3)& $<-4.93$ \\
	  J1939+2134   & 0&576 & 12&5 & 180 & 152 &    4&3(9) & $-4.58$ \\
	  J1744$-$1134 & 0&614 & 13&2 & 342 & 121 &    0&03(16)&$<-4.66$ \\
	  \\
	  J1600$-$3053 & 1&14   & 6&8 & 477 & 136 &    1&4(28)& $<-4.85$ \\
	  J0613$-$0200 & 1&54   & 8&2 & 190 & 320 & $-$6&1(22)& $<-4.56$ \\
	  J1824$-$2452 & 1&62   & 2&8 &  89 & 287 & 200&0(540)&
	  --\tnote{a}\\
	  J1022+1001   & 1&63   & 5&1 & 260 & 228 & $-$3&3(12)& $<-4.85$ \\
	  J2145$-$0750 & 1&81   &13&8 & 377 & 346 &  0&093(89)& $<-4.34$ \\
	  \\
	  J1603$-$7202 & 1&95 &12&4 &212 & 472 & 0&5(2)   & $<-4.06$ \\
	  J2129$-$5721 & 2&20 &12&5 &179 & 581 & 0&85(92) & $<-3.48$ \\
	  J1643$-$1224 & 2&51 &14&0 &241 & 605 & 1&2(7)   & $-3.82$ \\
	  J1730$-$2304 & 2&51 &14&0 &180 & 700 & 0&08(39) & $<-3.95$ \\
	  J1857+0943   & 2&91 & 3&9 &106 & 558 & $-$7&0(230)&$<-4.39$\\
	  \\
	  J0711$-$6830 & 3&24 &14&2 &227 & 810 & 0&2(6) & $<-4.00$ \\
	  J1732$-$5049 & 3&24 & 6&8 &129 & 744 & 6&2(62)& $-4.07$ \\
	  J2124$-$3358 & 4&03 &13&8 &416 & 925 & 0&01(61)&$<-4.14$ \\
	  J1024$-$0719 & 4&20 &12&1 &269 & 891 & $-$3&3(10)&$<-3.93$\\
	  J1045$-$4509 & 6&64 &14&1 &401 &1251 & 1&5(6) & $<-3.76$ \\
	  \hline
  \end{tabular}
  \begin{tablenotes}
  \item[a] The CPSR2 data on PSR J1824$-$2452 are only 2.8\,years long
    so insufficient data are available to determine a $\Delta_{\rm 8}$
    parameter.
  \end{tablenotes}
  \label{tab:summary}
\end{threeparttable}
\clearpage
\begin{landscape}
  \begin{table}
	\caption[Timing parameters for single PSRs J0711$-$6830,
	  J1024$-$0719, J1730$-$2304 and J1744$-$1134]{Timing parameters
	  for the single pulsars, PSRs J0711$-$6830, J1024$-$0719,
	  J1730$-$2304 and J1744$-$1134. Numbers in brackets give twice
	  the formal $1\sigma$ uncertainty in the last digit quoted.  Note
	  that these parameters are determined using \textsc{Tempo2},
	  which uses the International Celestial Reference System and
	  Barycentric Coordinate Time. As a result this timing model must
	  be modified before being used with an observing system that
	  inputs Tempo format parameters. See \citet{hem06} for more
	  information.}
	\begin{tabular}{lllll}
	  \hline
	  \multicolumn{5}{c}{Fit and data-set parameters}\\
	  \hline
	  Pulsar name\dotfill & 
	  J0711$-$6830        & J1024$-$0719      & 
	  J1730$-$2304        & J1744$-$1134      \\
	  \\
	  MJD range\dotfill   & 
	  49373.6$-$54546.4   & 50117.5$-$54544.6 & 
	  49421.9$-$54544.8   & 49729.1$-$54546.9 \\
	  
	  Number of TOAs\dotfill & 
	  227  & 269 & 180 & 342 \\

	  RMS timing residual ($\mu$s)\dotfill & 
	  3.24 & 4.20 & 2.51 & 0.614 \\

	  Reference epoch for P, $\alpha$, \\
	  $\delta$ and DM determination\dotfill & 
	  49800 & 53000 & 53300 & 53742 \\

	  \hline 
	  \multicolumn{5}{c}{Measured Quantities}\\
	  \hline
	  Right ascension, \\
	  $\alpha$ (J2000.0)\dotfill & 
	  07:11:54.22579(15)& 10:24:38.68849(4) & 
	  17:30:21.6612(3)& 17:44:29.403209(4) \\

	  Declination, $\delta$ (J2000.0)\dotfill & 
	  $-$68:30:47.5989(7)& $-$07:19:19.1696(11) & 
	  $-$23:04:31.28(7) & $-$11:34:54.6606(2)\\

	  Proper motion in $\alpha$, \\
	  $\mu_{\rm \alpha} \cos{\delta}$
	  (mas yr$^{-1}$)\dotfill & 
	  $-$15.55(9)        & $-$35.5(2) & 
	  20.27(6)          & 18.804(15) \\

	  Proper motion in $\delta$, \\
	  $\mu_{\rm \delta}$
	  (mas yr$^{-1}$)\dotfill & 
	  14.23(7)          & $-$48.6(3) & 
	  --                & $-$9.40(6) \\

	  Annual parallax, \\
	  $\pi$ (mas)\dotfill & 
	  --                & --  & 
	  --                & 2.4(2)  \\ 

	  Dispersion measure, \\
	  DM (cm$^{-3}$ pc)\dotfill & 
	  18.408(4)         & 6.486(4) & 
	  9.618(2)          & 3.1380(6) \\

	  Pulse frequency, $\nu$ (Hz)\dotfill & 
	  182.117234869347(4) & 193.715683568727(3) &
	  123.110287192301(2) & 245.4261197483027(5) \\

	  Pulse frequency derivative, \\
	   $\dot{\nu}$ ($10^{-16}\,$s$^{-2}$)\dotfill & 
	  -4.94406(15) & -6.9508(3) &
	  -3.05907(11) & -5.38188(4)\\
	  \hline
	\end{tabular}
	\label{tab:SinglePsrs1}
  \end{table}
\end{landscape}
\begin{landscape}
  \begin{table}
	\caption[Timing parameters for single PSRs J1824$-$2452,
	  J1939+2134 and J2124$-$3358]{Timing parameters for the single
	  pulsars, PSRs J1824$-$2452, J1939+2134 and J2124$-$3358. See
	  caption of Table \ref{tab:SinglePsrs1} for more information.}
	\begin{tabular}{lllll}
	  \hline
	  \multicolumn{4}{c}{Fit and data-set parameters}\\
	  \hline
	  Pulsar name\dotfill & 
	  J1824$-$2452      & J1939+2134        & J2124$-$3358 \\
	  \\
	  MJD range\dotfill & 
	  53518.8$-$54544.9 & 49956.5$-$54526.9 & 49489.9$-$54528.9 \\
	  
	  Number of TOAs\dotfill & 
	  89                & 180               & 416 \\

	  RMS timing residual ($\mu$s)\dotfill & 
	  0.990             & 0.576             & 4.03 \\

	  Reference epoch for P, $\alpha$,\\
	  $\delta$ and DM determination\dotfill & 
	  54219             & 52601             & 53174 \\

	  \hline
	  \multicolumn{4}{c}{Measured Quantities}\\
	  \hline
	  Right ascension, $\alpha$ (J2000.0)\dotfill & 
	  18:24:32.00797(5) & 19:39:38.561286(7)& 21:24:43.85347(3) \\

	  Declination, $\delta$ (J2000.0)\dotfill & 
	  $-$24:52:10.824(13)& +21:34:59.12913(15) & $-$33:58:44.6667(7) \\

	  Proper motion in $\alpha$, $\mu_{\rm \alpha} \cos{\delta}$
	  (mas yr$^{-1}$)\dotfill & 
	  0.1(7)            & 0.13(3)           & $-$14.12(13) \\

	  Proper motion in $\delta$, $\mu_{\rm \delta}$
	  (mas yr$^{-1}$)\dotfill & 
	  $-$11(15)         & $-$0.25(5)        & $-$50.34(25) \\

	  Annual parallax, $\pi$ (mas)\dotfill & 
	  --                & 0.4(4)            & 3.1(11) \\

	  Dispersion measure, DM (cm$^{-3}$ pc)\dotfill & 
	  120.502(3)        & 71.0227(9)        & 4.601(3) \\

	  Pulse frequency, $\nu$ (Hz)\dotfill & 
	  327.405594693013(7) & 641.928233559522(5) & 202.793893879496(2) \\

	  Pulse frequency derivative,
	   $\dot{\nu}$ ($10^{-16}\,$s$^{-2}$)\dotfill &
	  $-$1735.291(4) & $-$433.1100(5) & $-$8.4597(2) \\

	  Second frequency derivative, $\ddot{\nu}$ (s$^{-3}$)\dotfill &
	  $-$2.0(19) & -- & --\\

	  Third frequency derivative, $\dddot{\nu}$ (s$^{-4}$)\dotfill &
	  $-$2.6(12) & -- & -- \\

	  \hline
	\end{tabular}
	  \label{tab:SinglePsrs2}
  \end{table}
\end{landscape}
\begin{landscape}
  \begin{table}
	\caption[Timing parameters for binary PSRs J0613$-$0200,
	  J1045$-$4509 and J1643$-$1224]{Timing parameters for binary PSRs
	  J0613$-$0200, J1045$-$4509 and J1643$-$1224. See caption of
	  table \ref{tab:SinglePsrs1} for more information.}
	\label{tab:binpsrs1}
	\begin{tabular}{llll}
	  \hline
	  \multicolumn{4}{c}{Fit and data-set parameters}\\
	  \hline
	  Pulsar name\dotfill & 
	  J0613$-$0200      & J1045$-$4509     & J1643$-$1224 \\
	  \\
	  MJD range\dotfill & 
	  51526.6$-$54527.3 & 49405.5$-$54544.5 & 
	  49421.8$-$54544.7 \\
	  
	  Number of TOAs\dotfill & 
	  190               & 401 & 
	  241 \\
	  
	  RMS timing residual ($\mu$s)\dotfill & 
	  1.54              & 6.64 & 
	  2.51 \\

	  Reference epoch for P, $\alpha$, $\delta$ \\
	  and DM determination\dotfill & 
	  53114             & 53050 & 
      49524 \\
	  
	  \hline
	  \multicolumn{4}{c}{Measured Quantities}\\
	  \hline
	  Right ascension, $\alpha$ (J2000.0)\dotfill & 
	  06:13:43.975142(11)&10:45:50.18951(5) & 
	  16:43:38.15543(8) \\
	  
	  Declination, $\delta$ (J2000.0)\dotfill & 
	  $-$02:00:47.1737(4)& $-$45:09:54.1427(5)&
	  $-$12:24:58.735(5) \\
	  
	  Proper motion in $\alpha$, $\mu_{\rm \alpha} \cos{\delta}$ (mas
	  yr$^{-1}$)\dotfill& 
	  1.85(7)           & $-$6.0(2) & 
      6.0(1) \\

	  Proper motion in $\delta$, $\mu_{\rm \delta}$ (mas yr$^{-1}$)\dotfill&
	  $-$10.6(2)        & 5.3(2) & 
	  4.2(4) \\
		
	  Annual parallax, $\pi$ (mas)\dotfill & 
	  0.8(7)            & 3.3(38) & 
	  1.6(9) \\
	  
	  Dispersion measure, DM (cm$^{-3}$ pc)\dotfill & 
	  38.782(4)         & 58.137(6) & 
	  62.410(3) \\
		
		Pulse frequency, $\nu$ (Hz)\dotfill & 
		326.600562190182(4) & 133.793149594456(2) & 
		216.373337551615(7) \\

		Pulse frequency derivative, $\dot{\nu}$ 
		($10^{-16}$\,s$^{-2}$)\dotfill & 
		$-$10.2307(7) & $-$3.1613(3) & 
		$-$8.6439(2) \\

	  \\
	  Orbital period, $P_{\rm b}$ (days)\dotfill&
	  1.1985125753(1)   & 4.0835292547(9) & 
	  147.01739776(6) \\
	  
	  Epoch of periastron passage, $T_{\rm 0}$ (MJD)\dotfill&
	  53113.98(2)       & 53048.98(2) & 
	   49577.969(2) \\
	   
	   Projected semi-major axis, $x = a \sin{i}$ (s)\dotfill&
	   1.0914444(3)      & 3.0151325(10) & 
	   25.072614(2) \\

	   $\dot{x}$ ($10^{-14}$)\dotfill & 
	   --                & -- & 
	    $-$4.9(6) \\
		
	   Longitude of periastron, $\omega_{\rm 0}$ (deg)\dotfill&
	   54(6)             & 242.7(16) & 
	   321.850(4) \\
	   
	   Orbital eccentricity, $e$ ($10^{-5}$)\dotfill & 
	   0.55(6)           & 2.37(7) &
	   50.579(4) \\
	   \hline
	\end{tabular}
  \end{table}  
\end{landscape}

\begin{landscape}
  \begin{table}
	\caption[Timing parameters for binary PSRs J1022+1001,
	  J1600$-$3053 and J1857+0943]{Timing parameters for binary PSRs
	  J1022+1001, J1600$-$3053 and J1857+0943. See caption of Table
	  \ref{tab:SinglePsrs1} for more information.}
	\begin{tabular}{llll}
	  \hline
	  \multicolumn{4}{c}{Fit and data-set parameters}\\
	  \hline
	  Pulsar name\dotfill & 
	  J1022+1001 & J1600$-$3053 & J1857+0943 \\

	  \\
	  MJD range\dotfill & 
	  52649.7$-$54528.5 & 52055.7$-$54544.6 & 53086.9$-$54526.9 \\
	  
	  Number of TOAs\dotfill & 
	  260 & 477 & 106 \\
	  
	  RMS timing residual ($\mu$s)\dotfill & 
	  1.63 & 1.14 & 2.91 \\

	  Reference epoch for P, $\alpha$, $\delta$ \\
	  and DM determination\dotfill & 
	  53589 & 53283 & 53806 \\
	  
	  \hline
	  \multicolumn{4}{c}{Measured Quantities}\\
	  \hline
	  Right ascension, $\alpha$ (J2000.0)\dotfill & 
	  10:22:58.003(3) & 16:00:51.903798(12) & 18:57:36.39129(4) \\
	  
	  Declination, $\delta$ (J2000.0)\dotfill & 
	  +10:01:52.76(13) & $-$30:53:49.3407(5) & +09:43:17.225(1) \\
	  
	  Proper motion in $\alpha$, $\mu_{\rm \alpha} \cos{\delta}$ (mas
	  yr$^{-1}$)\dotfill& 
	  $-$17.02(14) & $-$1.06(9) & $-$2.4(5) \\

	  Proper motion in $\delta$, $\mu_{\rm \delta}$ (mas yr$^{-1}$)\dotfill&
	  -- & $-$7.1(3) & $-$5.7(9) \\
		
	  Annual parallax, $\pi$ (mas)\dotfill & 
	  1.8(6) & 0.2(3) & 2.8(23) \\
	  
	  Dispersion measure, DM (cm$^{-3}$ pc)\dotfill & 
	  10.261(2) & 52.3262(10) & 13.286(7) \\
		
		Pulse frequency, $\nu$ (Hz)\dotfill & 
		60.7794479762157(4) & 277.9377070984926(17) & 
		 186.494078441977(5) \\

		Pulse frequency derivative, $\dot{\nu}$ ($10^{-16}$\,s$^{-2}$)\dotfill &
		$-$1.6012(2) & $-$7.3390(5) & 
		$-$6.204(3) \\
	  \\
	  Orbital period, $P_{\rm b}$ (days)\dotfill&
	  7.8051302826(4) & 14.3484577709(13) & 12.327171383(7) \\
	  
	  Epoch of periastron passage, $T_{\rm 0}$ (MJD)\dotfill&
	  53587.3140(6) & 53281.191(4) & 53804.442(22) \\

	  Projected semi-major axis, $x = a \sin{i}$ (s)\dotfill&
	  16.7654074(4) & 8.801652(10) & 9.230778(4) \\

	  $\dot{x}$ ($10^{-14}$)\dotfill & 
	  1.5(10) & $-$0.4(4) & -- \\
		
	  Longitude of periastron, $\omega_{\rm 0}$ (deg)\dotfill&
	  97.75(3) & 181.85(10) & 276.8(6) \\
	   
	  Orbital eccentricity, $e$ ($10^{-5}$)\dotfill & 
	  9.700(4) & 17.369(4) & 2.21(4) \\
	  
	  Sine of inclination angle, $\sin{i}$\dotfill & 
	  0.73\footnotemark{} & 0.8(4) & 0.997(5) \\

	  Inclination angle, $i$ (deg)\dotfill & 
	  47\addtocounter{footnote}{-1}\footnotemark{} & -- & -- \\
	  
	  Companion mass, $M_{\rm c}$ ($M_{\odot}$)\dotfill  & 
	  1.05\addtocounter{footnote}{-1}\footnotemark{} & 0.6(15) &
        0.42(24) \\

	  \hline
	  \\
	  \multicolumn{4}{l}{\addtocounter{footnote}{-1}\footnotemark{}From
	  \citet{hbo06}} 
	\end{tabular}
	\label{tab:binpsrs2}
\end{table}
\end{landscape}

\begin{landscape}
  \begin{table}
	\caption[Timing parameters for binary PSRs J1603$-$7202,
	  J1732$-$5049 and J1909$-$3744]{Timing parameters for binary PSRs
	  J1603$-$7202, J1732$-$5049 and J1909$-$3744. See caption of
	  Table \ref{tab:SinglePsrs1} for more information.}
	\label{tab:binpsrs3}
	\begin{tabular}{llll}
	  \hline
	  \multicolumn{4}{c}{Fit and data-set parameters}\\
	  \hline
	  Pulsar name\dotfill & 
	  J1603$-$7202 & J1732$-$5049 & J1909$-$3744 \\

	  \\
	  MJD range\dotfill & 
	  50026.1$-$54544.7 & 52056.8$-$54544.8 & 
	  52618.4$-$54528.8\\
	  
	  Number of TOAs\dotfill & 
	  212 & 129 & 
	  893\\
	  
	  RMS timing residual ($\mu$s)\dotfill & 
	  1.95 & 3.24 & 
	  0.166\\
	  
	  Reference epoch for P, $\alpha$, $\delta$ \\
	  and DM determination\dotfill & 
	  53024 & 53300 & 
	  53631\\

	  \hline
	  \multicolumn{4}{c}{Measured Quantities}\\
	  \hline
	  Right ascension, $\alpha$ (J2000.0)\dotfill & 
	  16:03:35.67980(4) & 17:32:47.76686(4) & 
	  19:09:47.4366120(8)\\
	  
	  Declination, $\delta$ (J2000.0)\dotfill & 
	  $-$72:02:32.6985(3) & $-$50:49:00.1576(11) & 
	  $-$37:44:14.38013(3)\\
	  
	  Proper motion in $\alpha$, $\mu_{\rm \alpha} \cos{\delta}$ (mas
	  yr$^{-1}$)\dotfill& 
	  $-$2.52(6) & -- & 
	  $-$9.510(7)\\

	  Proper motion in $\delta$, $\mu_{\rm \delta}$ (mas yr$^{-1}$)\dotfill&
	  $-$7.42(9) & $-$9.3(7) & 
	  $-$35.859(19)\\
		
	  Annual parallax, $\pi$ (mas)\dotfill & 
	  -- & -- & 
	  0.79(4)\\
	  
	  Dispersion measure, DM (cm$^{-3}$ pc)\dotfill & 
	  38.060(2) & 56.822(6) & 
	  10.3934(2)\\
		
		Pulse frequency, $\nu$ (Hz)\dotfill & 
		67.3765811408911(5) &
		188.233512265437(3) &
		339.31568740949071(10)\\

		Pulse frequency derivative, $\dot{\nu}$&
		($10^{-16}$\,s$^{-2}$)\dotfill 
		-0.70952(5) &
		-5.0338(12) &
		-16.14819(5)\\

		\\ 
		Orbital period, $P_{\rm b}$ (days)\dotfill& 
		6.3086296703(7) & 5.262997206(13) & 1.533449474590(6)\\
	  
	  Orbital period derivative, $\dot{P}_{\rm b}$
	  ($10^{-13}$)\dotfill & 
	  -- & -- & 5.5(3) \\

	   Projected semi-major axis, $x = a \sin{i}$ (s)\dotfill&
	   6.8806610(4) & 3.9828705(9) & 
	   1.89799106(7) \\
	   
	   $\dot{x}$ ($10^{-14}$)\dotfill & 
	   1.8(5) & -- & 
	   $-$0.05(4)\\
	   
	   $\kappa = e \sin{\omega_{\rm 0}}$ ($10^{-6}$)\dotfill & 
	   1.61(14) & 2.20(5) & 
	   $-$0.4(4)\\

	   $\eta = e \cos{\omega_{\rm 0}}$ ($10^{-6}$)\dotfill & 
	   $-$9.41(13) & $-$8.4(4) & 
	   $-$13(2)\\

	   Ascending node passage, $T_{\rm asc}$ (MJD)\dotfill & 
	   53309.3307830(1) & 51396.366124(2) & 
	   53630.723214894(4)\\

	   Sine of inclination angle, $\sin{i}$\dotfill & 
	   -- & -- & 0.9980(2) \\
	   
	   Companion mass, $M_{\rm c}$ ($M_{\odot}$)\dotfill & 
	   -- & -- & 0.212(4) \\

	   \hline
	\end{tabular}
  \end{table}  
\end{landscape}

\begin{landscape}
  \begin{table}
	\caption[Timing parameters for binary PSRs J1713+0747,
	  J2129$-$5721 and J2145$-$0750]{Timing parameters for binary PSRs
	  J1713+0747, J2129$-$5721 and J2145$-$0750. See caption of Table
	  \ref{tab:SinglePsrs1} for more information.}
	\label{tab:binpsrs4}
	\begin{tabular}{llll}
	  \hline
	  \multicolumn{4}{c}{Fit and data-set parameters}\\
	  \hline
	  Pulsar name\dotfill & 
	  J1713+0747 & J2129$-$5721 & J2145$-$0750 \\

	  \\
	  MJD range\dotfill & 
	  49421.9$-$54546.8 & 49987.4 $-$54547.1 & 
	  49517.8$-$54547.1 \\
	  
	  Number of TOAs\dotfill & 
	  392 & 179 &
	  377 \\	  

	  RMS timing residual ($\mu$s)\dotfill & 
	  0.204 & 2.20 & 
	  1.81 \\

	  Reference epoch for P, $\alpha$, $\delta$ \\
	  and DM determination\dotfill & 
	  54312 & 54000 & 
	  53040 \\

	  \hline
	  \multicolumn{4}{c}{Measured Quantities}\\
	  \hline
	  Right ascension, $\alpha$ (J2000.0)\dotfill & 
	  17:13:49.532628(2) & 21:29:22.76533(5) & 
	  21:45:50.46412(3) \\

	  Declination, $\delta$ (J2000.0)\dotfill & 
	  +07:47:37.50165(6) & $-$57:21:14.1981(4) & 
	  $-$07:50:18.4399(13) \\
	  
	  Proper motion in $\alpha$, $\mu_{\rm \alpha} \cos{\delta}$ (mas
	  yr$^{-1}$)\dotfill& 
	  4.923(10) & 9.35(10) &  
	  $-$9.67(15) \\

	  Proper motion in $\delta$, $\mu_{\rm \delta}$ (mas yr$^{-1}$)\dotfill&
	  $-$3.85(2) & $-$9.47(10) & 
	  $-$8.8(4) \\

	  Annual parallax, $\pi$ (mas)\dotfill & 
	  0.94(11) & 1.9(17) & 
	  1.5(5) \\

	  Dispersion measure, DM (cm$^{-3}$ pc)\dotfill & 
	  15.9915(2) & 31.853(4) & 
	  8.9979(14) \\

	  Pulse frequency, $\nu$ (Hz)\dotfill & 
	  218.8118404414362(3) &
	  268.359227423608(3) &
	  62.2958878569665(6) \\
	  
	  Pulse frequency derivative, $\dot{\nu}$ ($10^{-16}$\,s$^{-2}$)\dotfill &
	  -4.08379(3) &
	  -15.0179(2) &
	  -1.15588(3) \\

	  \\
	  Orbital period, $P_{\rm b}$ (days)\dotfill&
	  67.825130964(16) & 6.625493093(1) &
	  6.83892(2) \\
	  
	  Orbital period derivative, $\dot{P}_{\rm b}$
	  ($10^{-13}$)\dotfill & 
	  41(20)& -- & 
	  4(3) \\

	  Epoch of periastron passage, $T_{\rm 0}$ (MJD)\dotfill&
	  54303.6328(8) & 53997.52(3) & 
	  53042.431(3)\\
	   
	  Projected semi-major axis, $x = a \sin{i}$ (s)\dotfill&
	  32.3424236(3) & 3.5005674(7) & 
	  10.1641080(3) \\
	  
	  $\dot{x}$ ($10^{-14}$)\dotfill & 
	  -- & 1.1(6) & 
	  $-$0.28(33) \\

	  Longitude of periastron, $\omega_{\rm 0}$ (deg)\dotfill&
	  176.190(4) & 196.3(15) & 
	  200.63(17) \\
	  
	  Orbital eccentricity, $e$ ($10^{-5}$)\dotfill & 
	  7.4940(3) & 1.21(3) & 
	  1.930(6)\\

	  Periastron advance, $\dot{\omega}$ (deg/yr)\dotfill & 
	  -- & -- & 
	  0.06(6) \\

	  Inclination angle, $i$ (deg)\dotfill & 
	  78.5(18) &  -- & 
	  -- \\

	  Companion mass, $M_{\rm c}$ ($M_{\odot}$)\dotfill & 
	  0.20(2) & -- & 
	  -- \\
	  
	  Longitude of ascending node, $\Omega$ (deg)\dotfill & 
	  68(17) & -- & 
	  -- \\
	  
	  \hline
	\end{tabular}
  \end{table}  
\end{landscape}


\section{Quantifying Low-Frequency Noise}\label{sec:stab}
In this Section, we determine some standard stability measures for our
data sets, to allow comparison with previous publications. The first
of these measures is the second time derivative of the spin
frequency. Since pulse frequency, $\nu$, and frequency derivative,
$\dot{\nu}$, are fitted as part of the timing model, a quadratic
functional form is effectively removed from the timing residuals. The
effect of any low-frequency process is therefore best characterised by
a cubic polynomial, as is clearly seen in the timing residuals of PSR
J1939+2134 in Figure \ref{fig:Residuals}. In order to quantify the
size of this instability, a second derivative of the pulsar spin
frequency with respect to time, $\ddot{\nu}$, can be fitted. The
$\ddot{\nu}$ values for all 20 MSPs of our sample are presented in
column six of Table \ref{tab:summary}. While the clear low-frequency
noise of PSRs J1939+2134 and J1824$-$2452 results in significant, high
values of $\ddot{\nu}$, the insignificance of the measurement for the
remaining pulsars renders this parameter ineffective for comparative
purposes.

The fact that $\ddot{\nu}$ effectively characterises the stability of
the data set on timescales of the data length, which differs between
data sets, further reduces the usefulness of this parameter. A partial
solution to this problem was presented by \citet{antt94} who
introduced the $\Delta_{\rm 8}$ parameter, which is proportional to
the logarithm of the average $|\ddot{\nu}|/\nu$ value measured over a
time span of $10^8$\,seconds ($\sim 3.16$\,years). The $\Delta_{\rm
8}$ parameters for our data are presented in column seven of Table
\ref{tab:summary}. As $\Delta_{\rm 8}$ measures stability on a given
timescale, it provides a more straightforward comparison between
pulsars. 

\begin{figure}
  \centerline{\psfig{angle=0.0,width=14cm,figure=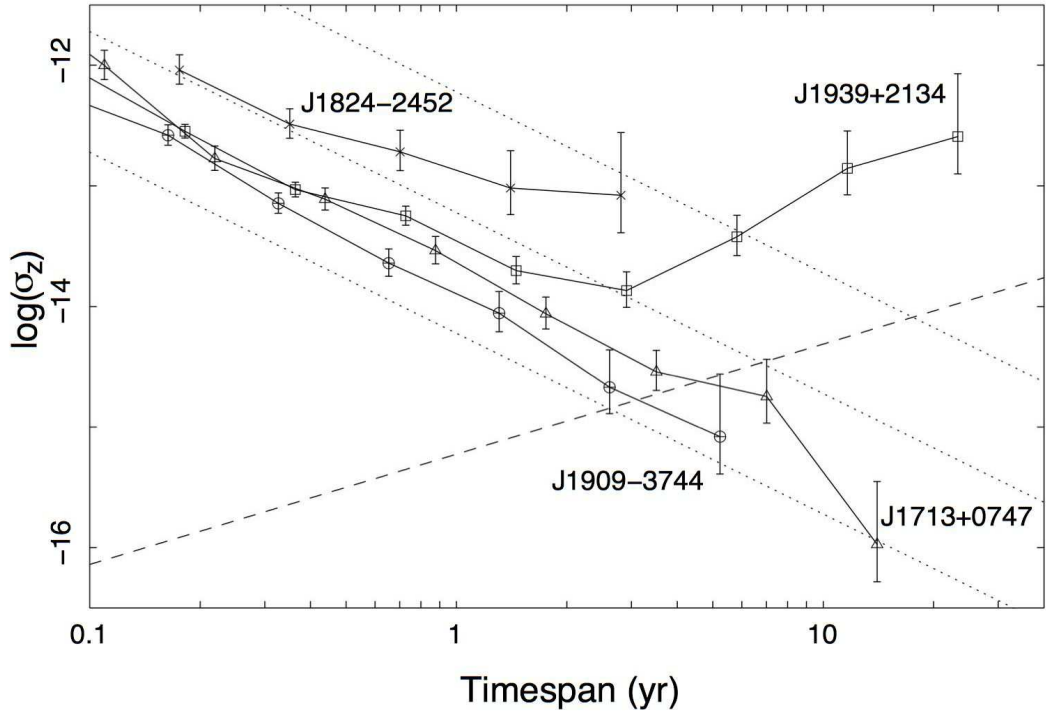}}
  \caption[$\sigma_{\rm z}$ graph for four representative
  MSPs]{$\sigma_{\rm z}$ stability parameter for the two most unstable
  (PSR J1939+2134: squares; PSR J1824$-$2452: crosses) and two of the
  most stable pulsars in our sample (PSR J1909$-$3744: circled
  plusses; PSR J1713+0747: triangles), against timescale. The dotted
  slanted lines represent white noise levels of (bottom to top)
  100\,ns, 1\,$\mu$s and 10\,$\mu$s; the dashed slanted line shows the
  steepness introduced by a hypothetical GWB (see \S\ref{sec:intro});
  pulsars whose curve is steeper than this line (like PSR J1939+2134),
  can therefore be expected to be of little use to PTA efforts. The
  specific line plotted here is for a GWB with $\Omega_{\rm gw} h^2 =
  10^{-9}$ and $\alpha = -1$ (i.e. $A = 1.26\times
  10^{-15}$). However, since this theoretical effect disregards
  sampling and model fitting effects, a bound on the GWB amplitude
  cannot be directly derived from this graph. To fully account for
  such effects, a method based on simulations of the GWB is presented
  in Chapter \ref{chap:GWBLimit}.}
  \label{fig:sigmaz}
\end{figure}

While $\Delta_{\rm 8}$ is of interest for comparison with previous
publications and as an initial measure of timing stability, a full
analysis at varying timescales requires a power spectrum. Because of
various pulsar timing-specific issues such as clustering of data,
large gaps in data and large variations in error bar size, however,
standard spectral analysis methods fail to provide reliable
results. An alternative is provided by the $\sigma_{\rm z}$ statistic,
as described by \citet{mte97}. The interpretation of this statistic -
plotted as a function of timescale in Figure \ref{fig:sigmaz} -
requires some attention. As presented by \citet{mte97}, a power
spectrum with spectral index $\beta$:
\[
P(\nu) \propto f^{\beta}
\]
would translate into a $\sigma_{\rm z}$ curve:
\[
\sigma_{\rm z}^2(\tau) \propto \tau^{\mu}
\]
where the spectral indices are related as:
\begin{equation}
  \mu = 
  \begin{cases}
	-(\beta+3) & {\rm if\ } \beta < 1\\
	-4         & {\rm otherwise.}
  \end{cases}
\end{equation}
This implies that spectra have different slopes in a $\sigma_{\rm z}$
graph than in a power spectrum. Figure \ref{fig:sigmaz} provides some
examples for guidance: lines with a slope of $-3/2$ represent
spectrally white data and a GWB with a spectral index $\alpha = -2/3$
in the gravitational strain spectrum, would have a positive slope of
$2/3$ in $\sigma_{\rm z}$.

Figure \ref{fig:sigmaz} clearly reveals the scale-dependent stability
of PSR J1939+2134: this pulsar is stable at sub-microsecond levels on
short timescales, but experiences a turnover on a timescale of
$\sim$2\,years, indicating poorer stability. PSR J1824$-$2452 shows
similar behaviour, with microsecond precision at timescales below one
year and increases at larger scales. Our longest high-precision data
set, on PSR J1713+0747, shows constant levels of stability up to
14\,years, which contradicts the analysis of \citet{sns+05}.

Overall, six pulsars show signs of a turnover or flattening off like
PSR J1939+2134. These six are PSRs J0613$-$0200, J1022+1001,
J1024$-$0719, J1045$-$4509, J1824$-$2452 and J1939+2134 itself. The
$\sigma_{\rm z}$ graphs for the 14 remaining pulsars all show
consistency with a white noise slope, implying stability at the level
of the residual RMS over all timescales shorter than the time span of
our data. While the graph for PSR J0437$-$4715 displays a slope
slightly less steep than that expected for pure white noise, this
excess of low-frequency noise \citei{first discussed in Chapter
\ref{chap:0437} and by}{vbv+08} is not strong enough to affect its
usefulness for GWB detection efforts since a GWB is expected to induce
a much steeper slope.

In summary, the $\sigma_{\rm z}$ graphs demonstrate that most of the
MSPs investigated in this analysis will prove useful in PTA projects,
provided the RMS is reduced and/or the stability is maintained over
longer time spans. Whilst stability on longer timescales cannot be
analysed without continued observing, the following sections will
assess the prospects for reduction of the residual RMS.

\begin{figure}
  \centerline{\psfig{angle=0.0,width=14cm,figure=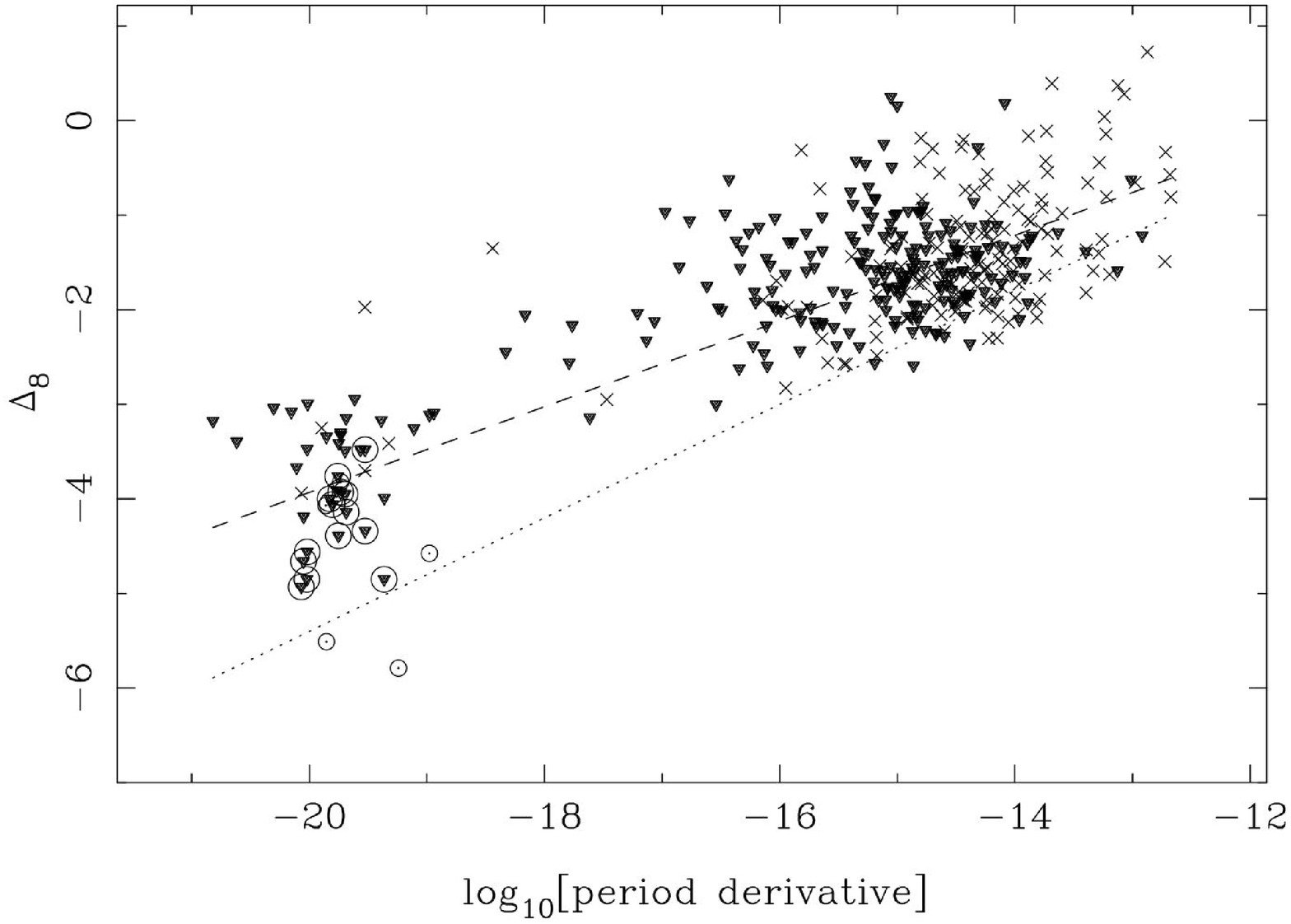}}
  \caption[$\Delta_{\rm 8}$ as a function of pulse
    period]{$\Delta_{\rm 8}$ stability parameter for a combination of
    the data presented in this paper (circled dots and triangles) and
    in the upcoming Hobbs, Lyne \& Kramer (2009; crosses and
    non-circled triangles). Inverted triangles present upper limits,
    crosses and dots show actual measured values. The dotted line
    corresponds to $\Delta_{\rm 8} = 6.6+0.6 \log{\dot{P}}$ as
    obtained by \citet{antt94}. Hobbs et al. (2009) suggest that this
    is low and obtain $\Delta_{\rm 8} = 5.1+0.5 \log{\dot{P}}$ (dashed
    line) instead.}
  \label{fig:delta8}
\end{figure}
\section{Timing Precision Analysis}\label{sec:stability}
For pulsar timing arrays to detect gravitational waves successfully,
MSPs must both be stable over long timescales and be able to be timed
with high precision. In the previous section, we have analysed some
standard stability measures for our data sets. In this section, we
will break down the timing RMS, $\sigma_{\rm Tot}$, into various
components in order to achieve an upper limit on intrinsic timing
noise and therefore on the ultimate timing precision of
MSPs\footnote{Since it is known that intrinsic timing noise differs
  strongly from pulsar to pulsar (see, e.g. \ref{sec:stab}), the upper
  limit we derive in this section will not necessarily hold for any
  MSPs that aren't included in this analysis. However, since this
  limit will imply that there is no inherent reason why MSPs would
  always time worse, we assume pulsar searches will discover MSPs with
  timing properties comparable to those of the pulsars investigated
  here.}.  In \S\ref{sec:theory}, we will estimate the level of
radiometer noise, $\sigma_{\rm Rad}$, in our timing. A new measure which
we dub ``the sub-band timing measure'', $\sigma_{\rm sb}$, will be
introduced in \S\ref{sec:ISM}; this new measure eliminates many
systematic effects to provide an indicator of theoretical
precision. In \S\ref{sec:stab:disc}, we discuss how these estimates
can be used to provide a simple upper limit on the precision with
which MSPs may be timed.  Because the timing of most pulsars in our
sample is strongly dominated by radiometer noise, we will perform this
analysis only on two of the most precisely timed pulsars, PSRs
J1909$-$3744 and J1713+0747, as well as on the most clearly unstable
pulsar, PSR J1939+2134. Also, given the large variation of systematic
effects one can expect for the different backends, this analysis will
be based only on the Parkes CPSR2 data, which implies that only
effects on timescales of five years or less will be estimated. While a
full analysis of lower spectral frequencies may be of interest in
itself, for our purpose of PTA feasibility assessment a five year
timescale is sufficient, as will be demonstrated in \S\ref{sec:PTA}.

\begin{table*}
  \begin{center}
	\caption[Breakdown of weighted timing residuals for three selected
	  pulsars]{Breakdown of weighted timing residuals for three selected
    pulsars. The timing RMS ($\sigma_{\rm Tot}$) is the RMS of the CPSR2
    timing residuals; the sub-band RMS ($\sigma_{\rm sb}$) is the RMS of
    the offset between the two frequency bands of the CPSR2 backend
    (divided by $\sqrt{2}$) and the radiometer limit ($\sigma_{\rm Rad}$)
    is the theoretical prediction of the timing precision, assuming that
    only radiometer noise is present in the data. The temporal and
    frequency components ($\sigma_{\tau}$ and $\sigma_{\nu}$
    respectively), are derived from the previous three quantities, as
    described in 
    \S\ref{sec:ISM}.}
	\label{tab:stability}
	\begin{tabular}{cccccc}
	  \hline
    Pulsar name & $\sigma_{\rm Tot}$ & $\sigma_{\rm sb}$ &
    $\sigma_{\rm Rad}$ & $\sigma_{\tau}$ & $\sigma_{\nu}$ \\
    & (ns) & (ns) & (ns) & (ns) & (ns) \\
	  \hline
	  J1909$-$3744 & 166 & 144 & 131 & 83 & 60 \\
	  J1713+0747   & 170 & 149 & 105 & 82 & 106 \\
	  J1939+2134   & 283 & 124 & \hspace{1ex}64 & 254 & 106 \\
	  \hline
	\end{tabular}
  \end{center}
\end{table*}

\subsection{Theoretical Estimation of Radiometer-Limited Precision}
\label{sec:theory}
The receiver noise present in pulsar timing data can be reduced
through longer observations, larger bandwidth or larger
telescopes. The larger bandwidth of new pulsar instrumentation will
therefore reduce this noise in future data. Considering timing
projects worldwide, the larger effective collecting area of several
telescopes with respect to Parkes will further limit the radiometer
noise present in future data sets. The advantage of these improved
technologies and larger telescopes, will be determined by the
proportion of our timing precision caused by this source of noise. The
amount of radiometer noise present in the timing residuals can be
determined based on the pulsar's observed pulsar profile shape and
brightness. 
Equation (13) of \citet{van06} provides the following
measure (notice we only consider the total intensity, $S_{\rm 0}$, to
allow direct comparison with our timing results):
\begin{equation}
\sigma = P\times \sqrt{V} = P\times \Bigg(4\pi^2 \sum_{m=1}^{N_{\rm
	max}\leq N/2} \nu_{\rm m}^2 \frac{S_{\rm 0,m}^2}{\varsigma_{\rm
	0}^2}\Bigg)^{-0.5},
\end{equation}
where $\nu_{\rm m}$ is the $m^{\rm th}$ frequency of the Fourier
transform of the pulse profile, $S_{\rm 0,m}^2$ is the total power at
that frequency, $\varsigma_{\rm 0}$ is the white noise variance of the
profile under consideration, $N$ is the total number of time bins
across the profile and $N_{\rm max}$ is the frequency bin where the
Fourier transform of the pulse profile reaches the white noise level,
$\varsigma_{\rm 0}$. $V$ is the expected variance in the phase-offset
or residual, $P$ is the pulse period and $\sigma$ is the residual RMS
predicted for the input pulse profile.

Applying this equation to the CPSR2 observations used in our timing,
provides a measure of the expected timing precision, assuming that
only radiometer noise affects hourly integrations. This measure is
listed in column 4 of Table \ref{tab:stability} for the three pulsars
considered. The consistency of these values with the formal errors
resulting from the TOA determination \citei{as described by}{tay92}
demonstrates the robustness of this method. The values show that even
the most precise timing data sets are dominated by white noise. For
more than half of our sample of 20, the estimated radiometer noise is
of the order of a microsecond or more, demonstrating the need for
longer integration times, larger bandwidth or larger collecting area.

\subsection{Estimating Frequency-Dependent Effects}\label{sec:ISM}
A second class of timing irregularities, which may be reduced by
improved modelling of the ISM, are frequency-dependent effects. By
measuring the offset between the timing residuals of the two
64\,MHz-wide frequency bands of the CPSR2 backend system (see
\S\ref{sec:Inst}), we achieve a combined measure of such
frequency-dependent effects and the earlier determined radiometer
noise. For ease of reference, we will call the RMS of the timing
offset between the two bands, divided by $\sqrt{2}$, the ``sub-band
RMS'' henceforth; beyond radiometer noise, it includes the effects of
frequency-dependent interstellar scattering, unmodelled dispersion
measure variations (to some degree) and possibly some
frequency-dependent calibration errors or profile variations. However,
errors that affect both bands equally will be excluded from the
sub-band RMS, while still contributing to the RMS of the normal timing
residuals. The difference between the timing and sub-band RMS must
therefore be composed of clock errors, most calibration imperfections,
errors in the pulsar and planetary ephemerides, intrinsic timing noise
and backend-induced instabilities.

The sub-band RMS for the three selected pulsars is presented in column
three of Table \ref{tab:stability}. Columns five and six of the same
Table provide the differences of the sub-band RMS with the timing RMS
and the radiometer noise, respectively and were derived by assuming
these effects all add in quadrature: 
\begin{equation}
  \sigma_{\nu} = \sqrt{\sigma_{\rm sb}^2-\sigma_{\rm Rad}^2}
\end{equation}
and
\begin{equation}
  \sigma_{\tau} = \sqrt{\sigma_{\rm Tot}^2-\sigma_{\rm sb}^2}.
\end{equation}

\subsection{Discussion}\label{sec:stab:disc}
The quantities derived in the previous sections allow a simple bound
to be placed on the potential timing precision that may be achieved
with new backends and larger telescopes. Specifically, assuming that
newly researched techniques \citep{hs08,wksv08} succeed in mitigating
frequency-dependent ISM effects, the time-dependent instabilities in
our timing will ultimately limit the timing precision. A simple
estimate of this bound is provided by the difference between the RMS
of the timing residuals and the sub-band RMS. This difference does not
only include intrinsic pulsar timing noise, but also correctable
corruptions such as errors in the clock corrections or planetary
ephemerides, instabilities in the observing system or changes in
hardware, incompleteness of the pulsar timing model and even
dispersion measure changes (since only the differential part of the DM
variations will be contained in the sub-band timing, while the largest
contribution affects both bands equally). The combination of these
effects implies the bound we derive on the time-dependent timing
variations (titled ``temporal systematic'' in Table
\ref{tab:stability}) is clearly a conservative limit on the achievable
timing precision. Nevertheless, our analysis of PSRs J1909$-$3744 and
J1713+0747 inspires confidence in the potential for pulsar timing at
$< 100\,$ns precision. We also note that, while both PSRs J1909$-$3744
and J1713+0747 are amongst the brightest MSPs in our sample, their
timing is dominated by white noise; this suggests that the timing of
most if not all of the weaker pulsars (which have inherently higher
levels of radiometer noise) can be readily enhanced by the adoption of
backends with larger bandwidth, longer integration times or future
larger telescopes.

It is interesting to ponder whether future very large X-ray telescopes
might be able to time MSPs accurately, thus removing the entire ISM
contribution to arrival time uncertainties. At present X-ray
observatories cannot compete with radio timing but future missions
might for selected objects. For example, X-ray timing profiles lack
the sharp features that make pulsars like PSR J1909$-$3744 such a
great timer, but this might change with increased
sensitivity. Ultimately, gravitational wave astronomy based on pulsar
timing might not only use data from a host of international
observatories, but also from different wavebands.

\section{Prospects for Gravitational Wave Detection}
\label{sec:PTA}
\begin{figure*}
  \psfig{angle=0.0,width=14cm,figure=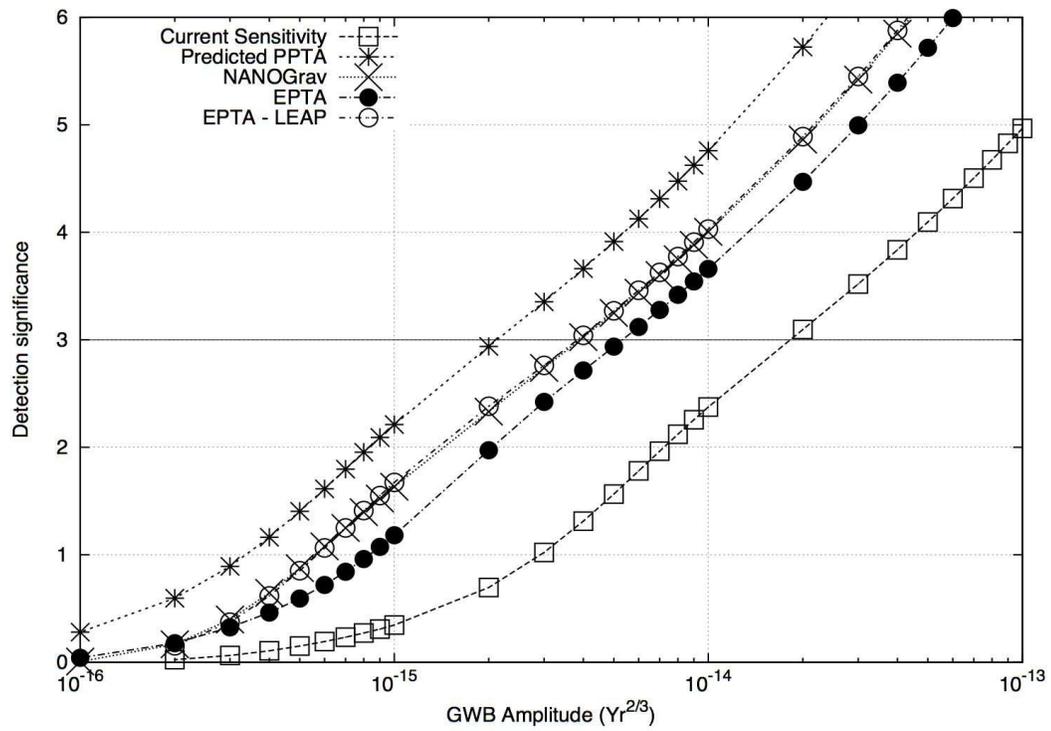}
  \caption[Sensitivity curves for current PTA efforts]{Sensitivity
  curves for different scenarios of PTA efforts. Notice the
  ``NANOGrav'' and ``EPTA -- LEAP'' curves are almost
  coincident. Gravitational waves are predicted to exist in the range
  $10^{-15}-10^{-14}$. See text and Table \ref{tab:PTAs} for more
  information.}
  \label{fig:PTAs}
\end{figure*}

\begin{figure*}
  \psfig{angle=0.0,width=14cm,figure=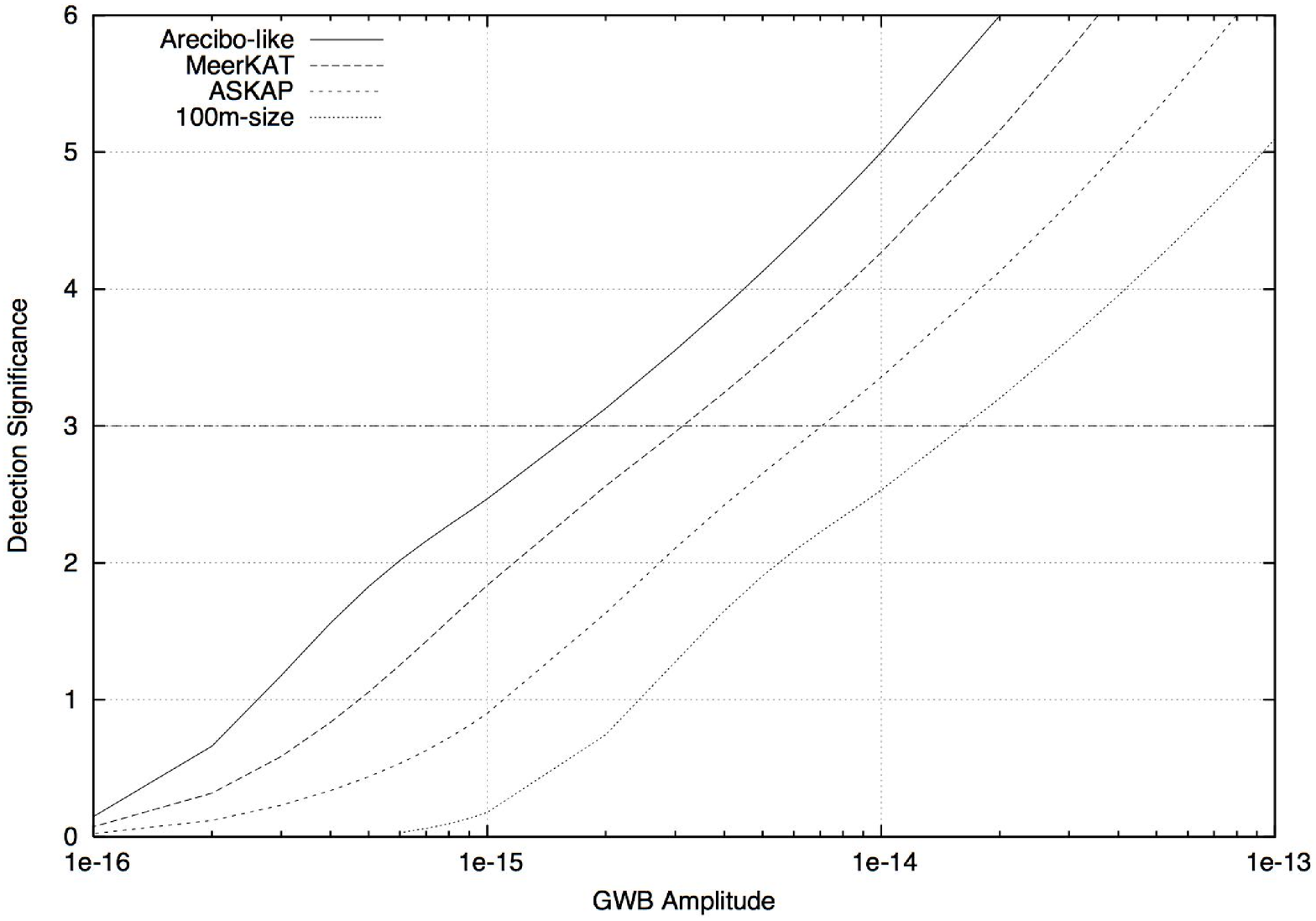}
  \caption[Sensitivity curves for future PTA efforts]{Sensitivity
  curves for the two main SKA pathfinders and for telescopes with
  collecting area equal to those of the Arecibo and Green Bank radio
  telescopes. Gravitational waves are predicted to exist in the range
  $10^{-15}-10^{-14}$. See discussion in \S\ref{sec:PTA} and Table
  \ref{tab:PTAs} for more information.}
  \label{fig:FutPTAs}
\end{figure*}

\citet{jhlm05} derived the expected sensitivity of a PTA to a GWB with
given amplitude, $A$, both for homogeneous arrays (where all pulsars
have comparable timing residuals) and inhomogeneous arrays. They also
pointed out the importance of prewhitening the residuals to increase
sensitivity at larger GWB amplitudes. We present a simpler derivation
of the sensitivity in Appendix \ref{app:Sens}, in a manner that
provides some guidance on analysing the data. We assume that the
prewhitening and correlation are handled together by computing
cross-spectra and we estimate the amplitude of the GWB directly rather
than using the normalised cross correlation function. We assume that
the noise is white, but can be different for each pulsar. Our results
are very close to those of \citet{jhlm05}. The analysis could be
easily extended to include non-white noise. In this section, we apply
this analysis to the current data and use it to make predictions for
ongoing and future PTA projects. The input parameters for the
different PTA scenarios considered are listed in Table \ref{tab:PTAs},
while the sensitivity curves are drawn in Figures \ref{fig:PTAs} and
\ref{fig:FutPTAs}.

We considered five ongoing PTA scenarios: \emph{Current} refers to the
data presented in this paper, using the shortest overlapping time span
of the sample: five years\footnote{This ignores the shorter sampling
of PSR J1824$-$2452, which may not prove useful in a PTA project
lasting longer than a few years. For the purpose of this simulation we
assume the timing precision of PSR J1824$-$2452 to remain identical,
while the time span is increased to five years.}. \emph{Predicted
PPTA} assumes the usage of 256\,MHz of bandwidth at Parkes, which
implies a four-fold bandwidth increase and therefore a two-fold timing
precision increase. In order to scale the RMS from our current data,
we assume an intrinsic noise floor of 80\,ns (as expected from Table
\ref{tab:stability}) and scale the remainder according to the
radiometer equation. This implicitly assumes that improvements in
technology (which reduce the radiometer noise) are equalled by
progress in calibration and ISM-correction methodologies (which
decrease the frequency-dependent noise). This scenario is also the
only one to be considered for more than five years, mainly in order to
show the large impact a doubling of campaign length can have, but also
because several years of high precision timing data with the given
bandwidth do already exist \citep{man08}. The \emph{NANOGrav} scenario
assumes Arecibo gain for the ten least well-timed pulsars and GBT gain
for the ten best-timed pulsars, in order to get a fairly equal RMS for
all 20 MSPs. \emph{EPTA} assumes monthly observations with five
100\,m-class telescopes \citep{jsk+08}. An alternative to this
scenario is presented in \emph{EPTA--LEAP}, which interferometrically
combines the five telescopes to form a single, larger one. This
decreases the number of observations, but increases the gain.

\begin{threeparttable}
  \caption[Assumed parameters for PTA projects]{Assumed parameters
	for future and ongoing PTA efforts.}
  \begin{tabular}{lccrlr}
	\hline
	Scenario & \multicolumn{1}{l}{N$_{\rm tel}$} & 
	\multicolumn{1}{l}{Relative} & \multicolumn{1}{l}{Dish} &
	Observing & \multicolumn{1}{l}{Project} \\ 
	name & & \multicolumn{1}{l}{bandwidth} &
	\multicolumn{1}{l}{diam. (m)} & regularity &
	\multicolumn{1}{l}{length (yrs)} \\ 
	\hline
	Current                & 1 & 1(=64\,MHz) & 64 & weekly & 5 \\
	Predicted PPTA         & 1 & 4 & 64 & weekly & 10 \\
	NANOGrav    & 2 & 4 & 305; 100 & monthly & 5 \\
	EPTA        & 5 & 2 & 100 & monthly & 5 \\
	EPTA - LEAP & 1\tnote{a} & 2 & 224 & monthly & 5\\
	\\
	Arecibo-like           & 1 & 8 & 305 & two-weekly & 5 \\
  100\,m-size            & 1 & 8 & 100 & two-weekly & 5 \\
	ASKAP                  & 40& 4 & 12 & weekly & 5 \\
	MeerKAT\tnote{b} & 80& 8 & 12 & weekly & 5 \\
	\hline
  \end{tabular}
  \begin{tablenotes}
	\item[a] Under the LEAP initiative, five 100\,m-class telescopes
	  will be combined into an effective 224\,m single telescope.
	\item[b] MeerKAT architecture from Justin Jonas,
	  private communication.
  \end{tablenotes}
  \label{tab:PTAs}
\end{threeparttable}
\vspace{0.5cm}\\
It must be noted that several of the pulsars under consideration
cannot be observed with most Northern telescopes, due to the
declination limits of these telescopes. Furthermore, this analysis
assumes the timing residuals to be statistically white and while more
than ten of the 20 pulsars under consideration already have shown
significant stability over ten years or more, for the remaining ten we
cannot confidently assume their stability on such timescales yet. We
therefore assume stable replacement pulsars to be discovered as
needed, especially for the few pulsars that have insufficient
stability over five years. As mentioned before, we also assume that
progress will be made in the mitigation of frequency-dependent
calibration and ISM effects. Finally, this analysis is based on the
Parkes data presented in this paper and therefore assumes systematic
effects to be at most at the level of the Parkes observing system
used.

Bearing all of this in mind, cautious optimism seems justified for GWB
detection through PTA experiments on timescales of five to ten years,
provided current models of gravitational wave backgrounds are
correct. As described in \S\ref{sec:GWBSources}, it must be noted that
there are a substantial number of badly determined inputs in these
models, especially those concerned with a GWB from SMBH mergers. It
is, for example, still unclear exactly what fraction of galaxy mass is
a result of merger events as opposed to accretion. Since only the
merging of galaxies results in binary black holes and hence
contributes to the GWB, this mass fraction is crucial for any reliable
prediction of GWB strength.

Given the scaling laws that can easily be derived from equation (12) of
\citet{jhlm05}, the GWB amplitude at which a 3\,$\sigma$ detection can
be made, scales as follows:
\begin{equation}\label{eq:PTAScaling}
  A_{S = 3} \propto \frac{\sigma}{T^{5/3}\sqrt{N_{\rm pts}}},
\end{equation}
where $T$ is the time span of the data set, $N_{\rm pts}$ is the
number of TOAs for each data set and $\sigma$ is the average RMS of a
data set. Since our analysis assumes an intrinsic noise floor of
80\,ns for each scenario, the potential for reduction of $\sigma$ is
limited, leaving only the regularity of observations and the length of
the campaign to dominate the sensitivity curve. This explains the
equivalence of the NANOGrav and LEAP scenarios. It also implies that
the sensitivity curves for the larger telescopes (i.e. all scenarios
except ``Current'' and ``PPTA'') are limited by our bound of 80\,ns -
if this assumption for MSP intrinsic instability is overly
pessimistic, then the actual sensitivity is expected to be higher. In
particular, the benefit of LEAP over the standard EPTA scenario is
strongly dependent upon this bound. Finally, the strong dependence on
$T$ underscores the importance of stability analysis over much longer
time spans and continued observing. While our $\sigma_{\rm z}$ graph
for J1713+0747 provides the first evidence for high stability over
timescales beyond 10 years, such stability must still be demonstrated
for many more MSPs.

With the completion of the Square Kilometre Array (SKA) pathfinders
expected in three years time, we consider the potential of both the
Australian SKA Pathfinder (ASKAP) and the South African Karoo Array
Telescope (MeerKAT) for PTA programmes. ASKAP is primarily designed
for H$\,${\sc i} surveys and therefore sacrifices point source
sensitivity for a wide field of view, whereas MeerKAT's design is
better suited for point source sensitivity over a more limited field
of view. The expected architecture for either telescope is listed in
Table \ref{tab:PTAs} - notice we assume phase-coherent combination of
the signals of all dishes, effectively resulting in a scenario
equivalent to a single telescope of diameter 107\,m for MeerKAT and
76\,m for ASKAP. The resulting sensitivity curves are drawn in Figure
\ref{fig:FutPTAs}, along with a hypothetical curve for the most
sensitive telescope currently operational, the Arecibo radio
telescope. This Figure clearly shows the advantage MeerKAT holds over
ASKAP for PTA work, both in number of dishes and in bandwidth. The
sensitivity of Arecibo is much higher than that of either prototype,
but its usefulness in reality is limited by the restricted sky
coverage and hence available pulsars. While both MeerKAT and ASKAP can
see large parts of the sky, the sky coverage of Arecibo as well as the
short transit time make an exclusively Arecibo-based PTA practically
impossible; however, its potential as part of a combined effort
(Figure \ref{fig:PTAs}) or in a global PTA, is undeniable if the level
of systematic errors is small compared to the radiometer noise.

\section{Conclusions}\label{sec:conclusions}
We have presented the first long-term timing results for the 20 MSPs
constituting the Parkes pulsar timing array. While two of these
pulsars show clear signs of unmodelled low-frequency noise (PSRs
J1824$-$2452 and J1939+2134), the remaining 18 pulsars show remarkable
stability on timescales of five to ten years. A stability analysis has
revealed that the overall level of systematic and pulsar-intrinsic
effects is estimated to be below 100\,ns for at least some of our
pulsars. We interpreted this result in the context of ongoing and
future pulsar timing array projects, demonstrating the realistic
potential for GWB detection through pulsar timing within five to ten
years, provided suitable replacements are found for the few unstable
pulsars currently in the sample and provided technical developments
evolve as expected. Given the location of currently known MSPs, the
prospects of the MeerKAT SKA pathfinder as a gravitational wave
detector are found to be particularly good.

\chapter[Spectral Analysis and the GWB]{Spectral Analysis of Pulsar
  Timing Residuals and a Limit on the GWB}
\label{chap:GWBLimit}
\noindent \textsf{It is difficult to make predictions, especially
  about the future.\\}
\vspace{0.25cm}
\textit{Niels Bohr and various others}
\vspace{1.5cm}

\section{Abstract}
Pulsar timing data has been used to place limits on the amplitude of
any potential GWB. The methods used for placing such limits have been
either statistically unreliable or had limited applicability due to
intricacies of real data sets (as discussed in
\S\ref{ssec:Limits}). We present a new and universally applicable
method that deals with many of these problems through the use of
Monte-Carlo simulations. The method is based on the power spectrum of
pulsar timing residuals, for the simulation of which we present a
method similar to that proposed by \citet{cf84}. We use the proposed
technique to bound the amplitude of the GWB based on the PSR
J1713+0747 data set presented in Chapter \ref{chap:20PSRS}. The
derived limit is: $A < 1.0\times10^{-14}$ for a background with a
spectral index of $\alpha = -2/3$.

\section{Introduction}
As described by \citet{jhv+06} and in \S\ref{ssec:Limits}, the power a
GWB introduces into pulsar timing residuals, is given by Equation
\ref{eq:GWBResPower}:
\[
  P(f) = \frac{A^2}{12\pi^2}\frac{f^{2\alpha-3}}{f_{\rm 0}^{2\alpha}},
\]
where $A$ and $\alpha$ are the dimensionless amplitude and spectral
index of the background and $f$ is the frequency in the timing
residuals. With spectral indices of $\alpha = -2/3$ or steeper (see
\S\ref{sec:GWBSources}), this results in a strongly low-frequency
dominated spectrum in the timing residuals: $\alpha_{\rm Res} =
2\alpha - 3 = -13/3$ or steeper. In contrast, most MSP timing
residuals have power spectra that approximate white noise
(i.e. spectral index of zero), as demonstrated in Figure
\ref{fig:sigmaz}. Thus, when the GWB is present at the limit of
detection, the GWB spectrum will rise above the white noise background
only at the lowest frequencies in the power spectrum. The lowest
frequencies of these power spectra can therefore be used as a
constraint on the amplitude of any GWB present in this data,
irrespective of the overall characteristics of the spectrum.

In order to achieve a proper estimate of the power present at these
lowest frequencies, one requires a spectral power estimator that is
capable of analysing unevenly sampled data with the potential presence
of large gaps and significant deviations in measurement errors between
data points. The spectral estimator must preserve the full spectral
resolution of the data set, i.e. it must be sensitive to power at $f =
1/T$, where $T$ is the length of the data set. The spectral estimator
should furthermore be capable of analysing both the steep red noise of
a GWB and the near-white noise of actual timing residuals. Some such
techniques have been proposed in literature. These include the
discrete Fourier transform \citei{DFT;}{sca89}, which has two major
drawbacks. Firstly, spectral leaking restricts this method to spectral
indices $\alpha_{\rm Res} \ge -2$; secondly, clustering of data can
significantly affect the resulting spectrum \citei{as e.g. shown in
  Figure 1d of}{sca89}. The Lomb-Scargle periodogram (LSP)
\citep{lom76,sca82} is an alternative method based on least-squares
fitting of sine waves to the data.  Spectral leakage also restricts
this method to spectral indices $\alpha_{\rm Res} \ge -2$. A different
approach was described by \citet{gro75b} and \citet{dee84} and
consists of fitting a series of orthonormal polynomials to data
subsets with different lengths. The major advantage of this approach
is its sensitivity to spectral indices down to $\alpha_{\rm Res} =
-7$, but it sacrifices spectral resolution and the frequency scaling
is not clearly defined. The $\sigma_{\rm z}$ stability parameter used
in Chapter \ref{chap:20PSRS} and defined based on the Allen variance
by \citet{mte97}, is similar to the previous one, as it is also
fundamentally defined in terms of third order polynomials (or
particularly, $\ddot{\nu}$, which has a cubic signature in timing
residuals). This method also lacks spectral resolution,
however. Bayesian methods of determining spectral properties (such as
spectral index and amplitude of spectral components) more directly,
are also being developed \citep{vlml08}.

In this chapter, we present a new method for limiting the power in the
GWB. The method is described in \S\ref{sec:GWBLimit:Overview} and is
based on spectral analysis of pulsar timing residuals. To this end, we
present a spectral analysis method for reliable estimation of
low-frequency power in pulsar timing residuals in
\S\ref{ssec:SpectralAnalysis}. The spectral estimates thus obtained
are subsequently added in a weighted sum. The optimal weighting
function is derived in \S\ref{ssec:WienerFilter}. In
\S\ref{sec:1713Limit}, this method is applied to the PSR J1713+0747
data set presented in Chapter \ref{chap:20PSRS}. In
\S\ref{sec:GWBL:Issues} we list a few lines of ongoing research which
may improve the sensitivity of the proposed method. Our findings are
summarised in \S\ref{chap:GWBL:conc}.


\section{Overview of the Method}\label{sec:GWBLimit:Overview}
Because the timing residuals are the sum total of all physical effects
that are not included in the pulsar timing model, the power in these
residuals can provide a limit on the strength of any GWB. Considering
the strong prevalence of low frequencies in the effect of a GWB on
timing residuals (as in Equation \ref{eq:GWBResPower}), the lowest
frequency bins of a power spectrum from timing residuals contain most
information and could be used to derive a stringent bound. The actual
power spectrum of pulsar timing residuals is, however, biased by
uneven sampling, fitting of model parameters and variable error bars
on residuals. This causes a discrepancy between the measured power
spectrum in timing residuals and the analytic prediction for the GWB
effect as given by Equation \ref{eq:GWBResPower}, since it is very
difficult to model these real-world biasing effects analytically. The
method presented here is therefore based on a comparison of the
spectrum of the actual timing residuals with the spectra of many
simulated timing residuals. The simulated timing residuals, consisting
of GWB plus white noise, are analysed in exactly the same way as the
actual timing residuals.

First, a weighted sum of the lowest frequency powers of the timing
data set is determined as the statistic $S_{\rm data}$ from which the
limit will be derived. Secondly, a series of stochastic GWBs at a
trial amplitude, $A_{\rm GWB}$, are added to statistically white
(radiometer) noise and sampled at the SATs of the real data set. After
performing the same parameter fitting as for the real data, the same
statistic is calculated from the simulated, GWB-affected data sets,
resulting in a distribution of statistics $S_{{\rm GWB},i}$. The
amplitude $A_{\rm GWB, 95\%}$ for which 95\% of the simulated
statistics $S_{{\rm GWB},i}$ are higher than $S_{\rm data}$, will be a
2\,$\sigma$ limit on the GWB amplitude.

The different steps of the method will be described
below. \S\ref{ssec:SpectralAnalysis} outlines the spectral analysis
method, \S\ref{ssec:WienerFilter} derives the optimal spectral weighting
formula for determination of the statistic. The Monte-Carlo
simulations are based on the code described in detail in
\citet{hjl+09}.  

\subsection{Spectral Analysis of Pulsar Timing Residuals}
\label{ssec:SpectralAnalysis}
Most commonly, power spectra are obtained based on the discrete
Fourier transform (DFT) of the data \citei{adapted
from}{bra00}\footnote{Notice that the normalisation by $1/N$ in
Equation \ref{eq:DFT} may be omitted depending on the definition
used.}:
\begin{equation}\label{eq:DFT}
  X(\nu_m) = {1\over N}\sum_{n=0}^{N-1} x(t_n) e^{-2\pi i t_n \nu_m}
\end{equation}
for a time series $x(t)$ with $N$ data points. Time and frequency are
discretised as follows:
\begin{eqnarray}
  t_n & = & n\,dt\notag \\
  \nu_m & = & \frac{m}{N\,dt}, 
\end{eqnarray}
with $m$ and $n$ integers running from $0$ to $N-1$. The power
spectral density $P(\nu)$ is then obtained as the squared amplitude of
$X(\nu)$. In order to allow comparison with a theoretical spectral
density, this function needs to be normalised. Furthermore, we are
interested only in the power at positive frequencies, because the
power spectrum is even ($X(\nu) = X(-\nu)$). This follows from the
hermitian property of Fourier transforms and the fact that the sampled
time-series $x(t_n)$ is real valued. Normalising the power spectral
density in case of a one sided power spectrum, is achieved through:
\begin{equation}
  P(\nu_m) = T \left| X(\nu_m)\right|^2
\end{equation}
with $T$ the length of the time series, $x(t)$.

The requirement for equal spacing between samples is not fulfilled in
pulsar timing data. A possible solution for this is provided by the
Lomb-Scargle periodogram \citep{lom76,sca82}, which estimates the
power spectrum of a time series based on least-squares fitting of
sines and cosines. An alternative method that holds our preference for
reasons to be outlined shortly, is to interpolate the data onto an
equally spaced grid \citei{as for example described in}{ptvf92}.

\subsubsection{Aliasing and Smoothing}
\begin{figure}
  \centerline{\psfig{figure=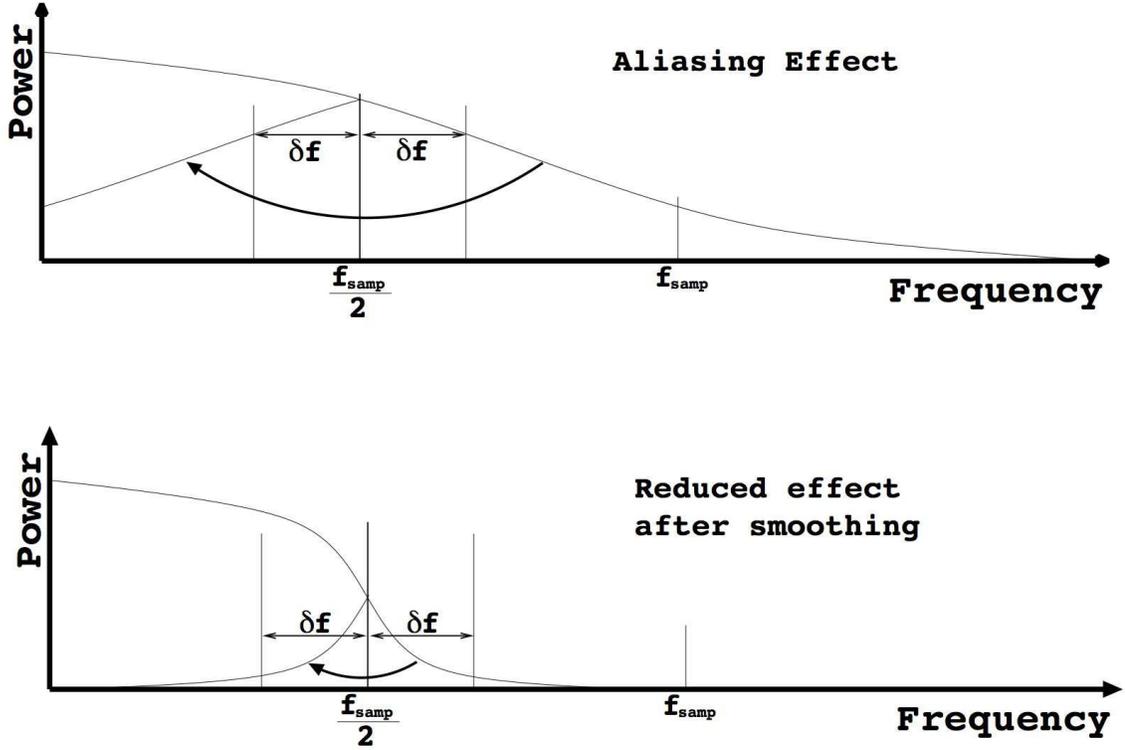,width=15cm,angle=0.0}}
  \caption[Aliasing effect in power spectra]{Aliasing in power
  spectral analysis. If a process contains power at frequencies higher
  than half the sampling frequency $f_{\rm samp}$, then that power
  will be mirrored into lower frequencies of the spectrum. This
  mirroring is displayed in the top image. This effect can be far
  reduced by smoothing the time signal at timescales up to $t_{\rm smooth} =
  2/f_{\rm samp}$, as shown in the bottom image. The effect of this
  smoothing will depend on the spectral characteristics of the
  smoothing filter and the spectral leakage properties of the spectral
  analysis method.}
  \label{fig:Aliasing}
\end{figure}

\begin{figure}
  \centerline{\psfig{figure=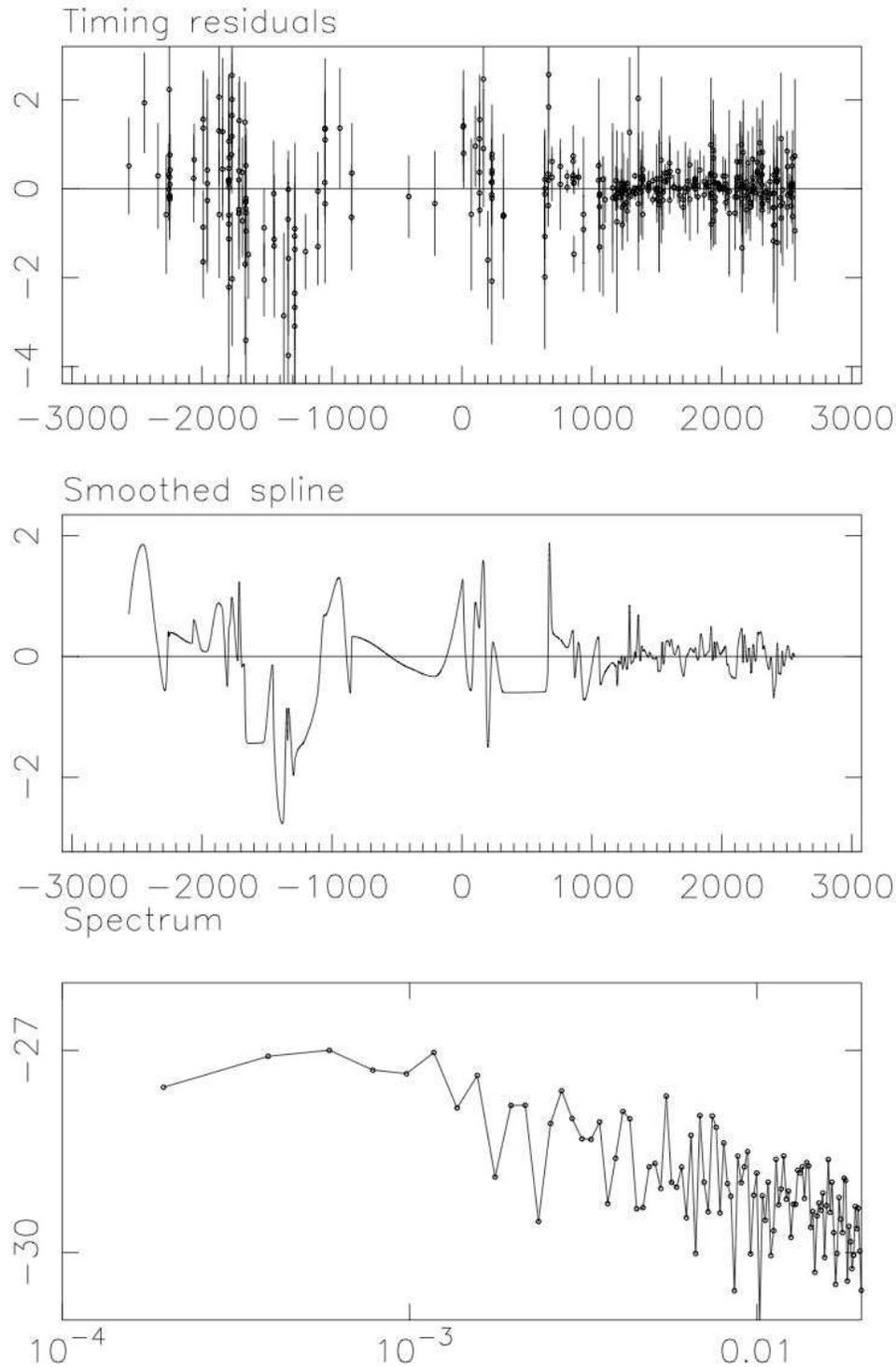,width=13cm,angle=0.0}}
  \caption[Spectral analysis of PSR J1713+0747 timing
    residuals]{Spectral analysis of timing residuals from PSR
    J1713+0747. Top: PSR J1713+0747 timing residuals. Middle: smoothed
    spline interpolation of the timing residuals shown above. Bottom:
    lowest frequencies of the power spectrum of PSR J1713+0747. Units
    of the X-axes are days from the centre for the top two figures,
    days$^{-1}$ for the bottom figure. Units of the Y-axes are $\mu$s
    for the top two figures and yr$^{3}$ for the bottom figure. Since
    this spectral analysis is unweighted, the power spectral estimates
    shown in the bottom plot are $\chi^2$ distributed with two degrees
    of freedom. This means that the uncertainties can be obtained by
    scaling of the displayed curve, as can e.g. be seen in Figure
    \ref{fig:CompSpec}.}
  \label{fig:1713Spec}
\end{figure}

\begin{figure}
  \centerline{\psfig{figure=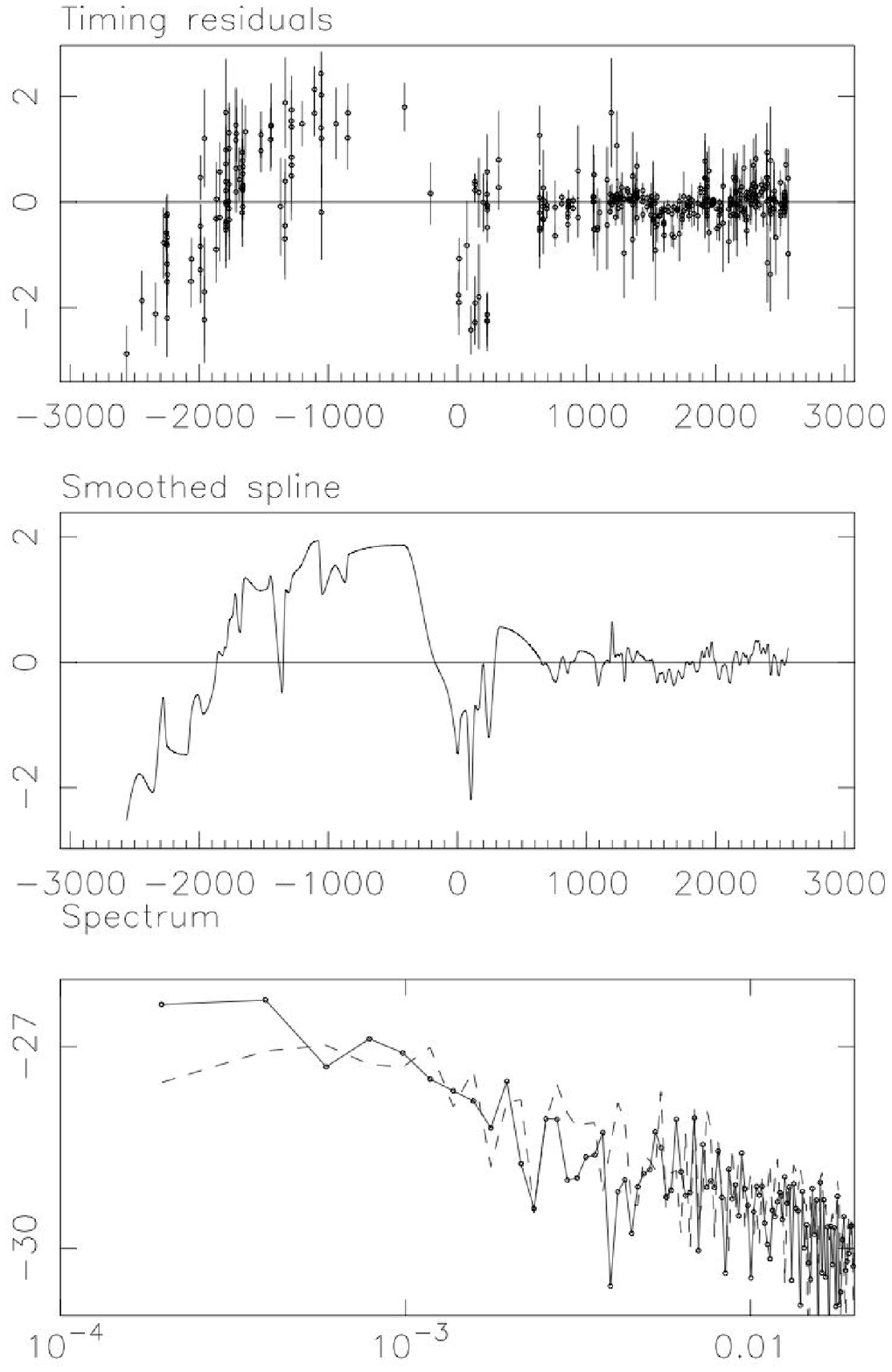,width=14cm,angle=0.0}}
  \caption[Spectral analysis of PSR J1713+0747 data plus simulated
	GWB]{Spectral analysis of the actual timing residuals from PSR
	J1713+0747 with a simulated GWB of $\alpha = -2/3$ and $A =
	1.1\times 10^{-14}$ added. Plots and units identical to those of
	Figure \ref{fig:1713Spec}. The dashed line in the bottom plot is
	the spectrum of the pulsar data shown in Figure
	\ref{fig:1713Spec}.}
  \label{fig:GWBSpec}
\end{figure}

According to the Nyquist sampling theorem, the sampling frequency
$f_{\rm samp}$ of a signal must be twice as high as the highest
frequency present in that signal. Pulsar timing data is typically only
sampled at weekly to monthly intervals, but has power at much higher
frequencies. In such an event, power at frequencies higher than
$f_{\rm samp}/2$ will be aliased to lower frequencies according to the
formula $f_{\rm measured} = |f_{\rm data} - f_{\rm samp}|$. By means
of example, power at a frequency $f_{\rm in} = f_{\rm samp}/2 + \delta
f$, will be aliased down to $f_{\rm measured} = f_{\rm samp}/2 -
\delta f$, as shown in the top image of Figure \ref{fig:Aliasing} and
more fully discussed in \citet{bra00}. This aliasing effect is only
significant at the high-frequency end of the spectrum since fitting
for pulse phase, pulse frequency and pulse frequency derivative
strongly reduce the power at the lowest frequencies; moreover, the
power spectrum is an even function as the time series is real
valued. To limit the aliasing effect at higher frequencies, we apply a
boxcar smoothing algorithm as described in \citet{ptvf92}. The
combined effect of this smoothing on the earlier mentioned
interpolation, can bee seen in Figure \ref{fig:1713Spec} for the
actual pulsar data and in Figure \ref{fig:GWBSpec} for a simulated
data set with a GWB included.


\subsubsection{Prewhitening and post-reddening}
Another difficulty in spectral analysis of pulsar timing residuals is
the potential spectral steepness of the data. While MSPs are mostly
spectrally white (i.e.  equal power at all frequencies - as shown in
Chapter \ref{chap:20PSRS}), the spectra of young pulsars can be much
steeper \citei{see, e.g.}{kop99} and residuals with a simulated GWB
are also expected to be steep, as shown in Equation
\ref{eq:GWBResPower}. Such steep spectra will cause problems with
spectral leakage \citep{bra00}. This arises as a consequence of the
discrete sampling of the time series. Effectively, sampling equals a
multiplication of the continuous time series with a set of delta
functions. As multiplication in the time domain is equivalent to
convolution in the Fourier domain, the effect of sampling on the
Fourier transform of the data (and by extension on the power spectrum)
is a convolution with a sinc-like function, the shape of which depends
on the actual characteristics of the sampling function. This
convolution causes power to be redistributed across the spectrum,
which has no significant effect in the case of a white noise
spectrum. In any other case however, power will seep from frequencies
with high power to frequencies with lower powers. In the case of a DFT
or LSP, this causes spectral features to be flattened to have a
maximal spectral slope of two. A possible remedy is to taper the
sampling function, but this reduces the effective length of the data
set and therefore the sensitivity to GWs.

An alternative solution is provided by the combined application of
(first or second order) prewhitening before spectral analysis and
post-reddening of the obtained spectrum.  In first order prewhitening,
data points are replaced by their difference with the next
point. Effectively the resulting data set is the first time derivative
of the time series and will therefore be less steep.  After the power
spectrum of this new data set is calculated, it will be artificially
steepened to undo the effect of the prewhitening.  In this way, the
DFT and Lomb-Scargle analyses are applicable to spectra with $-2 >
\alpha_{\rm Res} > -4$ when using first order prewhitening.  Second
order prewhitening applies the same differencing method twice,
extending the applicability of the spectral analysis to include $-4 >
\alpha_{\rm Res} > -6$, covering all spectral indices expected in the
analysis of GWBs and MSPs. While this method works well for spectrally
uniform data sets, a combination of (white) receiver noise and a (red)
GWB would require different levels of prewhitening at different
frequencies.  This could be accommodated through correct application
of low- and high-pass filters and subsequent combination of the
resulting spectra.  In the present case, however, the receiver noise
in the pulsar timing data is at or above the level of the GWB power,
even in the lowest frequency bins.  If a realistic level of white
noise is added to the simulated GWB and sampling and fitting effects
are taken into consideration, the effect of the GWB on the simulated
data set is insufficient to warrant any prewhitening in most (if not
all) cases.

\subsubsection{Power Spectrum}
After spline fitting, smoothing, resampling on a regular grid and
potential prewhitening, the power spectrum can be analysed with any
type of analysis technique. In order to avoid unneeded complexities,
we apply a DFT and subsequently add the real and imaginary parts in
quadrature to determine power. Because the data are now evenly spaced,
the LS approach provides identical results. An example of the PSR
J1713+0747 timing residual spectrum is given in the bottom plot of
Figure \ref{fig:1713Spec}. If a GWB with spectral index $\alpha = -2$
and amplitude $A = 10^{-14}$ is added to these residuals before
refitting the timing model and performing the spectral analysis, the
power spectrum in Figure \ref{fig:GWBSpec} is obtained. The comparison
of these two spectra shows that the background effect is very limited
and - if anything - restricted to the lowest frequency bin.

\subsection{Optimal Weighting Function for the Power Spectral Statistic}
\label{ssec:WienerFilter}
After the power spectrum of the timing residuals is determined, the power
spectral estimates of the lowest few frequency bins are combined into
a statistic that is optimally sensitive to the simulated GWB
effect. Theoretically this statistic could simply be the power in the
lowest frequency bin because the GWB would have its strongest
influence there. Optimally weighted addition of the power in several
frequency bins, however, will reduce the variance on our statistic and
therefore provide a more optimal limit. The optimal weighting can be
thought of as a series of two consecutive actions, namely:
\begin{description}
  \item[Prewhitening:]If the power spectrum of the pulsar timing
  residuals is not spectrally white but has excess low-frequency
  power, then the detection statistic $S$ will be dominated by the
  power in the lowest frequency bin alone. This will limit the
  usability of any pulsar timing data set with some form of red noise
  - whether this noise is due to intrinsic timing irregularities, the
  ISM, the instrumentation or anything else. To normalise the relative
  influence of frequency bins of the pulsar timing data, we prewhiten
  the data by dividing the power spectrum by a model of the residual
  power spectrum of the pulsar data, $\tilde{P}_{\rm model}(\nu)$
  (Note: $\tilde{P}(\nu)$ is the power spectrum after sampling and
  fitting, where $P(\nu)$ is the theoretical input power spectrum. If
  $H$ is the \textsc{Tempo2} transfer function, then $\tilde{P}(\nu) =
  H P(\nu)$.)
  \item[Filtering:]The prewhitened power spectral estimates are
  multiplied by a frequency-dependent filtering function $F(\nu)$
  before being added into the detection statistic $S$. This implies
  $S$ is defined as: 
  \begin{equation}
	S = \sum_{i=0}^{N} \left(
	\frac{\tilde{P}(\nu_i)}{\tilde{P}_{\rm model}(\nu_i)} F(\nu_i)
	\right)
  \end{equation} 
  where $N$ is the number of frequency bins to be added.
\end{description}

The derivation of the optimal filter $F(\nu)$ follows below. Since
this filter optimises the sensitivity to a GWB, it is identical for
the current purpose of limiting the GWB strength and for actual
detection of the GWB. It therefore also provides the optimal weighting
function used in Equation \ref{eq:Weights} in Appendix \ref{app:Sens}.

The goal of filtering is to optimise the contribution of the GWB
component in the observed pulsar timing residuals. The filtering
function $F(\nu)$ will therefore be defined so that $\tilde{P}_{\rm
obs}(\nu)F(\nu)$ will approach $\tilde{P}_{\rm GWB}(\nu)$. Based on a
standard $\chi^2$ minimisation, this means that
\begin{equation}
  \sum_{i=0}^{\infty}\left[ \tilde{P}_{\rm obs}(\nu_i) F(\nu_i) - 
	\tilde{P}_{\rm GWB}(\nu_i)\right]^2
\end{equation}
has to be minimised. Applying prewhitening as described above, this
minimisation becomes:
\begin{equation}\label{eq:WFilterBegin}
  \sum_{i=0}^{\infty}\left[ \frac{\tilde{P}_{\rm
  obs}(\nu_i)}{\tilde{P}_{\rm model}(\nu_i)} F(\nu_i) - \frac{\tilde{P}_{\rm
  GWB}(\nu_i)} {\tilde{P}_{\rm model}(\nu_i)}\right]^2.
\end{equation}
Derivation of this equation with respect to $F$ at each frequency
gives:
\begin{equation}
  2\left( \frac{\tilde{P}_{\rm obs}(\nu_i)}{\tilde{P}_{\rm model}(\nu_i)}
  F(\nu_i) - \frac{\tilde{P}_{\rm GWB}(\nu_i)}{\tilde{P}_{\rm
  model}(\nu_i)}\right) \frac{\tilde{P}_{\rm obs}(\nu_i)}{\tilde{P}_{\rm
  model}(\nu_i)} = 0,
\end{equation}
which results in:
\begin{equation}
  F(\nu) = \frac{\tilde{P}_{\rm GWB}(\nu)}{\tilde{P}_{\rm obs}(\nu)}.
\end{equation}
Effectively, this is a Wiener filter as derived in \citet{ptvf92}.

Combining both the prewhitening and the filtering factors, gives the
effective weighting function:
\begin{equation}
  W(\nu) = \frac{\tilde{P}_{\rm GWB}(\nu)}{\tilde{P}_{\rm
  obs}(\nu) \tilde{P}_{\rm model}(\nu)},
\end{equation}
which will be applied to the power spectrum of the pulsar timing
residuals as well as to the power spectra of the post-fit residuals
from the simulated GWBs. 

Practically, $\tilde{P}_{\rm GWB}(\nu)$ is the mean post-fit residual
power spectrum of the Monte-Carlo simulations of the
GWB. $\tilde{P}_{\rm obs}(\nu)$ cannot directly be obtained, since
only a single instance of the timing residual spectrum can be
obtained. We therefore assume the underlying post-fit power spectrum
of the data to be smooth and sufficiently approximated by an analytic
model, i.e. $\tilde{P}_{\rm obs}(\nu) \approx \tilde{P}_{\rm
model}(\nu)$. This results in:
\begin{equation}\label{eq:FinalWeights}
  W(\nu) = \frac{\tilde{P}_{\rm GWB}(\nu)}{\tilde{P}_{\rm
  model}^2(\nu)},
\end{equation}
which is of the same form as the optimal weighting function used for
detection of a GWB, used in Equation \ref{eq:Weights}. The detection
statistic now becomes simply:
  \begin{equation}\label{eq:DetectionStatistic}
	S = \sum_{i=0}^{N} \left( \tilde{P}(\nu_i) W(\nu_i) \right).
  \end{equation} 

This weighting works identically to the whitening method proposed by
\citet{jhlm05} and can easily be understood intuitively. Consider a
residual spectrum with power $\tilde{P}(\nu) = \tilde{P}_{\rm
GWB}(\nu)+ \tilde{P}_{\rm WN}(\nu)$, where $\tilde{P}_{\rm GWB}$ is
the GWB contribution to the pulsar spectrum and $\tilde{P}_{\rm WN}$
is the Gaussian (white) noise in the data. Now define the corner
frequency $\nu_{\rm c}$ at which $\tilde{P}_{\rm GWB}(\nu_{\rm c}) =
\tilde{P}_{\rm WN}(\nu_{\rm c})$. Then, due to the steepness of the
GWB, $\tilde{P}_{\rm GWB}(\nu) \gg \tilde{P}_{\rm WN}(\nu)$ for all
$\nu < \nu_{\rm c}$ and therefore:
\begin{eqnarray}
  \tilde{P}(\nu) \approx & \tilde{P}_{\rm GWB}(\nu) & {\rm if\ }\nu
  \ll \nu_{\rm c}\\ 
  \tilde{P}(\nu) \approx & \tilde{P}_{\rm WN}(\nu) & {\rm if\ }\nu \gg
  \nu_{\rm c}
\end{eqnarray}
This implies that the detection statistic can be split into two sums:
\begin{equation}
  S = \sum_{0}^{\nu_{\rm c}} \tilde{P}_{\rm GWB}(\nu) W(\nu) +
  \sum_{\nu_{\rm c}}^{\infty} \tilde{P}_{\rm N}(\nu) W(\nu).
\end{equation}
Which further simplifies to:
\begin{equation}
  S = \sum_{0}^{\nu_{\rm c}} \frac{\tilde{P}_{\rm
  GWB}^2}{\tilde{P}_{\rm GWB}^2}
  + \sum_{\nu_{\rm c}}^{\infty}\frac{\tilde{P}_{\rm
  GWB}}{\tilde{P}_{\rm WN}}.
\end{equation}
Effectively, this means that any frequency channel where the GWB power
is dominant adds as unity, while any other channel only adds according
to its SNR. Without weighting or whitening, however, each channel
simply adds its power, which always results in a strong domination of
the lowest frequency bin and consequentially a saturation of the
detection significance for increasing GWB amplitudes.

\subsection{Measurement Uncertainty of the Detection
  Statistic}\label{sec:SUnc}

As described at the start of \S\ref{sec:GWBLimit:Overview}, the
detection statistic derived from the data ($S_{\rm data}$) will be
compared to the Monte-Carlo-derived distribution of statistics
($S_{\rm sim, GWB}$). In performing this comparison, no direct
measurement uncertainty of $S_{\rm data}$ is determined. The
justification for this is that the measurement $S_{\rm data}$ is
treated as a single realisation of a distribution. Given this
realisation, we can assess the likelihood of a given distribution of
detection statistics resulting from the Monte-Carlo
simulations. Because our analysis treats the timing residuals in an
unweighted way, each detection statistic calculated from the
simulations will have identical error bars to those applicable to the
measurement of the actual data and therefore the shape of the
distribution from which $S_{\rm data}$ is derived, will be identical
to the shape of the distribution derived from $S_{\rm sim, GWB}$,
irrespective of the sources of timing residuals used in the
simulations.

A related point concerns the inclusion of TOA uncertainties in the
Monte-Carlo simulations. Since the limit method proposed in this
chapter effectively compares power levels and since any additional
source of timing residuals only increases these power levels, the most
conservative limit on the GWB will be obtained from simulations that
contain nothing but a GWB as source of timing residuals. As will be
outlined shortly, we will add white noise at a fraction of the TOA
error bars to this GWB. This addition will strengthen the limit since
a lower level of GWB will be needed to achieve the same power levels -
or, by extension, detection statistic. We will, however, be careful to
only add white noise at a level that is well below the level that can
reasonably be expected to be present in the data, in order to avoid
putting an overly optimistic limit on the GWB. While this implies the
level of white noise in the simulations is lower than that in the real
data, this will not affect the shape of the distribution of detection
statistics, but will only shift it to lower values - pushing the GWB
amplitude limit higher.

\section{A New Limit on the Amplitude of the GWB}\label{sec:1713Limit}
In this section we apply the limiting method described in the
preceding section to the most precise long-term timing data set
presently available - the PSR J1713+0747 timing presented in Chapter
\ref{chap:20PSRS}. While the method consists of iterating over a
series of GWB amplitudes to determine the amplitude that results in a
95\% confidence limit, throughout this section we will illustrate the
procedure using a GWB with amplitude $A = 1.1\times 10^{-14}$ and
spectral index $\alpha = -2/3$.

The spectrum of the PSR J1713+0747 data set after smoothing and
interpolating can be seen in Figure \ref{fig:1713Spec}. The relative
steepness of this spectrum at higher frequencies is mainly due to a
combination of the applied smoothing, \textsc{Tempo2} fitting and
sampling effects. At the lower frequencies that are presently of
interest, however, this spectrum is fairly well modelled with a
uniform (i.e. white) spectrum, so that $\tilde{P}_{\rm model}$ in the
weighting function becomes inconsequential.

After determining the spectral model of the pulsar data, we create
fake data sets that contain a GWB with a given amplitude, to which we
apply the same model fitting and spectral analysis methods as were
applied to the pulsar data. However, because of the prior knowledge
that radiometer noise exists in our data and because the least-squares
fitting routines may not function properly in the presence of the
steep red noise of the GWB, we add some amount of Gaussian scatter to
the timing residuals of the simulated GWB. Because this addition will
increase the spectral power levels and therefore lower the limit on
the GWB amplitude, we have to make sure that the level of added white
noise is realistic, if not conservative. To that purpose, we add white
noise at half the level of the TOA error bar. This amount can be
considered conservative when taking into account the analysis of
\S\ref{sec:stability}, which separated the effects of radiometer noise
from the total timing RMS. As an example, Figure \ref{fig:GWBSpec}
shows the post-fit residuals and spectrum of a GWB added to the actual
PSR J1713+0747 data set. The white noise in the simulations used to
derive the limit on the GWB amplitude, will be half of what is present
in the residuals shown here.

\begin{figure}
  \centerline{\psfig{figure=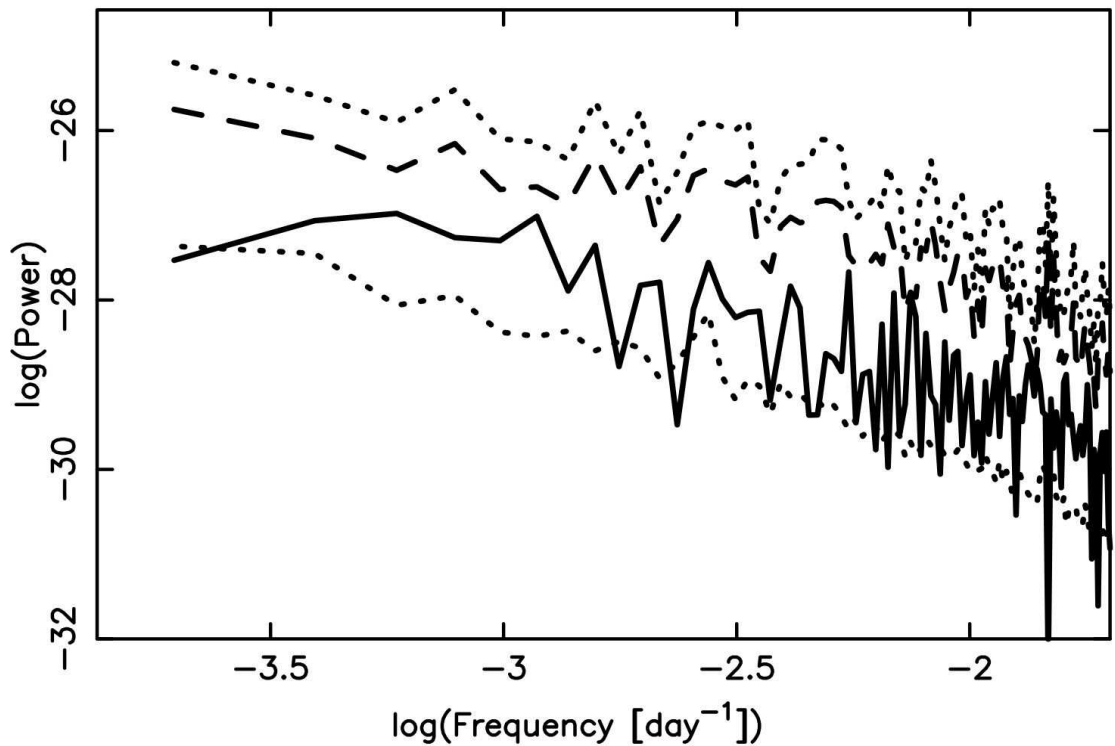,width=15cm,angle=0.0}}
  \caption[Comparison of data spectrum with average GWB
  spectrum]{Comparison of the power spectrum from the data with the
  average power spectrum of simulated data sets with a GWB
  introduced. The full line shows the spectrum of the PSR J1713+0747
  data set previously shown in Figure \ref{fig:1713Spec}. The dashed
  line shows the average post-fit spectrum of 5000 simulated pulsar
  timing data sets with sampling and model fitting identical to that
  of the PSR J1713+0747 data set and with the effects of a GWB with
  amplitude $A = 1.1\times 10^{-14}$ and spectral index $\alpha =
  -2/3$ included. The dotted lines show the 90\% confidence interval
  on this spectrum. This implies that in each frequency bin, 5\% of
  realisations fall below the lowest dotted line, which can therefore
  be used as a lower bound at 95\% certainty on the power spectral
  densities of the simulated data. These spectra are based on the DFT,
  after spline interpolation and smoothing on a timescale of 30
  days. White noise at half the level of the TOA error bars was added
  to the simulated data sets, but this is not visible in this graph
  due to the combined effects of leakage, model fitting and
  smoothing.}
  \label{fig:CompSpec}
\end{figure}

\begin{figure}
  \centerline{\psfig{figure=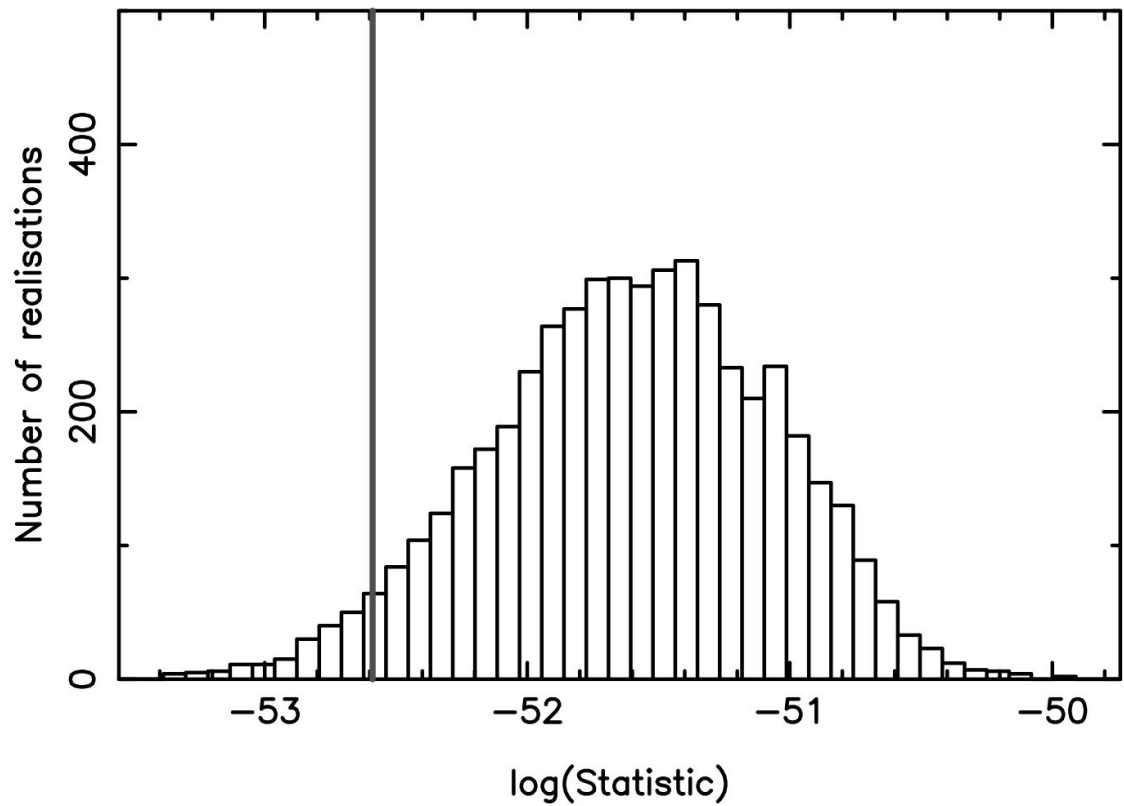,width=15cm,angle=0.0}}
  \caption[Histogram of detection statistics]{Histogram of the
  detection statistics from 5000 simulated pulsar timing data sets
  that contain the same GWB as described in the caption of Figure
  \ref{fig:CompSpec}. The vertical line indicates the pulsar
  statistic. 96\% of the statistics from simulated data sets are
  higher than this pulse statistic.}
  \label{fig:StatHist}
\end{figure}

\begin{figure}
  \centerline{\psfig{figure=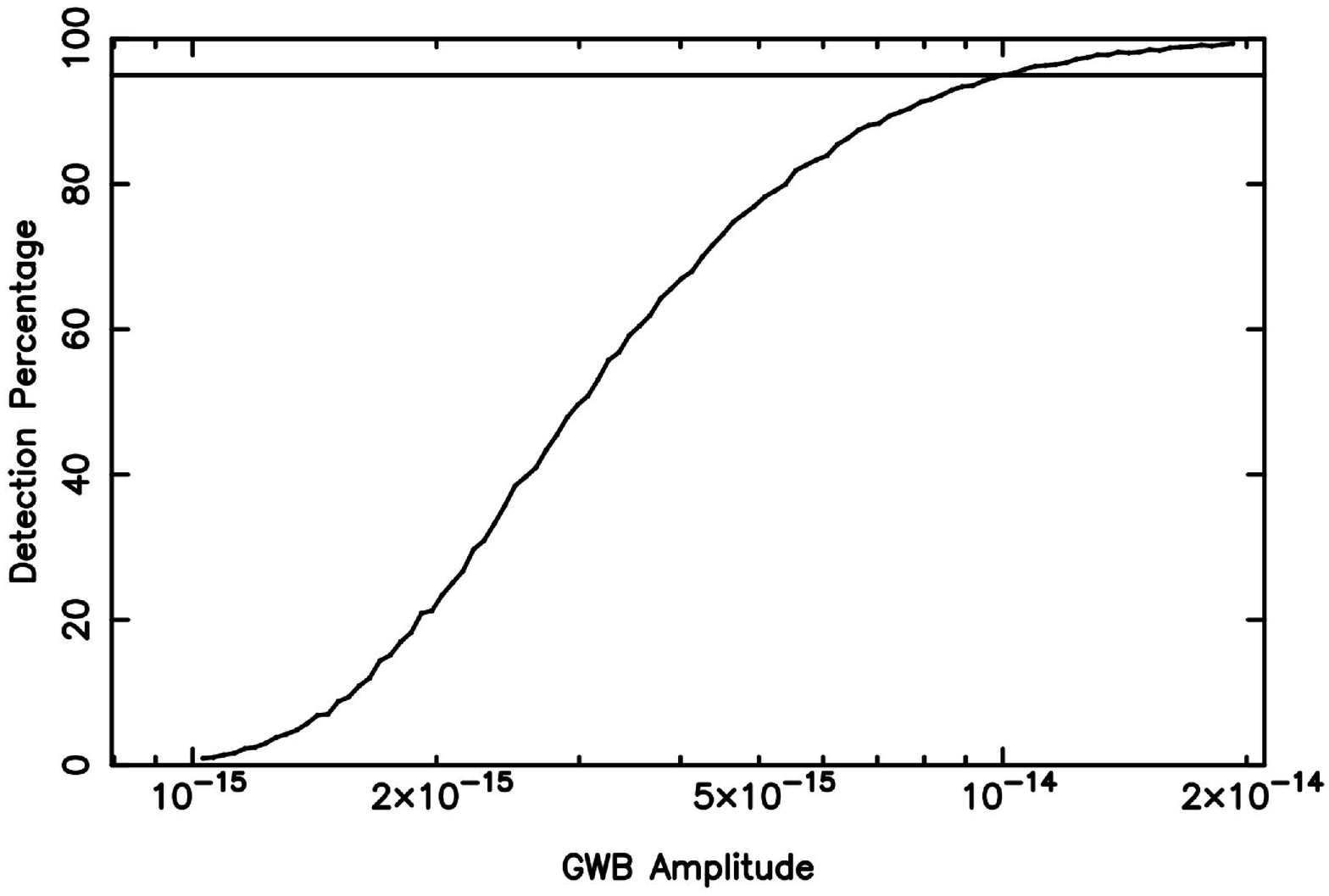,width=15cm,angle=0.0}}
  \caption[Detection statistic versus amplitude]{Detection statistic
	versus amplitude. As the GWB amplitude increases, the number of
	simulations that result in a detection increases. This increase
	is smooth, suggesting the uncertainty in the detection percentages
	is small. The curve crosses the 95\% threshold at an amplitude
	around $1.0\times 10^{-14}$.}
  \label{fig:DetPct}
\end{figure}

The simulations of a large number of GWB-affected timing residual data
sets is followed by the derivation of an average spectrum for the
post-fit GWB spectrum, $\tilde{P}_{\rm GWB}$, along with its
confidence interval. For our example GWB, such a spectrum is shown in
Figure \ref{fig:CompSpec}, along with the spectrum from the original
PSR J1713+0747 data set. Using this average spectrum, we can now
determine the optimal weighting function defined in Equation
\ref{eq:FinalWeights} and calculate the detection statistic from
Equation \ref{eq:DetectionStatistic} for both the real data set and
the simulated GWB-affected data sets. A histogram of these statistics
can be drawn for each GWB amplitude (see, e.g. Figure
\ref{fig:StatHist}). The detection percentage at a given GWB amplitude
is the fraction of Monte-Carlo realisations that results in a statistic
that is larger than the data statistic. The amplitude for which
$S_{\rm data} < S_{\rm sim,GWB}$ for 95\% of the time, will be our
limit. In Figure \ref{fig:DetPct}, the detection percentage is plotted
for a range of GWB amplitudes, demonstrating the smooth relation
between detection percentage and GWB amplitude. This figure also shows
the detection percentage reaches 95\% at an amplitude of $1.0\times
10^{-14}$ - which is just lower than the limit of $1.1\times 10^{-14}$
placed by \citet{jhv+06}, but still outside the predicted range of
$10^{-15} - 10^{-14}$. 


\section{Ongoing Research}\label{sec:GWBL:Issues}
The previous section demonstrated the basic functionality of the limit
method presented. The method may, however, still be optimised in a few
ways. These will be discussed briefly below.

\begin{description}
\item[Spectral analysis method:]{The analysis performed in the
  previous section used a DFT-based spectral analysis as described in
  \S\ref{ssec:SpectralAnalysis}. However, since the PSR J1713+0747
  data set is very long and has reasonable sampling regularity, the
  effects of aliasing may be limited. This would imply smoothing is
  not required and that the LSP could be used without the need for
  interpolation of the data. This approach could affect the resulting
  limit, but the difference between the two approaches is difficult to
  analyse analytically. A quantification of the relative merits based
  on simulations may therefore be in order.}
\item[Pulsar model spectrum:]{Modelling the noisy power spectrum of a
  pulsar can be a relatively complicated exercise once the spectrum is
  not (close to) white as it is in the case of PSR
  J1713+0747. Formulations based on sums of exponential functions and
  white noise floors provide a reasonable analytic basis for such
  models \citei{such functions have been used to model pulsar timing
  noise, see e.g.}{cd85,kop99}, but these ignore the fact that model
  fitting for pulse phase, frequency and frequency derivative always
  depresses the power in the lowest frequency bin.}
\item[Accurate estimation of timing residual sources:]{As noted in
  \S\ref{sec:1713Limit}, every additional source of noise that can be
  added to the Monte-Carlo simulations, will lower the derived limit
  on the GWB amplitude. As an illustration, the limit derived from
  adding white noise at 50\% of the TOA error bars is $1.0\times
  10^{-14}$, which drops to $0.9\times 10^{-14}$ when 75\% of the
  error bars is included and to $0.8\times 10^{-14}$ when the TOA
  error bars are taken at face value. While the relative difference
  between these values may look insignificant at this stage, the
  increase of timing precision expected from the new generation of
  observational hardware may in a few years push these values down to
  levels close to the bottom of the predicted range, making every
  justifiable decrease important. Beyond addition of receiver noise,
  accurate spectral estimates of interstellar effects may also be
  determined and included in the simulations (when comparing to
  DM-uncorrected data such as e.g. the uncorrectable archival data),
  further lowering the limit.}
\item[Weighted spectrum:]{While the uncertainties on the pulsar timing
  residuals are not uniform, the spectrum analysis techniques we
  presented generally treat them as such. This does not invalidate the
  technique, since the treatment of the GWB-affected data is identical
  (and the white noise added to those TOAs is also not constant), but
  a weighted approach may provide a more sensitive limit. Again, an
  objective quantification of the difference in these approaches is
  not trivial.}
\item[Combination of pulsar statistics:]{The limit we derived is
  identical to that from \citet{jhv+06}, although those authors use a
  combination of seven pulsars, as opposed to only PSR J1713+0747. It
  is conceivable that an optimal way of combining the detection
  statistics of several pulsars will further improve our bound, but
  such a derivation is statistically complex because the pulsars are
  statistically heterogeneous.}
\end{description}
 
\section{Conclusions}\label{chap:GWBL:conc}
Ultimately the most reliable upper bound on the GWB amplitude will be
obtained through cross correlation of timing residuals from different
pulsars, as implicitly intended in PTA projects. The sensitivity to
the GWB, however, is a very strong function of the length of data
sets. This makes long time series currently more sensitive than
correlation analysis of more, but shorter data sets.

We have therefore presented a new and conceptually simple method to
use pulsar timing data for placing a limit on the amplitude of the
GWB. We have also applied this method to one of the longest and most
precise data sets presented in Chapter \ref{chap:20PSRS}, on PSR
J1713+0747. As opposed to the method proposed by \citet{jhv+06}, our
method does not make any assumptions about the timing data and can
therefore be applied to any data set. As this method is based on the
power spectrum of pulsar timing residuals, we have described a
rigorous algorithm that allows the lowest frequencies of timing
residuals to be analysed, notwithstanding potential issues caused by
sampling effects and excess red noise. A Monte-Carlo simulation of the
influence of a GWB on the pulsar timing residuals lies at the core of
the technique, simplifying statistical arguments that caused problems
for earlier methods. Our application of this method to the PSR
J1713+0747 data set resulted in a limit of $A = 1.0\times 10^{-14}$
for a background with $\alpha = -2/3$, which is just below the
strongest limit placed to date, by \citet{jhv+06}. A few ongoing lines
of research that may improve the sensitivity of the technique, have
been proposed.

\chapter{Discussion and Conclusion}
\label{chap:conclusion}
\noindent \textsf{... to boldly go where no man has gone before.\\}
\vspace{0.25cm}
\textit{William Shatner (as Captain Kirk), Star Trek, 1966-1969}
\vspace{1.5cm}

\section{Introduction}
This thesis considers the stability of MSPs over timescales of five to
fifteen years, with the broader aim of feasibility studies for
gravitational wave (GW) detection through pulsar timing arrays
(PTAs). In this chapter, our most important conclusions are summarised
(\S\ref{sec:Conc:Conc}) in close combination with proposed extensions
and improvements to this research (\S\ref{sec:Conc:FutRes} and
\S\ref{sec:Conc:Stat}). We also list some of the more exciting
prospects that can be expected in light of this research
(\S\ref{sec:Conc:FutRes}). We end in \S\ref{sec:Conc:Close} with the
answer to the question we set out to ask: whether MSPs will enable
direct detection of GWs.

\section{Summary of Conclusions}\label{sec:Conc:Conc}
Concerning the broader aim of GW detection, we have demonstrated the
following:
\begin{description}
  \item[Timing precision:]{At least some pulsars can be timed at
	precisions of $\sim200$\,ns over time spans of 5 to
	14\,years. Specifically, we achieved 166\,ns over 5.2\,years on
	PSR J1909$-$3744; 199\,ns over 9.9\,years on PSR J0437$-$4715 and
	204\,ns over 14\,years on PSR J1713+0747. Following the analysis
	by \citet{jhlm05} and the scaling laws derived from that analysis,
	this implies that GWB detection over a timescale of a decade or
	less is possible, provided this high level of timing precision can
  be achieved on a high enough number of pulsars.}
  \item[Timing stability:]{A majority of the pulsars analysed (14 out
	of 20) show no convincing signs of instability in a $\sigma_{\rm
	z}$ analysis\footnote{$\sigma_{\rm z}$ is a statistic related to
	the Allen variance. It analyses the stability of time series by
	determining the power present at different timescales.} on
	timescales of 12\,years on average. This ensures sensitivity to a
	GWB will continue to increase as observing with new,
	high-bandwidth backends is continued up to a decade or
	longer. Such stability then enables PTA-style detection of GWs
	over long timescales, even if not enough bright pulsars can be
	found to achieve timing residuals well below $1\,\mu$s, as
	needed for detection efforts over shorter terms.}
  \item[PTA prospects:]{The sensitivity of the radio telescope used
	drastically decreases the duration of the PTA project needed for a
	potential detection. This effect is, however, strongly dependent on
	the ultimate timing precision that might be achievable, but does
	underscore the importance of large telescopes for PTA efforts.}
  \item[Ultimate timing precision:]{Over a period of five years, PSRs
	J1909$-$3744 and J1713+0747 place an upper bound of $\sim80$\,ns
	on intrinsic limitations to timing precision. This demonstrates
	that improved algorithms for mitigation of interstellar effects as
	well as new observing hardware and calibration schemes, may well
	enable sub-100\,ns timing over a time span of five years for the
	brightest pulsars in our sample.}
  \item[Gravitational]{{\bf wave background (GWB) from supermassive black
	hole (SMBH) coalescence:} In order to become sensitive to the
	entire predicted amplitude range from SMBH binary mergers, a PTA
	project will need to be maintained for well over five years, given
	current levels of timing precision and simple scaling laws (as
	shown in Figure \ref{fig:PTAs}). Five years of observations with
	the world's largest telescopes might, however, already provide
	sensitivity to a large fraction of the predicted range.}
\end{description}

A few limits and measurements were made that in themselves or in
combination with results from other authors, provide interesting input
into some scientific discussions and investigations. These are
summarised below:
\begin{itemize}
\item Limit on the variability of Newton's gravitational constant
  (Equation \ref{eq:GdotLimitJoris}): $\dot{G}/G =
  \left(-5\pm18\right)\times 10^{12}\,{\rm yr}^{-1}$. In combination
  with the VLBI parallax of PSR J0437$-$4715 measured by
  \citet{dvtb08}, this limit was improved to $\dot{G}/G =
  \left(-5\pm26\right)\times 10^{-13}$\,yr$^{-1}$.
\item Mass of PSR J0437$-$4715: $M_{\rm psr} =
  1.76\pm0.20\,M_{\odot}$.
\item Limit on the GWB strength from PSR J1713+0747, for a background
  with $\alpha = - 2/3$: $A < 1.0\times 10^{-14}$. This is very close
  to the limit derived by \citet{jhv+06}.
\item Anomalous Solar System acceleration: $| a_{\odot}/c | \leq
  1.5\times 10^{-18}$\,s$^{-1}$ (95\% certainty). This excludes the
  presence of any Jupiter-mass trans-Neptunian objects (TNOs) within
  117\,AU along the line of sight from the Earth to PSR J0437$-$4715.
\item The formal measurement uncertainties reported by \textsc{Tempo2}
  for the timing model parameters were underestimated by factors
  between 1.3 and 3.7 for binary parameters and factors around ten for
  non-binary parameters. While these are probably conservative
  estimates, they do demonstrate the large impact low-frequency noise
  can have on the reliability of timing results.
\end{itemize}

\section{Lines of Further Research}\label{sec:Conc:FutRes}
In order to expand and improve the analysis presented in this thesis,
several lines of future and ongoing research are proposed below.
\begin{description}
\item[Number of millisecond pulsars (MSPs) in a PTA:]{The standard
  scenario proposed by \citet{jhlm05} required a minimum of 20 MSPs to
  construct a timing array and assumed the timing precision on all
  these pulsars to be equal. This thesis has provided some insight
  into the stability and relative timing precision achievable on the
  20 MSPs of the Parkes PTA. Furthermore, prewhitening and optimised
  weighting methods effectively remove the strict requirement of 20
  MSPs. In order to properly distinguish correlations due to clock
  errors, solar system ephemeris errors and GWB effects, some minimal
  number of pulsars and related distribution across the sky, will
  still be needed, however.  An analysis into optimal inhomogeneous
  PTA scenarios could provide more clarity in this matter and as such
  aid in more optimally allocating the limited resource of observing
  time to a dedicated set of pulsars.}
\item[Surveys:]{The Southern sky has been thoroughly surveyed for all
  kinds of pulsars throughout the late nineties and early
  two-thousands. Nevertheless, PSR J1909$-$3744 - currently the most
  precisely timed pulsar in the PPTA sample - was only discovered in
  2003. Given the computational complexity involved in discovering
  binary pulsars with high spin frequencies and the continuous
  increases in computational power, it can therefore be expected that
  even more bright and stable MSPs that may be of great use to PTA
  efforts, could be found in new and ongoing surveys. This is
  particularly true in the Northern hemisphere, where past surveys
  have either been insensitive to MSPs, or were badly affected by RFI.
  The greater sensitivity of the new generation of broadband pulsar
  backends as well as new, large telescopes such as the 500\,m
  aperture spherical telescope (FAST) and (eventually) the SKA, will
  increase the sensitivity of these surveys, making exotic detections
  of distant but interesting pulsars more likely than ever
  before. Such pulsars will undoubtedly be weaker than many that
  already exist, unless they have large dispersions that have hidden
  them from earlier generation surveys. If they are to feature in
  future timing arrays then it may be necessary to either use very
  large telescopes or use some form of methodology that eliminates or
  attentuates the impact of variable scattering on pulsar timing, such
  as those being pursued by \citet{hs08} and \citet{wksv08}.}
\item[International pulsar timing array (IPTA):]{The data presented in
  this thesis were based on Parkes observations with typically hourly
  integrations. The timing precision of future data sets will be
  improved because of the increased bandwidth of the new generation of
  observing backends. Beyond that, timing precision could further be
  improved by using longer integration times, or by using larger
  telescopes. With current observing intensities of around 40\,hours
  per fortnight for the PPTA (20\,hours of multi-frequency
  observations for DM determination and an additional 20\,hours for
  the actual timing observations), a further increase in observing
  time is unlikely to be granted given oversubscription rates of radio
  telescopes. As PTA-type projects are being undertaken at many
  different observatories across the world, however, a global joining
  of force would have access to a multiple of both the observing time
  and telescope sensitivity currently available to any one of the
  projects individually. This would imply that international
  collaboration may well be the fastest and surest means of securing a
  direct detection of the GWB.}
\item[Ultimate timing precision:]{With the anticipated commissioning
  of several large telescopes in 2012 and 2013 (FAST in the Northern
  hemisphere and MeerKAT and ASKAP in the Southern hemisphere) and
  with the proposed interferometric combination of the five major
  European telescopes under the LEAP project, telescope sensitivity in
  pulsar timing may be expected to reach unprecedented levels. This
  should drastically increase our knowledge of the intrinsic stability
  levels of MSPs and the highest possible timing precision that could
  be achieved. This, in turn, will provide more accurate predictions
  of realistic PTA sensitivity levels and provide strong bounds on
  timing noise in MSPs, which may aid the understanding of neutron
  star interiors and magnetospheres.}
\item[Jumps:]{In combining data from different observing systems or
  telescopes, arbitrary phase offsets are introduced to remove any
  differences in instrumental delays. In the case where data from
  different generations of backend systems are combined, there is
  often no overlapping data available, strongly limiting the accuracy
  with which these offsets can be determined. While attempts at
  achieving reliable jump values between old data sets may not
  succeed, future instrumental changes may take heed and ensure
  sufficient overlap of data sets. At present attempts are being made
  to determine the difference in instrumental delays through
  simultaneous observation of an artificial nanosecond pulse, which
  may - if successful - provide an alternative to long periods of
  overlap.}
\item[Parameter uncertainties:]{In Chapter \ref{chap:0437}, we have
  demonstrated the effect small amounts of low-frequency noise can
  have on the parameter uncertainties that are returned by standard
  pulsar timing software. The cause of this is that the $\chi^2$
  fitting routines currently in use assume that the timing residual
  data is statistically white and therefore don't take any covariances
  between the measurement points into account. A revision of the
  fitting procedure to include such correlation effects is possible
  through correct whitening of the data before attempting to fit the
  timing model. While a computationally expensive Monte Carlo-based
  alternative was used in Chapter \ref{chap:0437}, a more automated
  approach for wider use, is under development.}
\item[Spectral analysis:]{In Chapter \ref{chap:20PSRS}, we have hinted
  at the value reliable spectral analysis of pulsar timing residuals
  can have. The spectral analysis technique presented in Chapter
  \ref{chap:GWBLimit} only provides reliable measures of low-frequency
  power, but has already been applied to place bounds on the GWB. In a
  similar way limits could be placed on pulsar timing instabilities or
  pulsar planets. A different use was presented in Chapter
  \ref{chap:0437}, where the residual spectrum was employed to
  simulate residuals and obtain more reliable parameter
  uncertainties. An investigation of MSP spectra with a variety of
  spectral analysis tools, much like \citet{dee84} and \citet{cd85}
  performed on normal pulsars, may be in order. Also, the expansion of
  the spectral analysis method presented in Chapter
  \ref{chap:GWBLimit} to include the TOA uncertainties, may prove
  useful.}
\item[Bayesian analysis:]{\citet{vlml08} presented a Bayesian approach
  to GW detection with pulsar timing data. While a comparative study
  of the sensitivity of this method and the frequentist approach
  advocated in \citet{jhlm05} and \citet{abc08} would be useful, the
  Bayesian approach may be continued well beyond that. Specifically,
  Bayesian analysis of timing residuals might provide an independent
  and radically different means of estimating model parameters and
  their uncertainties as well as properties of pulsar timing noise or
  even the residual power spectrum. A comparison or partial
  assimilation of Bayesian and traditional methods \citei{as most
  recently described in}{hem06}, may well uncover some unexpected
  results.}
\item[Calibration:]{In Chapter \ref{chap:0437}, we applied the
  polarimetric calibration modelling (PCM) and matrix template
  matching (MTM) techniques developed by \citet{van04a} and
  \citet{van06} to the most recent years of data on PSR
  J0437$-$4715. Application of these methods reduced the residual RMS
  by a factor of about two. While such dramatic improvements are not
  predicted for all pulsars, the wider and more automated application
  of these schemes in future research, should be encouraged.}
\end{description}

\section{Increasing the Statistical Significance}\label{sec:Conc:Stat}
The scientific interest of some of our results would be greatly
increased if they were to make out part of a larger scientific sample
or if observing is continued. This holds specifically for the
measurements listed below.
\begin{description}
\item[Newton's gravitational constant:]{Our limit on the variability
  of Newton's gravitational constant as presented in \citet{vbv+08},
  has already been improved by the VLBI parallax to the PSR
  J0437$-$4715 system and is now within a factor of three to the best
  limit available \citei{from lunar laser ranging
  (LLR);}{wtb04}. Because of the smaller time span over which our
  limit is obtained (ten years as opposed to forty), continued
  observing can be expected to improve this limit beyond that of
  LLR. Given the scaling of our measurement uncertainty with
  $T^{-5/2}$ where $T$ is the time span of the experiment, a further
  decade of data should bring the precision of the 2\,$\sigma$ bound
  on $|\dot{G}/G|$ below $7\times 10^{-13}$\,yr$^{-1}$\footnote{The
  uncertainty in the $\dot{G}/G$ value is currently influenced equally
  much by the $\dot{P}_{\rm b}$ value derived from pulsar timing and
  by the VLBI distance. However, the uncertainty in the VLBI distance
  will scale with the inverse square root of the number of observing
  epochs and is therefore not directly dependent on time.}. Such a
  limit would demonstrate that the measured variability of the AU
  \citep{kb04} is due (at least in part) to systematic effects.}
\item[Pulsar masses:]{The mass of PSR J0437$-$4715, which we
  determined at $1.76\pm0.20\,M_{\rm \odot}$, suggests certain classes
  of equations of state for dense nuclear matter can be
  disproven. This can be substantiated partly by continued observing
  of this pulsar and partly by the discovery of new pulsars, which
  should increase the sample size of pulsars with known (and heavy)
  masses, if these exist. One such discovery has been made since the
  PSR J0437$-$4715 mass was published: the mass of PSR J1903+0327 was
  determined to be $1.74\pm0.04\,M_{\rm \odot}$ \citep{crl+08}.}
\item[Excess accelerations:]{In Chapter \ref{chap:0437}, we have
  placed a bound on the anomalous acceleration of the Solar System in
  the direction of PSR J0437$-$4715. Provided correct error analysis
  is undertaken, the directional sensitivity of this bound can now be
  increased by adding the timing residuals from pulsars presented in
  Chapter \ref{chap:20PSRS}. We interpreted our limit on the apparent
  acceleration in terms of a bound on the existence and mass of
  TNOs. The same principle can, however, be applied to the existence
  of Earth-mass dark-matter haloes that are predicted to exist within
  our Galaxy \citep{dms05}. While the event rate of one of these
  passing close to the Solar System is extremely low, any bound would
  not only restrict their presence near the Solar System, but also in
  areas around the neutron stars concerned.}
\end{description}

\section{Closing Remarks}\label{sec:Conc:Close}
As PTA timing data sets grow in both length and precision, their
combination with new or improved processing techniques will allow
timing of MSPs to grow ever stronger as an astrophysical tool to probe
fundamental physics. Even before completion of the SKA, our knowledge
and understanding of fundamental gravitational theories can be
expected to evolve considerably. Specifically, we note that a
detection of the GWB through pulsar timing looks achievable, since the
timing precision and stability of MSPs is provably sufficient. One
caveat that goes beyond the scope of this thesis, though, is the
existence of this background, since any prediction is only as good as
the assumptions that go into it. If SMBH binaries stall rather than
merge or if the predicted properties of the GWBs are significantly
off, then clearly no detection may be made, though pulsar timing may
be used to prove this point. In case the predicted GWBs do exist, the
timescale for a detection may well be a decade or less, though a
precise value will be dictated by the quantity and quality of new
pulsar discoveries and the efficiency of international collaboration.


\clearpage
\addcontentsline{toc}{chapter}{References}
\bibliographystyle{mnras}
\bibliography{journals,modrefs,psrrefs,crossrefs}


\appendix
\chapter{PTA Sensitivity}\label{app:Sens}
In this appendix we derive a simplified formalism for estimating the
sensitivity of a pulsar timing array (PTA) to a gravitational wave
background (GWB) of given amplitude, $A$. This derivation produces
results equivalent to those resulting from equation (14) of
\citet{jhlm05}, but is more readily implemented and inherently treats
optimal weighting (or prewhitening) of the pulsar power spectra.

The detection statistic is the sample cross covariance of the
residuals of two pulsars $i$ and $j$, separated by an angle
$\theta_{i,j}$:
\begin{equation}\label{eq:R}
  R(\theta_{i,j}) = \frac{1}{N_{\rm s}} \sum_{t = 0}^{T}
  T_{{\rm res},i}(t)\times T_{{\rm res},j}(t)
\end{equation}
(where $N_{\rm s}$ is the number of samples in the cross covariance
and $T$ is the data span.) The expected value of $R(\theta_{i,j})$ is
the covariance of the clock error, which is 100\% correlated, plus the
cross covariance of the GWB, $\sigma_{\rm GW}^2
\zeta(\theta_{i,j})$. The clock error can be included in the fit, but
one must also include its variance in the variance of the detection
statistic. It is better to estimate the clock error and remove it,
which also removes its ``self noise''. So in subsequent analysis we
neglect the clock noise. We model the pulsar timing residuals as a GWB
term and a noise term: $T_{\rm res}(t) = T_{\rm GW}(t) + T_{\rm
N}(t)$, with variances $\sigma_{\rm G}^2$ and $\sigma_{\rm
N}^2$. $\zeta(\theta_{i,j})$ is the cross-correlation curve predicted
by \citet{hd83}, as a function of the angle between the pulsars,
$\theta_{i,j}$:
\[\zeta(\theta_{i,j}) = \frac{3}{2}x \log x - \frac{x}{4} +
\frac{1}{2} \]
in which $x=(1-\cos{\theta_{i,j}})/2$.

Since the detection significance will be limited by the variance in
the sample cross covariance, we consider 
\begin{multline}
  {\rm Var}(R(\theta_{i,j})) \\
  \shoveleft{= {\rm Var}\Bigg(\sum\Big( 
	\frac{(T_{{\rm GW},i}+T_{{\rm N},i})(T_{{\rm GW},j}+T_{{\rm
  N},j})}{N_{\rm s}}\Big)\Bigg)}\\	
  \shoveleft{= \sigma_{{\rm G},i}^2 \sigma_{{\rm G},j}^2 
	            \frac{(1+\zeta(\theta_{i,j})^2)}
				     {N_{\rm s}}
    + \frac{\sigma_{{\rm N},i}^2 \sigma_{{\rm G},j}^2 +
	        \sigma_{{\rm G},i}^2 \sigma_{{\rm N},j}^2}
	       {N_{\rm s}}
    + \frac{\sigma_{{\rm N},i}^2 \sigma_{{\rm N},j}^2}{N_{\rm s}}}.\\
\end{multline}
Which, after prewhitening, becomes (notice our notation 
$\sigma_{\rm PW} = \varsigma$):
\begin{multline}\label{eq:VarSum}
  {\rm Var}(R_{\rm PW}(\theta_{i,j})) \\
  \shoveleft{= \varsigma_{\rm G}^4 \frac{(1+\zeta(\theta_{i,j})^2)}{N_{\rm s}}
	+ \varsigma_{\rm G}^2 \frac{(\varsigma_{{\rm N},i}^2+
	  \varsigma_{{\rm N},j}^2)}{N_{\rm s}} + \frac{\varsigma_{{\rm N},i}^2 
	  \varsigma_{{\rm N},j}^2}{N_{\rm s}}}.\\
\end{multline}
in which we have used $\varsigma_{{\rm G},i}^2 = \varsigma_{{\rm
G},j}^2 = \varsigma_{{\rm G}}^2$, which will be proven shortly.

We derive the GWB power from equations (\ref{eq:h_c}) and
(\ref{eq:GWBPower}), for a GWB with spectral index $\alpha = -2/3$:
\begin{equation}\label{eq:PGWB}
  P_{\rm GWB}(f) = K (f/f_{\rm ref})^{-13/3},
\end{equation}
with $K$ a constant proportional to the amplitude of the GWB and
$f_{\rm ref} = 1\,yr^{-1}$.

Defining the corner frequency, $f_{\rm c}$, as the frequency at which
the gravitational wave power equals the noise power, enables us to use
equation (\ref{eq:PGWB}) to determine the noise power: $P_{\rm Noise} = K
(f_{\rm c}/f_{\rm ref})^{-13/3}$. 

As illustrated by \citet{jhlm05}, the steep spectral index of
GWB-induced residuals implies that large gains in sensitivity can be
achieved through optimal prewhitening of the data. Assesssment of the
variance of both the GWB and noise components of the residuals after
prewhitening, can most easily be done through integration of the
spectral powers, multiplied by the whitening filter, $W(f)$, which is
a type of Wiener filter, designed to minimize the error in the
estimation of $\sigma_{\rm G}$ and is of the form (as derived in \S\ref{ssec:WienerFilter}): 
$W(f) = P_{\rm GWB}/(P_{\rm GWB}+P_{\rm Noise})^2$. Rescaling the
weighting function thus defined, we get:
\begin{equation}\label{eq:Weights}
  W(f) = C \frac{\big(f/f_{\rm ref}\big)^{-13/3}}
  {\big(1+(f/f_{\rm c})^{-13/3}\big)^2}
\end{equation}
with $C$ a normalisation constant chosen for convenience to be:
\begin{equation}\label{eq:C}
  C = \Bigg(\sum_f{\frac{\big(f/f_{\rm ref}\big)^{-26/3}}{\big(1+(f/f_{\rm
  c})^{-13/3}\big)^2}}\Bigg)^{-1}
\end{equation}
The prewhitened variances then become:
\begin{eqnarray}
  \varsigma_{\rm G}^2 &=& \sum_f K (f/f_{\rm ref})^{-13/3} C \frac{
  (f/f_{\rm ref})^{-13/3}}{\big(1+(f/f_{\rm c})^{-13/3}\big)^2}\notag\\
  &=& K \label{eq:sigmaG}\\
  \varsigma_{\rm N}^2 &=& \sum_f K (f_{\rm c}/f_{\rm ref})^{-13/3} C \frac{
	(f/f_{\rm ref})^{-13/3}}{\big(1+ (f/f_{\rm c})^{-13/3}\big)^2}\notag\\
  &=& K C \sum_f{\frac{\big(f_{\rm c} f/f_{\rm
  ref}^2\big)^{-13/3}}{\big(1+(f/f_{\rm c})^{-13/3}\big)}}
\end{eqnarray}
which justifies our choice for $C$ and shows that, based on our
weighting scheme, $\varsigma_{{\rm G},i}^2 = \varsigma_{{\rm G},j}^2 = K$, as
used earlier.

Since the spectra are effectively bandlimited to $f_{\rm c}$ after
prewhitening, both the GWB and noise will have the same number of
degrees of freedom, namely: $N_{\rm dof} = 2 T_{\rm obs}f_{\rm c}-1$,
where $T_{\rm obs}$ is the length of the data span and therefore the
inverse of the lowest frequency, implying there are $T_{\rm obs}f_{\rm
c}$ independent frequencies measured below $f_{\rm c}$. Since each
frequency adds a real and imaginary part, there are twice as many
degrees of freedom as there are independent frequency samples;
quadratic fitting removes a single degree of freedom from the
total. Notice that $N_{\rm dof}$ is the number of independent samples
in the cross-covariance spectrum and therefore replaces $N_{\rm s}$ in
equations (\ref{eq:R}) and (\ref{eq:VarSum}).

The optimal least-squares estimator for $K$ (and hence for the
amplitude of the GWB), based on a given set $R_{\rm PW}(\theta_{i,j})$
with unequal errors, is (from equations \ref{eq:R} and
\ref{eq:sigmaG}) :
\begin{equation}\label{eq:Estimator}
  \tilde{K} = \frac{\sum{R_{\rm PW}(\theta_{i,j})\zeta(\theta_{i,j})}
	/{\rm Var}(R_{{\rm
	PW,}i,j})}{\sum{\zeta(\theta_{i,j})^2/{\rm Var}(R_{{\rm PW,}i,j})}}
\end{equation}
The variance of this estimator is:
\begin{equation}\label{eq:EstVar}
  {\rm Var}(\tilde{K}) = 
  \frac{1}{\sum{\zeta(\theta_{i,j})^2/{\rm Var}(R_{{\rm PW,}i,j})}}
\end{equation}

We can now write the expected signal-to-noise of a given timing array
as the square root of the sum over all pulsar pairs of equation
(\ref{eq:sigmaG}) divided by the square root of equation (\ref{eq:EstVar})
\begin{equation}
  S = \sqrt{\sum_{i=1}^{N_{\rm psr}-1}
    \sum_{j=i+1}^{N_{\rm psr}}
	\frac{\varsigma_{\rm G}^4 \zeta^2 N_{\rm dof}}{\varsigma_{\rm G}^4
	  (1+\zeta^2)+\varsigma_{\rm G}^2(\varsigma_{{\rm N},i}^2+\varsigma_{{\rm
	N},j}^2)+\varsigma_{{\rm N},i}^2 \varsigma_{{\rm N},j}^2}}.
\end{equation}
Rewriting leads to:
\begin{equation}
  S = \sqrt{\sum_{i=1}^{N_{\rm psr}-1}\sum_{j=i+1}^{N_{\rm psr}}
	\frac{\zeta^2 N_{\rm dof}}
		 {1+\zeta^2+
		   \left(\varsigma_i^{\prime}\right)^2+
		   \left(\varsigma_j^{\prime}\right)^2+
		   \left(\varsigma_i^{\prime}
		   \varsigma_j^{\prime}\right)^2}}
\end{equation}
where $\varsigma_i^{\prime} = \varsigma_{{\rm N},i}/\varsigma_{\rm G}$.
\vspace{0.5cm}

\clearpage
\singlespacing
\chapter{Nomenclature}
\label{app:nomen}
 
The following glossary defines the various mathematical symbols and
acronyms used throughout the thesis.

\begin{tabbing}
\hspace{3cm} \= \\
$A$ \> Dimensionless amplitude of the GWB\\
$a$ \> Semi-major axis of the binary orbit \\
$\alpha$ \> Right ascension, RA\\
$A_{\rm e}$ \> True anomaly of the binary orbit \\
$a_{\odot}$ \> Acceleration of the Solar System \\
$a_{\rm i}$ \> Amplitude of the electric field in direction i. \\
A/D \> Analogue-to-digital converter \\
AFB \> Analogue filter bank (also ``FB'') \\
AOP \> Annual-orbital parallax \\
ASKAP \> Australian SKA pathfinder \\
AU \> Astronomical unit ($1\,$AU $= 149597870\,$km) \\
AXP \> Anomalous X-ray pulsar \\
$B$ \> Bandwidth \\
$B$ \> Source brightness \\
$B_{\rm 0}$ \> Magnetic field strength at the surface of the pulsar
(Gauss) \\
$\vec{b}$ \> Vector pointing from the BB to the pulsar \\
BAT \> Barycentric arrival time \\
BB \> Binary barycentre \\
$c$ \> speed of light ($= 3\times 10^8\,$m/s) \\
CPSR/CPSR2 \> Caltech-Parkes-Swinburne recorder versions one and two.\\
$D$ \> Dispersion constant, $D = 4.15\times
10^3$\,MHz$^2$pc$^{-1}$cm$^3$s \\
$\vec{d}$ \> Vector pointing from the SSB to the pulsar or to the BB
\\
$\Delta$ \> Timing delay \\
$\delta$ \> Declination, dec \\
$D_{\rm k}$ \> Kinematic distance \\
$D_{\rm \pi}$ \> Parallax distance \\
$\delta_{x}$ \> Kronecker delta ($\delta_x = 1$ if $x=0$; $\delta_x =
0$ if $x\neq0$).\\
DFT \> discrete Fourier transform \\
$DM$ \> Dispersion measure or integrated electron density \\
$\Delta_{\rm 8}$ \> Stability parameter \\
$E$ \> Fourier transform of $e$ \\
$e$ \> Orbital eccentricity \\
$\vec{e}$ \> Electric field vector, decomposed into $e_{\rm x}$,
$e_{\rm y}$ and $e_{\rm z} = 0$. \\
EOS \> Equation of state \\
EPTA \> European pulsar timing array \\
$f$ \> Observing frequency \\
$f_{\rm ref}$ \> Reference frequency, $f_{\rm ref} = 1$\,yr$^{-1}$.\\
$\phi$ \> Phase \\
FAST \> Five hundred metre aperture spherical telescope \\
FFT \> Fast Fourier transform \\
FPTM \> Fast pulsar timing machine \\
$G$ \> Gain \\
$G$ \> Newton's gravitational constant ($G = 6.67259\times
10^{-11}$\,Nm$^2$kg$^{-2}$) \\
$\gamma$ \> Gravitational redshift parameter \\
GBT \> Green Bank telescope \\
GC \> Globular cluster \\
GW \> Gravitational wave \\
GWB \> Gravitational wave background \\
GR \> General relativity \\
$H_{\rm 0}$ \> Hubble constant \\
$h$ \> Planck's constant ($h = 6.6260755\times 10^{-34}\,$Js) \\
$h_{\rm c}$ \> Characteristic strain spectrum of the GWB \\
$I$ \> moment of inertia of the pulsar ($\approx 10^{45}\,$g cm$^2$)
\\
$i$ \> Inclination angle of the binary orbit. $i = 0^{\circ}$ is seen
as a clockwise rotation; \\
\> $i = 90^{\circ}$ is an edge-on orbit; $i = 180^{\circ}$ is
counter-clockwise rotation. \\
IF \> Intermediate frequency \\
IPTA \> International pulsar timing array \\
ISM \> Interstellar medium \\
Jy \> Jansky, unit of flux density ($1\,$Jy $= 10^{-26}\,$W
m$^{-2}$Hz$^{-1}$) \\
$k$ \> Boltzmann's constant ($k = 1.380658\times 10^{-23}\,$J/K) \\
$\lambda$ \> Wavelength \\
LEAP \> Large European array for pulsars - a PTA based on
interferometric coupling \\
\>of the radio telescopes that form the EPTA. \\
LLR \> Lunar laser ranging \\
LNA \> Low noise amplifier \\
LO \> Local oscillator \\
$M$ \> Mass of the binary system ($M = M_{\rm psr} + M_{\rm c}$) \\
$M_{\rm psr}$ \> Pulsar mass \\
$M_{\rm c}$ \> Mass of the binary companion \\
$M_{\odot}$ \> Solar mass ($1.989\times 10^{30}\,$kg) \\
$\mu$ \> Proper motion (often decomposed in RA and dec: $\mu_{\rm
  \alpha}$; $\mu_{\rm \delta}$) \\
MeerKAT \> Extended Karoo array telescope. The South African SKA prototype \\
MSP \> Millisecond pulsar \\
MTM \> Matrix template matching \\
$\nu$ \> Observing frequency ($\nu$ = $f$) \\
$\ddot{\nu}$ \> Second time derivative of the pulsar spin frequency \\
$n_{\rm e}$ \> Electron density (cm$^{-3}$) \\
$N_{\rm p}$ \> Number of polarisations \\
NANOGrav \> North American nanohertz observatory for gravitational
waves \\
\>(North American pulsar timing array) \\
$P$ \> Pulse period \\
$P$ \> Power \\
$\dot{P}$ \> Spin period derivative, spindown \\
$\vec{p}$ \> Vector pointing from the telescope to the pulsar \\
$\pi$ \> Parallax, PX \\ 
$P_{\rm b}$ \> Binary period (days) \\
$\dot{P}_{\rm b}$ \> First derivative of the binary period, orbital
decay \\
PCM \> Polarimetric calibration modelling \\
PPTA \> Parkes pulsar timing array \\
PTA \> Pulsar timing array \\
$R$ \> Pulsar radius \\
$r$ \> Shapiro delay range \\
$\vec{r}$ \> Vector pointing from the telescope to the SSB \\
$\bar{\rho}$ \> Coherency matrix. Contains the coherency products. \\
RF \> Radio frequency \\
RRAT \> Rotating radio transient \\
$S$ \> Stokes parameters. Contains four components: $I,\ Q,\ U,\
V$. \\
$S$ \> PTA sensitivity (in standard deviations) to a GWB.\\
$S_{\rm 0}$ \> Total intensity \\
$S_{\rm peak}$ \> Brightness of the pulse peak \\
$s$ \> Shapiro delay shape \\
$\sigma$ \> Standard deviation, RMS \\
$\varsigma_{\rm 0}$ \> White noise variance of a pulse profile \\
$\varsigma_{\rm G}$ \> Timing RMS due to the GWB (or due to other
noise sources \\
\>in case of $\varsigma_{\rm N}$, after prewhitening.\\
$\sigma_{\rm G}$ \> Timing RMS due to the GWB. \\
$\sigma_{\rm N}$ \> Timing RMS due to noise sources other than the GWB.\\
$\sigma_{\rm z}$ \> Stability parameter based on Allen variance \\
SAT \> Site arrival time \\
SGR \> Soft gamma repeater \\
SKA \> Square kilometre array \\
SMBH \> Supermassive black hole \\
SNR \> Signal to noise ratio \\
sr \> Steradian, unit of solid angle \\
SSB \> Solar system barycentre \\
SSE \> Solar system ephemerides \\
$T$ \> Time span of the data set, length of observational campaign. \\
$T_{\rm 0}$ \> Time of periastron passage \\
$T_{\rm A}$ \> Antenna temperature \\
$T_{\rm b}$ \> Brightness temperature \\
$T_{\rm N}$ \> Noise temperature \\
$\theta_{i,j}$ \> Angular separation between pulsars $i$ and $j$.\\
$\tau_{\rm c}$ \> Characteristic age \\
TAI \> Temps atomique international - international atomic time \\
TNO \> Trans-Neptunian object \\
TOA \> (pulse) time-of-arrival \\
$u$ \> Eccentric anomaly of the binary orbit \\
$\vec{v}$ \> Pulsar (or binary) velocity. Note that proper motion
$\mu$ \\
\> is the projection of $\vec{v}$ onto the plane of the sky. \\
VLBI \> Very long baseline interferometry \\
$W(f)$ \> Wiener filter for prewhitening of residuals.\\
$\Omega$ \> Longitude of ascending node. Defined from North through
East. \\
$\Omega_{\rm gw}$ \> Energy density of the GWB per unit logarithmic
frequency interval.\\
$\omega$ \> Longitude of periastron \\
$\dot{\omega}$ \> Periastron advance \\
$x = a \sin{i}$ \> Projected semi-major axis of the binary orbit \\
$\zeta(\theta)$ \> Hellings \& Downs correlation at separation
$\theta$. \\
\end{tabbing}


\clearpage
\pagestyle{empty}
\addcontentsline{toc}{chapter}{Publications}
\noindent
{\Large\bf Publications} \\

The following publications appear as chapters in this thesis: \\

\textbf{Precision Timing of PSR J0437$-$4715: An Accurate Pulsar
Distance, a High Pulsar Mass, and a Limit on the Variation of Newton's
Gravitational Constant}
Verbiest, J. P. W.; Bailes, M.; van Straten, W.; Hobbs, G. B.;
Edwards, R. T.; Manchester, R. N.; Bhat, N. D. R.; Sarkissian, J. M.;
Jacoby, B. A. \& Kulkarni, S. R.
\textit{The Astrophysical Journal}
Volume 679, Issue 1, pp. 675$-$680, May 2008 \\

The following papers were prepared with the assistance of the candidate
during the course of this thesis: \\

\textbf{Upper Bounds on the Low-Frequency Stochastic Gravitational
Wave Background from Pulsar Timing Observations: Current Limits and
Future Prospects}
Jenet, F. A.; Hobbs, G. B.; van Straten, W.; Manchester, R. N.;
Bailes, M.; Verbiest, J. P. W.; Edwards, R. T.; Hotan, A. W.;
Sarkissian, J. M. \& Ord, S. M.
\textit{The Astrophysical Journal}
Volume 653, Issue 2, pp. 1571$-$1576, December 2006 \\

\textbf{Dispersion measure variations and their effect on precision
pulsar timing}
You, X. P.; Hobbs, G.; Coles, W. A.; Manchester, R. N.; Edwards, R.;
Bailes, M.; Sarkissian, J.; Verbiest, J. P. W.; van Straten, W.;
Hotan, A.; Ord, S.; Jenet, F.; Bhat, N. D. R. \& Teoh, A.
\textit{Monthly Notices of the Royal Astronomical Society}
Volume 378, Issue 2, pp. 493$-$506, June 2007 \\

\textbf{Using pulsars to limit the existence of a gravitational wave
background} 
Hobbs, G.; Jenet, F.; Lommen, A.; Coles, W.; Verbiest, J. P. W.\&
Manchester, R. 
\textit{40 Years of Pulsars: Millisecond Pulsars, Magnetars and
More. AIP Conference Proceedings}
Volume 983, pp. 630$-$632, February 2008 \\

\textbf{Gravitational-radiation losses from the pulsar$-$white-dwarf
binary PSR J1142$-$6545}
Bhat, N. D. R.; Bailes, M. \& Verbiest, J. P. W.
\textit{Physical Review D}
Volume 77, Issue 12, id. 124017, June 2008 \\

\textbf{Extremely High Precision VLBI Astrometry of PSR J0437$-$4715
and Implications for Theories of Gravity}
Deller, A. T.; Verbiest, J. P. W.; Tingay, S. J. \& Bailes, M.
\textit{The Astrophysical Journal}
Volume 685, Issue 1, pp. L67$-$70, September 2008 \\

\textbf{Gravitational wave detection using pulsars: status of the
Parkes Pulsar Timing Array project}
Hobbs, G. B.; Bailes, M.; Bhat, N. D. R.; Burke-Spolaor, S.; Champion,
D. J.; Coles, W.; Hotan, A.; Jenet, F.; Kedziora-Chudczer, L.; Khoo,
J.; Lee, K. J.; Lommen, A.; Manchester, R. N.; Reynolds, J.;
Sarkissian, J.; van Straten, W.; To, S.; Verbiest, J. P. W.; Yardley,
D. \& You, X. P.
\textit{Publications of the Astronomical Society of Australia}
Accepted for publication in Volume 26, 2009 \\

\textbf{TEMPO2, a new pulsar timing package. III: Gravitational wave
simulation} 
Hobbs, G.; Jenet, F.; Lee, K. J.; Verbiest, J. P. W.; Yardley, D.;
Manchester, R.; Lommen, A.; Coles, W.; Edwards, R. \& Shettigara, C.
\textit{Monthly Notices of the Royal Astronomical Society}
Accepted for publication in 2009 \\

\end{document}